\documentclass[preprint,11pt]{aastex}
\singlespace
 
\newcommand{\be}{  \begin{eqnarray} }
\newcommand{\ee}{  \end{eqnarray} }

\def\crad{c_{\rm r}}
\def\cgas{c_{\rm g}}
\def\vaz{v_{{\rm A}z}}
\def\vaphi{v_{{\rm A}\phi}}
\def\vph{v_{\rm ph}}
\def\msun{{\rm M}_\odot}
\def\lsun{{\rm L}_\odot}

\def\kappaj{\kappa_{\rm J}}
\def\kappap{\kappa_{\rm P}}
\def\kappat{\kappa_{\rm T}}
\def\kappaf{\kappa_{\rm F}}
\def\sigmat{\sigma_{\rm T}}
\def\kb{k_{\rm B}}
\def\omegaa{\omega_{\rm a}}
\def\omegath{\omega_{\rm th}}
\def\kf{{\bf k}\cdot{\bf F}}
\def\thetarho{\Theta_\rho}
\def\thetatg{\Theta_{T{\rm g}}}
\def\thetatr{\Theta_{T{\rm r}}}
\def\thetat{\Theta_T}
\def\spose#1{\hbox to 0pt{#1\hss}}
\def\lta{\mathrel{\spose{\lower 3pt\hbox{$\mathchar"218$}}
     \raise 2.0pt\hbox{$\mathchar"13C$}}}
\def\gta{\mathrel{\spose{\lower 3pt\hbox{$\mathchar"218$}}
     \raise 2.0pt\hbox{$\mathchar"13E$}}}
\font\syvec=cmbsy10                        %for boldface nabla
\font\gkvec=cmmib10                         %for boldface lowercase Greek
\def\bnabla{\hbox{{\syvec\char114}}}       %bold face nabla
\def\bepsilon{\hbox{{\gkvec\char15}}}      %bold face epsilon
\def\bxi{\hbox{{\gkvec\char24}}}           %bold face xi
\begin{document}
 
\title{Local Radiative Hydrodynamic and Magnetohydrodynamic Instabilities
in Optically Thick Media}
 
\author{Omer Blaes and Aristotle Socrates}
\affil{Department of Physics, University of California, Santa Barbara,
CA 93106}

\begin{abstract}
We examine the local conditions for radiative damping and driving of short
wavelength, propagating hydrodynamic and magnetohydrodynamic (MHD) waves in
static, optically thick, stratified equilibria.
We show that so-called strange modes in stellar oscillation
theory and magnetic photon bubbles are intimately related and are both
fundamentally driven by the background radiation flux acting on
compressible waves.  We identify the necessary
criteria for unstable driving of these waves, and show that this
driving can exist in both gas and radiation pressure dominated media, as
well as pure Thomson scattering media in the MHD case.
The equilibrium flux acting on opacity fluctuations can drive both
hydrodynamic acoustic waves and magnetosonic waves unstable.
In addition, magnetosonic waves can be driven unstable by a combination
of the equilibrium flux acting on density fluctuations and changes in
the background radiation pressure along fluid displacements.
We briefly describe the conditions under which these instabilities
might be manifested in both main sequence stellar envelopes and accretion disks.
\end{abstract}

 
\section{Introduction}

It is well known that the presence of a substantial equilibrium radiation
pressure gradient can destabilize optically thick astrophysical media.
Prendergast \& Spiegel (1973) first speculated that compressible fluid
flow subject to a large radiative flux would be unstable to the formation
of buoyant rarefied regions of enhanced radiation pressure, better known as
``photon bubbles.''  The existence of these instabilities has never been
fully demonstrated in hydrodynamic systems, but radiation pressure acting
in stellar envelopes with opacity peaks can drive certain oscillation modes
(now known as ``strange modes'') unstable (e.g. Wood 1976; Saio, Wheeler,
\& Cox 1984; Gautschy 1993; Kiriakidis, Fricke, \& Glatzel 1993).  Strange
modes have also been found to exist in hydrodynamic models of accretion disks
(Glatzel \& Mehren 1996, Mehren-Baehr \& Glatzel 1999).

Gautschy \& Glatzel (1990) showed
that strange mode instabilities satisfy a ``non-adiabatic reversible''
(NAR) approximation, in which the modes occur in a medium with effectively
zero specific heat capacity, resulting in a vanishing luminosity perturbation.
A physical model for strange mode instabilities within the NAR approximation
was established by Glatzel (1994).  He noted that opacity peaks act as a
reflection barrier for high overtone acoustic modes, sonically decoupling
different radial regions in the star.  The basic mechanism for the instability was
the relative phase shift between the pressure and density perturbations,
resulting from the constraint of having zero luminosity perturbation.
Further aspects of the physics of strange modes have been elucidated by
Papaloizou et al. (1997) and Saio, Baker, \& Gautschy (1998).
In numerical studies, Shaviv (2001) has found that both standing and
propagating acoustic waves are unstable in atmospheres with sufficient
radiation pressure support, even when the opacity is pure Thomson
scattering.

Radiation pressure driven instabilities occur in magnetized
systems as well.  Arons (1992) identified and investigated photon bubble
instabilities in strongly magnetized atmospheres of accreting X-ray pulsars.
He showed that such instabilities are caused by the enhanced buoyancy that
occurs when radiation diffuses into relatively rarefied regions in compressible
perturbations.  With accretion disk applications in mind, Gammie (1998)
performed a linear, local stability analysis of
a static, magnetized, stratified medium and found an instability over a
finite range of wavenumbers that he identified as being similar in nature
to photon bubbles.  Using the simplifying assumption that gas and radiation
were coupled together just by Thomson scattering, Blaes \& Socrates (2001,
hereafter BS01)
found that Gammie's (1998) instability could be extended to arbitrarily short
wavelengths where it manifested itself as an overstable slow mode.
In addition, the fast magnetosonic mode could
also be unstable for sufficiently strong radiation fluxes, albeit with
smaller growth rate.\footnote{BS01 also found that Alfv\'en waves could be
destabilized, an effect that is almost certainly due to the fact that they
considered a rotating equilibrium and these waves then have a small
compressible component because rotation couples them
to slow modes.}  Begelman (2001) has constructed a periodic shock
train solution that may represent one possible nonlinear outcome of the slow
mode instability.

The purpose of this paper is to investigate the local, WKB version of these
compressive instabilities in a unified manner.  We focus on the
local driving of propagating waves, and do not address here the issue of whether
and how this local driving can manifest itself as a global instability of
a standing, normal mode of the object in question.  We treat the
thermodynamics of the medium
quite generally, making no restrictive assumptions on the equality of the
gas and radiation temperature in the perturbed state or using the
NAR approximation.  We find that
hydrodynamic instabilities driven by radiative diffusion exist under a wide
range of conditions, for {\it both} gas and radiation pressure dominated media.
However, a medium supported purely by Thomson scattering opacity exhibits no
local driving of acoustic wave instabilities, in contrast to the conclusions
of Shaviv (2001).  On the other hand, the anisotropic
stress produced by magnetic tension widens the applicability of these
instabilities to even broader classes of equilibria, even those which have
opacities given by Thomson scattering only.  The MHD instabilities can exist
even when the equilibrium magnetic pressure is {\it less than} either the
gas or radiation pressure.

All these instabilities may grow on the dynamical
time scale or even faster.  The instability mechanism
originates from
the interaction of the equilibrium radiative flux with density and opacity
fluctuations in the perturbed flow, as well as changes in radiation pressure
along fluid displacements.  The ultimate source of free energy is
provided by the stratified radiation field.

This paper is organized as follows.  In section 2, we describe our basic
equations and assumptions, concentrating in particular on the thermodynamics
of the coupled gas and radiation fluids, and also derive the general dispersion
relation for waves in a stratified, magnetized medium within the local
(WKB) limit.  In section 3 we discuss the solutions of this dispersion relation
in the hydrodynamic limit where there is no magnetic field, generalizing
and extending the work of previous authors.  We discuss in some detail the
physics underlying these instabilities in section 4.  We then incorporate
the effects of magnetic fields in section 5, and in section 6 discuss how
the physics of magnetoacoustic wave instabilities is related to the
hydrodynamic instabilities of the previous sections.
In section 7 we briefly discuss astrophysical applications of
these instabilities to accretion disks and stars, deferring more detailed
applications to observed phenomena for future work.
Section 8 summarizes our main conclusions.  Our WKB analysis in the
main body of the paper is not completely rigorous, and we provide more
solid mathematical justification for our conclusions for two particular
cases in an appendix.  We also include an additional
appendix where we examine the effects of radiative diffusion on the
magnetorotational (MRI) instability in differentially rotating flows,
generalizing our previous work (BS01) to include the more generic
thermodynamics we use here.  Readers not
interested in the details of the linear perturbation theory may wish to
skip ahead to sections 7 and 8 directly.  Readers interested in obtaining a
basic physical understanding of the causes of these instabilities may wish to
focus on sections 4 and 6.

\section{Equations and Assumptions}

The basic equations of radiation magnetohydrodynamics (RMHD) have been
discussed by Stone, Mihalas, \& Norman (1992), and we adopt these equations
here with slight changes in notation as well as some changes in the physics.
The fluid equations we consider are
\begin{equation}
{\partial\rho\over\partial t} +{\bnabla\cdot}(\rho{\bf v})=0,
\label{eqcont}
\end{equation}
\begin{equation}
\rho\left({\partial{\bf v}\over\partial t}+{\bf v\cdot\bnabla v}\right)
=-{\bnabla}p+\rho{\bf g}+{1\over4\pi}({\bf\bnabla\times B}){\bf\times B}+
{\kappaf\rho\over c}{\bf F},
\label{gasmom}
\end{equation}
\begin{equation}
{\partial u\over\partial t}+{\bf v\cdot\bnabla}u+\gamma u{\bf \bnabla\cdot v}
=\kappaj\rho cE-\kappap\rho caT_{\rm g}^4-\kappat\rho c
\left({4\kb T_{\rm g}\over m_{\rm e}c^2}-{h\bar\nu\over m_{\rm e}c^2}\right)E,
\label{gasen}
\end{equation}
\begin{equation}
{\partial E\over\partial t}+{\bf v\cdot\bnabla}E+
{4\over3}E{\bf \bnabla\cdot v} = -{\bf \bnabla\cdot F}-\kappaj\rho cE+
\kappap\rho caT_{\rm g}^4+\kappat\rho c
\left({4\kb T_{\rm g}\over m_{\rm e}c^2}-{h\bar\nu\over m_{\rm e}c^2}\right)E,
\label{raden}
\end{equation}
\begin{equation}
0=-{1\over3}{\bnabla}E-{\kappaf\rho\over c}{\bf F},
\label{fdiff}
\end{equation}
and
\begin{equation}
{\partial{\bf B}\over\partial t}={\bnabla\times}({\bf v\times B}).
\label{dbdt}
\end{equation}
Here $\rho$, $p$, $T_{\rm g}$, and $u$ are the density, pressure,
temperature, and
energy density in the gas, respectively.  These quantities are related to
each other by
\begin{equation}
u={p\over\gamma-1}
\label{equgas}
\end{equation}
and
\begin{equation}
p={\rho\kb T_{\rm g}\over\mu},
\label{gasstate}
\end{equation}
where $\gamma$ is the ratio of specific heats in the gas and $\mu$ is the
mean molecular mass of the gas.  We assume throughout this paper that
$\gamma$ and $\mu$ are constant, and thereby neglect the effects of
composition gradients, ionization, and recombination.

Other symbols in equations (\ref{eqcont})-(\ref{gasstate}) have their usual
meaning: $c$ is the speed of light, $a$ is the radiation density constant,
$\kb$ is Boltzmann's constant, and $m_{\rm e}$ is the electron mass.  The
vector ${\bf v}$ is the fluid velocity, ${\bf B}$ is the magnetic field,
and ${\bf g}$ is the gravitational acceleration.  We assume that ${\bf g}$
is time-independent throughout this paper, either because the gravitational
field is due to some fixed external source (as is the case for a
non-self-gravitating accretion disk), or because we choose to consider only
short-wavelength perturbations that are well-described by the Cowling
approximation.

The radiation energy
density $E$ and the radiation flux ${\bf F}$ are defined in terms of
frequency integrals over the radiation spectrum as measured in the local
fluid rest frame:
\begin{equation}
E\equiv{4\pi\over c}\int_0^\infty d\nu J_\nu\equiv{4\pi\over c}J,
\end{equation}
where $J_\nu$ is the angle-averaged mean specific intensity in the local
fluid rest frame, and
\begin{equation}
{\bf F}\equiv\int_0^\infty d\nu{\bf F}_\nu.
\end{equation}
We have simplified the RMHD equations by assuming that the radiation stress
tensor is isotropic in the local fluid rest frame, i.e. that the tensor
variable Eddington factor defined by Stone et al. (1992) is
one third times the identity matrix.  We have therefore neglected the effects
of photon viscosity.  We have also neglected terms corresponding to radiation
inertia in the radiation momentum equation (\ref{fdiff}), leaving us with just
the radiation diffusion equation.  Such terms are generally negligible for
the low frequency instabilities we explore in this paper (e.g. BS01).

The two opacities $\kappaj$ and $\kappaf$ are also defined by integrals
over the radiation spectrum in the local rest frame:
\begin{equation}
\kappaj\equiv{1\over\rho J}\int_0^\infty d\nu
\chi_\nu^{\rm th}(\rho,T_{\rm g})J_\nu,
\label{kappaJ}
\end{equation}
and
\begin{equation}
\kappaf{\bf F}\equiv{1\over\rho}\int_0^\infty d\nu
[\chi_\nu^{\rm th}(\rho,T_{\rm g})+n_{\rm e}\sigmat]{\bf F}_\nu.
\label{kappaF}
\end{equation}
Here $\chi_\nu^{\rm th}(\rho,T_{\rm g})$ is the thermal absorption
coefficient at frequency
$\nu$, $n_{\rm e}$ is the electron number density, and $\sigmat$ is the
Thomson cross-section.  The Planck mean opacity is
\begin{equation}
\kappap\equiv{4\pi\over\rho acT_{\rm g}^4}\int_0^\infty
d\nu\chi_\nu^{\rm th}(\rho,T_{\rm g})B_\nu(T_{\rm g}),
\label{kappaP}
\end{equation}
where $B_\nu(T_{\rm g})$ is the Planck function at the gas temperature.
The Thomson opacity is
\begin{equation}
\kappat\equiv{n_e\sigmat\over\rho}
\label{kappaT}
\end{equation}
and is constant due to our neglect of ionization/recombination processes
and composition gradients.  Note that we have extended the RMHD equations of
Stone et al. (1992) to include electron scattering.  Coherent (Thomson)
scattering merely allows the gas and radiation to exchange momentum, not
energy.  However, we have also allowed for incoherent (Compton) scattering
in the last terms on the right hand side of the gas and radiation energy
equations (\ref{gasen}) and (\ref{raden}).  We do this by treating Compton
scattering in the frequency diffusion (Kompaneets) limit, neglecting
anisotropies in the radiation field in the local fluid rest frame.  A
derivation of the Compton scattering terms can be found, e.g., in section
3.1 of Hubeny et al. (2001).  The quantity $\bar\nu$ is an average frequency
of the radiation field in the local rest frame of the fluid, defined by
\begin{equation}
\bar\nu\equiv{1\over J}\int_0^\infty d\nu \nu J_\nu
\left(1+{J_\nu c^2\over2h\nu^3}\right).
\label{nubar}
\end{equation}
The second term in parentheses represents the effects of
stimulated scattering, and guarantees that there is no net heat exchange
between the gas and the radiation when thermal equilibrium is established
with the radiation field being blackbody at the local gas temperature.

In principle, evaluation of the frequency integrals in the opacities and
$\bar\nu$ requires a full solution of the radiative transfer equation
coupled to the fluid flow.  Rather than do that here, we adopt some
simplified prescriptions that we believe will not alter the essential physics
we wish to explore.  First, we adopt the Eddington
closure relation between the zeroth and second moments of the comoving
frame specific intensity, consistent with our earlier assumption that the
tensor variable Eddington factor is one third times the identity matrix.
Second, because we will only consider perturbations about equilibria that
are in local thermodynamic equilibrium (LTE), we assume that non-LTE
departures in the perturbed flow can be adequately parameterized by
a blackbody radiation spectrum at a temperature $T_{\rm r}$, which may be
different from the gas temperature
$T_{\rm g}$.\footnote{Departures from LTE will of course
generally not maintain a blackbody radiation field in the perturbed flow.
Strictly speaking, we are defining $T_{\rm r}$ as the effective temperature
of the radiation field, and we are assuming that the various frequency moments
we need are related to each other in the perturbed flow in approximately the
same way as they would be if the radiation field were blackbody.}

Under these assumptions, equation (\ref{nubar}) implies that
\begin{equation}
\bar\nu={4\kb T_{\rm r}\over h}.
\label{nubarbb}
\end{equation}
Equation (\ref{kappaJ}) gives an expression for $\kappaj$
which more closely resembles the Planck mean opacity $\kappap$,
\begin{equation}
\kappaj={4\pi\over\rho acT_{\rm r}^4}\int_0^\infty
d\nu\chi_\nu^{\rm th}(\rho,T_{\rm g})B_\nu(T_{\rm r}).
\label{kappaJbb}
\end{equation}
In addition, equation (\ref{kappaF}) can be replaced with an expression
analogous to the Rosseland mean opacity,
\begin{equation}
\kappaf=\left(\int_0^\infty d\nu{dB_\nu\over dT_{\rm r}}\right)
\left\{\int_0^\infty d\nu\left[{\chi_\nu^{\rm th}(\rho,T_{\rm g})\over\rho}+
\kappat\right]^{-1}{dB_\nu\over dT_{\rm r}}\right\}^{-1}.
\label{kappaFbb}
\end{equation}
Unlike $\kappap$, which is a function only of the gas density and
temperature, $\kappaj$ and $\kappaf$ are also functions of the radiation
temperature.
Finally, the blackbody assumption implies that $T_{\rm r}$ can be related
to the radiation energy density through the equation of state
\begin{equation}
E=aT_{\rm r}^4.
\label{radstate}
\end{equation}

To summarize, equations (\ref{eqcont})-(\ref{gasstate}), (\ref{kappaP}), and
(\ref{nubarbb})-(\ref{radstate}) are the basic dynamical equations we
will use throughout the rest of this paper.

\subsection{Equilibrium}

For the majority of this paper we will consider the dynamical stability
of static equilibria in which the equilibrium fluid velocity is zero.
As mentioned in the previous section, we also assume that the equilibrium is
in LTE, i.e. that the unperturbed gas and radiation temperatures are identical,
\begin{equation}
T_{\rm g}=T_{\rm r}\equiv T.
\label{eqlte}
\end{equation}
This can be somewhat problematic for applications to accretion disk
models, which in some cases can be effectively thin.  However, we
do not presume to model the turbulent dissipation that must exist in
these flows.  In fact, our purpose in that application is to consider
equilibria that might be useful starting points for exploring the development
of this turbulence by simulation.  Lacking an explicit treatment of dissipation,
we are forced to consider equilibria that are in LTE and in local radiative
equilibrium.

Under these assumptions, the only nontrivial partial differential equations
describing our equilibrium are those expressing hydrostatic equilibrium,
\begin{equation}
0=\rho{\bf g}
-{\bnabla}p+{1\over4\pi}({\bf\bnabla\times B}){\bf\times B}+
{\kappaf\rho\over c}{\bf F},
\label{mechequil}
\end{equation}
radiative equilibrium,
\begin{equation}
0={\bnabla}\cdot{\bf F},
\label{radeq}
\end{equation}
and radiative diffusion,
\begin{equation}
0=-{1\over3}{\bnabla}E-{\kappaf\rho\over c}{\bf F}.
\label{raddiff}
\end{equation}
Throughout this paper we will assume that the equilibrium magnetic field
is uniform, so that the Lorentz force term in equation (\ref{mechequil})
vanishes.  Equation
(\ref{mechequil}) then has the immediate consequence that all thermodynamic
variables are constant on horizontal surfaces, i.e. those surfaces that
are perpendicular to the gravitational acceleration ${\bf g}$.  We
define $z$ to be a vertical coordinate that increases upward, and
$\hat{\bf z}$ to be the corresponding upward unit vector. Then
${\bf g}\equiv-g\hat{\bf z}$, with $g>0$.

\subsection{The Linear Perturbation in Total Pressure}

When we consider the behavior of linear perturbations about the equilibrium
just discussed, the problem naturally divides into two parts: the thermodynamics
of the perturbed flow, and the coupling of this with the perturbed magnetic and
velocity fields.  We examine the former problem first by determining the
perturbed total (gas plus radiation) pressure.

We begin by considering expressions for the perturbations in $\kappaj$
and $\kappap$.  The functional dependence of these two opacities on $\rho$,
$T_{\rm g}$ and $T_{\rm r}$ immediately implies that
\begin{equation}
{\delta\kappaj\over\kappaj}=
{\partial\ln\kappaj\over\partial\ln\rho}{\delta\rho\over\rho}+
{\partial\ln\kappaj\over\partial\ln T_{\rm g}}{\delta T_{\rm g}\over T_{\rm g}}+
{\partial\ln\kappaj\over\partial\ln T_{\rm r}}{\delta T_{\rm r}\over T_{\rm r}}
\end{equation}
and
\begin{equation}
{\delta\kappap\over\kappap}=
{\partial\ln\kappap\over\partial\ln\rho}{\delta\rho\over\rho}+
{\partial\ln\kappap\over\partial\ln T_{\rm g}}{\delta T_{\rm g}\over T_{\rm g}}.
\end{equation}
(We use Eulerian perturbations throughout this paper, i.e. $\delta Q$ is
the Eulerian perturbation in the quantity $Q$.)
Here all opacity derivatives are evaluated in the equilibrium state which
is in LTE with $T_{\rm r}=T_{\rm g}$ and $\kappap=\kappaj$.  Because of this,
equations (\ref{kappaP}) and (\ref{kappaJbb}) immediately imply
\begin{equation}
{\partial\ln\kappap\over\partial\ln\rho}=
{\partial\ln\kappaj\over\partial\ln\rho}
\label{kdrho}
\end{equation}
and
\begin{equation}
{\partial\ln\kappap\over\partial\ln T_{\rm g}}=
{\partial\ln\kappaj\over\partial\ln T_{\rm g}}+
{\partial\ln\kappaj\over\partial\ln T_{\rm r}}.
\label{kdt}
\end{equation}

These two conditions on the absorption opacity derivatives greatly simplify
the mathematical form of the thermal coupling between the gas and radiation
in the perturbed flow.  Linearizing the gas energy equation (\ref{gasen}) and
the radiation energy equation (\ref{raden}), and using equations
(\ref{nubarbb}), (\ref{radstate}), and (\ref{kdrho})-(\ref{kdt}), we obtain
\begin{equation}
{\partial\delta u\over\partial t}+\delta{\bf v}\cdot{\bnabla}u+\gamma u
{\bnabla}\cdot\delta{\bf v}=
\omegaa E{4(\delta T_{\rm r}-\delta T_{\rm g})\over T}
\label{eqdu}
\end{equation}
and
\begin{equation}
{\partial\delta E\over\partial t}+\delta{\bf v}\cdot{\bnabla}E+
{4\over3}E{\bnabla}\cdot\delta{\bf v} = -{\bnabla}\cdot\delta{\bf F}-
\omegaa E {4(\delta T_{\rm r}-\delta T_{\rm g})\over T}.
\label{eqde}
\end{equation}
Here $\omegaa$ is an angular frequency describing the rate at which radiation
and matter are thermally coupled in the perturbed flow, defined by
\begin{equation}
\omegaa\equiv\left[\kappap\left(1+{1\over4}{\partial\ln\kappaj\over\partial
\ln T_{\rm r}}\right)+\kappat{\kb T\over m_{\rm e}c^2}\right]\rho c\equiv
\kappa_{\rm a}\rho c.
\label{omegaabs}
\end{equation}

Equations (\ref{eqdu}) and (\ref{eqde}) can be further manipulated in a
physically revealing manner.  We eliminate ${\bnabla}\cdot\delta{\bf v}$
from them using the perturbed continuity equation (\ref{eqcont}),
\begin{equation}
{\partial\delta\rho\over\partial t}+\rho{\bnabla}\cdot\delta{\bf v}+
\delta{\bf v}\cdot{\bnabla}\rho=0.
\label{deqcont}
\end{equation}
In addition, we eliminate $\delta{\bf F}$ using the perturbed radiative
diffusion equation (\ref{fdiff}), which can be written as
\begin{equation}
\delta{\bf F}=-{c\over3\kappaf\rho}{\bnabla}\delta E-{\bf F}\left(
{\delta\rho\over\rho}+{\delta\kappaf\over\kappaf}\right).
\label{eqdeltaf}
\end{equation}
Eliminating $\delta u$ using equation (\ref{equgas}), equation (\ref{eqdu})
becomes
\begin{equation}
{\partial\delta p\over\partial t}-\cgas^2{\partial\delta\rho\over\partial t}
=4(\gamma-1)\omegaa E{(\delta T_{\rm r}-\delta T_{\rm g})\over T}-
(\gamma-1)\rho T\delta{\bf v}\cdot{\bnabla}S_{\rm g},
\label{ddpgdt}
\end{equation}
where $\cgas$ is the adiabatic sound speed in the gas,
\begin{equation}
\cgas\equiv\left({\gamma p\over\rho}\right)^{1/2},
\end{equation}
and $S_{\rm g}$ is the entropy per unit mass in the gas,
\begin{equation}
S_{\rm g}\equiv{\kb\over\mu(\gamma-1)}\ln(p\rho^{-\gamma})+{\rm constant}.
\end{equation}
Equation (\ref{eqde}) can be written in a very similar fashion,
\begin{equation}
{\partial\over\partial t}\left({1\over3}\delta E\right)-\crad^2
{\partial\delta\rho\over\partial t}
=-{4\over3}\omegaa E{(\delta T_{\rm r}-\delta T_{\rm g})\over T}-
{1\over3}\rho T\delta{\bf v}\cdot{\bnabla}S_{\rm r}
+{\bnabla}\cdot\left({c\over3\kappaf\rho}{\bnabla}{1\over3}\delta E\right)
+{1\over3}{\bf F}\cdot{\bnabla}\left({\delta\rho\over\rho}+
{\delta\kappaf\over\kappaf}\right),
\label{eqde2}
\end{equation}
where $\crad$ is the adiabatic sound speed in the radiation,
\begin{equation}
\crad\equiv\left({4E\over9\rho}\right)^{1/2},
\end{equation}
and $S_{\rm r}$ is the entropy per unit mass in the radiation,
\begin{equation}
S_{\rm r}={4E\over3\rho T}+{\rm constant}.
\end{equation}
Note that we have simplified the terms involving the equilibrium flux ${\bf F}$
in equation (\ref{eqde2}) by using the radiative equilibrium equation
(\ref{radeq}).

Thus far, we have not made any assumptions about the wavelengths of the
perturbations we are considering, and we could proceed from this point
by doing a full global perturbation analysis.  We wish to focus on local
instabilities in this paper, however.  Hence from now on, we invoke the WKB
ansatz by adopting a plane wave spacetime dependence
$\propto\exp[i({\bf k}\cdot{\bf r}-\omega t)]$ for all perturbations.  Here
${\bf r}$ is the position vector of the point of interest,
${\bf k}$ is the perturbation wave vector,
and $\omega$ is the perturbation angular frequency.  As a result, the time
derivative operator $\partial/\partial t$ in equations (\ref{ddpgdt}) and
(\ref{eqde2}) is replaced by $-i\omega$, and the spatial derivative
operator ${\bnabla}$ in the last two terms of equation (\ref{eqde2}) is
replaced by $i{\bf k}$, i.e.
\begin{equation}
{\bnabla}\cdot\left({c\over3\kappaf\rho}{\bnabla}{1\over3}\delta E\right)
+{1\over3}{\bf F}\cdot{\bnabla}\left({\delta\rho\over\rho}+
{\delta\kappaf\over\kappaf}\right)\rightarrow -{ck^2\over3\kappaf\rho}
\left({\delta E\over3}\right)+{1\over3}i{\bf k}\cdot{\bf F}
\left({\delta\rho\over\rho}+{\delta\kappaf\over\kappaf}\right).
\label{wkbdiff}
\end{equation}
The diffusive instabilities in which we are interested are driven by
background gradients, in particular the presence of a nonzero equilibrium
radiative flux ${\bf F}=-(c/3\kappaf\rho){\bnabla}E$.  Hence this last step
may at first sight appear to be dangerous: gradients in $c/(3\kappaf\rho)$
arising from the first term in expression (\ref{wkbdiff}) might be as
important as the second term that depends on ${\bf F}$ as we take the short
wavelength limit $k\rightarrow\infty$.  As may be verified {\it a posteriori},
however, it turns out that the instabilities have $\delta E$ scaling as
$k^{-1}\delta p$ at high $k$, so that our ordering is in fact consistent.
This makes physical sense, because at short wavelengths radiative diffusion
is very fast and perturbations in the radiation temperature are smoothed out.

After some algebra,
equations (\ref{gasstate}), (\ref{radstate}), (\ref{ddpgdt}), and
(\ref{eqde2}) can be used to derive an equation for the total
(gas plus radiation) pressure perturbation in terms of the density
and velocity perturbations,
\begin{equation}
\delta p+{1\over3}\delta E=({\cal A}\cgas^2+{\cal B}\crad^2+{\cal C})
\delta\rho-{i\rho T\over\omega}\delta{\bf v}\cdot\left[(\gamma-1){\cal A}
{\bnabla}S_{\rm g}+{1\over3}{\cal B}{\bnabla}S_{\rm r}\right].
\label{dptot}
\end{equation}
Here
\begin{equation}
{\cal A}\equiv{\omega\over{\cal D}}
\left[\omega+{ick^2\over3\kappaf\rho}+i\omegaa
\left(1+{4E\over3p}\right)+\kf\left({\thetatr\over4E}-{\thetatg\over3p}
\right)\right],
\label{cala}
\end{equation}
\begin{equation}
{\cal B}\equiv{\omega\over{\cal D}}\left[\omega+i\omegaa(\gamma-1)\left(3+
{4E\over p}\right)\right],
\label{calb}
\end{equation}
\begin{eqnarray}
{\cal C}&\equiv&{1\over{\cal D}}
\Biggl\{{4i\omegaa E\over\rho}\left[\left(\gamma-
{4\over3}\right)\omega+{ick^2\over3\kappaf\rho}(\gamma-1)\right]\nonumber\\
& &-{\kf\over\rho}\biggl[{\omega\over3}\left(1+\thetarho-\thetatg\right)
+i(\gamma-1)\omegaa\left(1+\thetarho-\thetatg-\thetatr\right)\nonumber\\
& &+i(\gamma-1)\omegaa\left({4E\over3p}\right)\left(1+\thetarho\right)
\biggr]\Biggr\},
\label{calc}
\end{eqnarray}
and
\begin{equation}
{\cal D}\equiv\left(\omega+{ick^2\over3\kappaf\rho}+{\kf\over4E}\thetatr\right)
\left[\omega+{4i(\gamma-1)\omegaa E\over p}\right]+
i\omegaa\left[\omega+(\gamma-1){\kf\over p}\thetatg\right].
\label{cald}
\end{equation}
The $\Theta$ quantities are defined as logarithmic derivatives of the flux
mean opacity with respect to the variable in the subscript, i.e.
\begin{equation}
\thetarho\equiv{\partial\ln\kappaf\over\partial\ln\rho},\,\,\,\,\,
\thetatg\equiv{\partial\ln\kappaf\over\partial\ln T_{\rm g}},\,\,\,\,\,
{\rm and}\,\,\,\,\,
\thetatr\equiv{\partial\ln\kappaf\over\partial\ln T_{\rm r}}.
\label{thetadefs}
\end{equation}

Equation (\ref{dptot}) expresses all the
thermal physics and is the one we shall use throughout the rest of the paper.

\subsection{Coupling to Density, Velocity, and Magnetic Field Perturbations}

Now that we have determined the total pressure perturbation, the only work
remaining is to couple this to the perturbed continuity, gas momentum,
and flux-freezing equations.
Employing the WKB ansatz, the perturbed continuity equation (\ref{deqcont})
becomes
\begin{equation}
-i\omega\delta\rho + i\rho {\bf k}\cdot\delta {\bf v} +
\delta {\bf v}\cdot{\bnabla}\rho=0.
\label{dcont}
\end{equation}

The perturbed gas momentum
equation and flux freezing equations may be written as
\begin{equation}
-i\omega\rho\delta{\bf v}=-i{\bf k}\left(\delta p+{1\over3}\delta E\right)+
{\bf g}\delta\rho+{i\over4\pi}({\bf k}\times\delta{\bf B})\times{\bf B}
\label{dgasmom}
\end{equation}
and
\begin{equation}
-i\omega\delta{\bf B}=i{\bf k}\times(\delta{\bf v}\times{\bf B}),
\label{dfluxfreezing}
\end{equation}
respectively.

\subsection{The Dispersion Relation}

After some algebra, equations (\ref{dptot}) and
(\ref{dcont})-(\ref{dfluxfreezing}) may be combined to give a dispersion
relation for short wavelength modes on a static, stratified and magnetized
equilibrium:

\begin{eqnarray}
0&=&\tilde{\omega}^2\left[\omega^4 -\omega^2k^2v_{\rm A}^2-\tilde{\omega}^2
k^2({\cal A}\cgas^2+{\cal B}\crad^2+{\cal C})\right]\nonumber\\
&+&\tilde{\omega}^2(k^2-k_z^2)\left({\cal A}\cgas^2 N_{\rm g}^2+{\cal B}\crad^2
N_{\rm r}^2\right)\nonumber\\
&+&i\tilde{\omega}^2\left[\omega^2 {\bf k} -k^2 ({\bf k}\cdot{\bf v}_{\rm A})
{\bf v}_{\rm A}\right]\cdot
\left[\frac{1}{\rho}{\bnabla}p\left({\cal A}-1\right)
+\frac{1}{3\rho}{\bnabla}E\left({\cal B}-1\right)+{\cal C}
{\bnabla}\ln\rho\right]\nonumber\\
&-&\left[\omega^4-\omega^2k^2
v_{\rm A}^2-2\omega^2k_zv_{{\rm A}z}{\bf k}\cdot{\bf v}_{\rm A}+
\omega^2k_z^2v_{\rm A}^2+k^2({\bf k}\cdot{\bf v}_{\rm A})^2v_{{\rm A}z}^2
\right]{\bf g}\cdot{\bnabla}\ln\rho,
\label{staticdisp}
\end{eqnarray}
where
\begin{equation}
\tilde{\omega}^2\equiv\omega^2-({\bf k}\cdot{\bf v}_{\rm A})^2,
\end{equation}
${\bf v}_{\rm A}\equiv{\bf B}/(4\pi\rho)^{1/2}$ is the vector Alfv\'en speed,
\begin{equation}
N_{\rm g}^2\equiv-{\bf g}\cdot\left({1\over\rho\cgas^2}{\bnabla p}-{\bnabla}
\ln\rho\right)=-{(\gamma-1)\rho T\over\gamma p}{\bf g}\cdot{\bnabla}S_{\rm g}
\end{equation}
is the Brunt-V\"ais\"al\"a frequency in the gas,
\begin{equation}
N_{\rm r}^2\equiv-{\bf g}\cdot\left({1\over3\rho\crad^2}{\bnabla E}-{\bnabla}
\ln\rho\right)=-{3\rho T\over4E}{\bf g}\cdot{\bnabla}S_{\rm r}
\end{equation}
is the Brunt-V\"ais\"al\"a frequency in the radiation,
and the quantities ${\cal A}$, ${\cal B}$, and ${\cal C}$ are defined by
equations (\ref{cala})-(\ref{cald}).

Equation (\ref{staticdisp}) is a dispersion relation for eight modes.
We originally had nine first order, time-dependent perturbuation equations
(one continuity, three momentum, two energy, and three flux-freezing), and
so we expect nine modes.  However, one of these modes has zero frequency
and is inconsistent with the additional constraint of Gauss' Law that
${\bnabla}\cdot\delta{\bf B}=0$.
In sections 3 and 5, we provide a full discussion of the unstable waves
contained in equation (\ref{staticdisp}).

\section{Hydrodynamic Instabilities}

Before considering the full effects of MHD, it is useful to first explore
radiative diffusion instabilities in the hydrodynamic case.
%Such instabilities
%have appeared in a number of guises in the literature (Hearn 1972, OTHERS),
%and our analysis here generalizes and extends this previous work, although we
%restrict consideration to the local (WKB) limit.
Setting ${\bf v}_{\rm A}=0$
in the general dispersion relation (\ref{staticdisp}), we obtain the
following equation for hydrodynamic modes:
\begin{eqnarray}
0&=&\omega^4-\omega^2
k^2({\cal A}\cgas^2+{\cal B}\crad^2+{\cal C})
+(k^2-k_z^2)\left({\cal A}\cgas^2 N_{\rm g}^2+{\cal B}\crad^2
N_{\rm r}^2\right)\nonumber\\
&+&i\omega^2 {\bf k}\cdot
\left[\frac{1}{\rho}{\bnabla}p\left({\cal A}-1\right)
+\frac{1}{3\rho}{\bnabla}E\left({\cal B}-1\right)+{\cal C}
{\bnabla}\ln\rho\right]
-\omega^2{\bf g}\cdot{\bnabla}\ln\rho.
\label{staticdisphydro}
\end{eqnarray}
This equation can be written as a sixth order polynomial in $\omega$, which
is to be expected given that it arises from six time-dependent hydrodynamic
perturbation equations.  However, one power of $\omega$ can be factored
out, indicating that one of the modes always has zero frequency.  This
mode has zero total pressure perturbation $\delta p+\delta E/3$, at least
when the wave vector is not entirely vertical.

\subsection{Short Wavelength Limit}

In the short wavelength limit, the remaining five modes described by equation
(\ref{staticdisphydro}) can be easily factored:
\begin{equation}
0=\left(\omega^2-k^2\cgas^2\right)\left(\omega+{ick^2\over3\kappaf\rho}\right)
\left\{\omega^2+{4i\omegaa E(\gamma-1)\over\gamma p}\omega-
\left[1-\left(\hat{\bf k}\cdot\hat{\bf z}\right)^2\right]N_{\rm g}^2\right\}.
\label{highkhydro}
\end{equation}
From left to right, the three factors respectively correspond to adiabatic
acoustic waves in the gas, a purely damped radiative diffusion mode, and
gravity waves in the gas modified by damping of
gas temperature fluctuations by radiative emission and absorption.
Note that radiation pressure and radiative buoyancy have been lost due
to the rapid radiative diffusion at short wavelengths.

The behavior and character of the gravity waves depends on the relative
magnitude of the Brunt-V\"ais\"al\"a frequency and a characteristic gravity
wave thermal coupling frequency
\begin{equation}
\omega_{\rm thg}\equiv{2\omegaa E(\gamma-1)\over\gamma p}.
\label{omegathg}
\end{equation}
Solving equation (\ref{highkhydro}) for the gravity mode frequencies,
\begin{equation}
\omega=-i\omega_{\rm thg}
\pm\left\{-\omega_{\rm thg}^2+N_{\rm g}^2\left[1-(\hat{\bf k}\cdot\hat{\bf z})^2
\right]\right\}^{1/2}.
\end{equation}
If $N_{\rm g}^2[1-(\hat{\bf k}\cdot\hat{\bf z})^2]>\omega_{\rm thg}^2$, then
the gravity waves are damped.  If
$0<N_{\rm g}^2[1-(\hat{\bf k}\cdot\hat{\bf z})^2]<\omega_{\rm thg}^2$, then
this damping is so strong that the modes lose their wavelike character and
have purely negative imaginary frequencies.  If $N_{\rm g}^2<0$, then one of
the gravity modes becomes convectively unstable, but with a growth rate that
is reduced compared to ideal hydrodynamic convection.  This reduction is
due to the fact that emission and absorption damp gas temperature
fluctuations in the wave, relative to the nearly uniform radiation temperature.

The acoustic waves are more interesting.  Expanding equation
(\ref{staticdisphydro}) to first order about the high $k$ limit, we find
\be
\omega=\pm k\cgas-i\frac{\kappaf}{2c\cgas}\left\{{4E\cgas\over3}+
\frac{4Ec\omegaa(\gamma-1)^2}{\kappaf\rho\cgas}\mp\left({\hat{\bf k}}\cdot
{\bf F}\right)
\left[\thetarho+(\gamma-1)\thetatg\right]\right\}
+{\cal O}(k^{-1}).
\label{omeganocouphydro}
\ee
The first two terms within curly brackets represent damping by radiative
diffusion and emission/absorption, respectively.
The third term, which depends on the equilibrium radiative flux ${\bf F}$,
can give rise to instability if it dominates the first two.
For a Kramers type opacity law in a gas with $\gamma=5/3$,
$\thetarho+(\gamma-1)\thetatg<0$, and equation (\ref{omeganocouphydro}) then
implies that {\it downward} propagating sound waves are potentially unstable,
while upward propagating waves are damped.  On the other hand, if there
is no opacity perturbation, $\thetarho=\thetatg=0$, and there is no
acoustic instability.  Such is the case for a medium with pure Thomson
scattering opacity.

Instability requires that the driving term dominate the two damping terms
in equation (\ref{omeganocouphydro}), so that a rough, order of magnitude
instability criterion is
\begin{equation}
F\tilde{\Theta}\gta E\times
{\rm max}\left[\cgas,\left({\omegaa\over\kappaf\rho c}
\right)\left({c^2\over\cgas}\right)\right],
\label{nocoupstabcrit}
\end{equation}
where $\tilde{\Theta}\equiv|\thetarho+(\gamma-1)\thetatg|$.  The first part
of this criterion has an intuitive interpretation:  if
$\tilde{\Theta}$ is of order unity, short wavelength acoustic waves will
be unstable if the radiative flux is transporting the local radiation energy
density faster than the sound speed in the gas.

As we will see in section 3.3 below, the maximum growth rate of the instability,
when it exists, occurs for wavenumbers such that the first term in equation
(\ref{omeganocouphydro}) exceeds the second, i.e. for
$k\gta\kappaf F\tilde{\Theta}/(c\cgas^2)$.  In a radiation pressure dominated
medium, this implies that $k\gta\tilde{\Theta}g/\cgas^2$, i.e. wavelengths
shorter than the {\it gas} scale height will have maximal growth rates.  This
growth rate itself will be $\sim\tilde{\Theta}(g/\cgas)$, {\it faster} than
the reciprocal of the local free fall time by the ratio
$\tilde{\Theta}\crad/\cgas$.

On the other hand, in a gas pressure dominated medium, the growth rate is
$\sim\tilde{\Theta}E/(H_T\rho\cgas)$, where $H_T$ is the
temperature scale height.  Taking this to be comparable to the pressure
scale height $\sim\cgas^2/g$, we find a much smaller growth rate
$\sim\tilde{\Theta}(g/\cgas)(E/p)$, {\it slower} than the reciprocal
of the local free fall time by the ratio $\tilde{\Theta}(\crad/\cgas)^2$.
This growth rate occurs for wavenumbers $\gta\tilde{\Theta}(g/\cgas^2)(E/p)$,
a threshold which is in violation of our WKB requirement that $k\gg g/\cgas^2$
for gas pressure dominated media.  If the gas and radiation temperatures
are not tightly locked, it appears difficult to produce vigorous acoustic
wave instabilities in gas pressure dominated media by this mechanism.

An expression similar to equation (\ref{omeganocouphydro}) appears to have
been first derived by Hearn (1972; his equations 25, 26 and 30 and surrounding
discussion), the only difference being that he does not have the radiative
diffusion damping term [the first term in curly brackets in equation
(\ref{omeganocouphydro})].  This is because Hearn (1972) was interested in
radiative amplification of acoustic waves in optically {\it thin} media,
and therefore did not include the dynamical evolution of the radiation field in his
analysis.\footnote{It is worth noting that Hearn's (1972) analysis had
an error in his treatment of the gas energy equation (his equation 8).
In this equation the partial time derivatives should be Lagrangian derivatives.
It turns out, however, that his results are correct because he also neglected
gas pressure gradient contributions to the equilibrium hydrostatic balance.
These two errors cancelled one another.}

\subsection{Short Wavelength Limit With $T_{\rm g}=T_{\rm r}$}

If we first take the $\omegaa\rightarrow\infty$ limit in equation
(\ref{staticdisphydro}), so that the gas and radiation are tightly thermally
coupled, then a mode is eliminated from the dispersion relation which is now only
fifth order, including the zero frequency mode.  The fact that we have lost
a mode makes sense, as this limit corresponds to replacing the two time
dependent gas and radiation energy equations with a total energy equation
and the time-{\it independent} condition that the gas and radiation
temperatures be equal.  On taking the short wavelength limit, the four
modes with nonzero frequencies factor as follows:
\begin{eqnarray}
0&=&\left(\omega^2-k^2c_{\rm i}^2\right)\left[\omega+{ick^2\over3\kappaf\rho}
\left({4(\gamma-1)E\over p+4(\gamma-1)E}\right)\right]\nonumber\\
&\times&\left\{\omega+{i3\kappaf\rho\over ck^2}
\left[1-\left(\hat{\bf k}\cdot\hat{\bf z}\right)^2\right]
\left(1+{4E\over3p}\right)\left[{\gamma p\over4(\gamma-1)E}N_{\rm g}^2+
{1\over3}N_{\rm r}^2\right]\right\},
\label{highkhydrolocked}
\end{eqnarray}
where
\begin{equation}
c_{\rm i}\equiv\left({p\over\rho}\right)^{1/2}={\cgas\over\gamma^{1/2}}
\end{equation}
is the isothermal sound speed in the gas.

These short wavelength modes resemble those of equation (\ref{highkhydro}).
Acoustic waves propagate at the isothermal sound speed, because the gas
temperature is locked to a radiation temperature which is made nearly uniform
by the rapid radiative diffusion.
The two gravity modes have collapsed to a single mode in the last factor of
equation (\ref{highkhydrolocked}).  This mode has a purely imaginary
frequency, and has therefore lost its wavelike character.  It is unstable if
\begin{equation}
{\gamma p\over4(\gamma-1)E}N_{\rm g}^2+
{1\over3}N_{\rm r}^2<0.
\end{equation}
However, radiative diffusion strongly diminishes the growth rate of this
instability in the short wavelength limit, with
$|\omega|\propto k^{-2}$ as $k\rightarrow\infty$.  This is because buoyancy
is strongly suppressed: a perturbed parcel of gas
always has the same density as its surroundings, because it has the same
pressure and is forced to have the same temperature due to the rapid
radiative diffusion and rapid thermal coupling with the radiation.

The acoustic waves are again unstable when
first order corrections to their frequencies are made:
\begin{equation}
\omega=\pm kc_{\rm i}-i{\kappaf\over2cc_{\rm i}}\left(1+{3p\over4E}\right)
\left[\left({4E\over3}+p\right)c_{\rm i}
\mp\left({\hat{\bf k}}\cdot{\bf F}\right)\thetarho\right]+{\cal O}(k^{-1}).
\label{omegalockhydro}
\end{equation}
The first term inside square brackets again represents damping by radiative
diffusion, while the second term will drive instability if it is larger.
For, e.g., Kramers law type opacities, $\Theta_\rho>0$ and now the {\it
upward} propagating wave is unstable while the downward propagating wave
is damped.  Once again, a pure Thomson scattering medium exhibits no
acoustic wave instability.

An order of magnitude instability criterion from equation (\ref{omegalockhydro})
is
\begin{equation}
F\thetarho\gta{\rm max}[E,p]c_{\rm i},
\label{lockstabcrit}
\end{equation}
which should be contrasted with equation (\ref{nocoupstabcrit}).  Note that
there is no threshold from emission and absorption because the gas and
radiation temperatures are equal.  Here,
acoustic waves are unstable if the local radiative flux is transporting
the local thermal energy density, whether it be dominated by gas or radiation,
faster than the gas sound speed.

In a radiation pressure dominated equilibrium, equation (\ref{omegalockhydro})
implies that the growth rate of the acoustic wave instability is
$\sim\kappaf F\thetarho/(cc_{\rm i})\sim \thetarho g/c_{\rm i}$ from
hydrostatic equilibrium.  In a gas pressure dominated medium, the growth
rate is $\sim\kappaf F\rho c_{\rm i}\thetarho/(cE)\sim c_{\rm i}\thetarho/H_T$.
Hence the growth rate in a gas pressure dominated medium is {\it also}
$\sim \thetarho g/c_{\rm i}$.  In contrast to the previous case where gas and
radiation did not exchange heat rapidly, radiation pressure support is
{\it not} required to obtain high instability growth rates when the gas and
radiation temperatures are the same.  In a gas pressure dominated equilibrium,
the small radiation pressure fluctuations produced by the damping and driving
forces produce gas pressure fluctuations that are $\sim(p/E)$ times
larger just by the fact that the gas and radiation temperatures are locked
together by rapid absorption and emission.

Hearn (1972; equations 22 and 23) was also the first to derive an
expression similar to equation (\ref{omegalockhydro}), although again
without the first damping term because he was interested in optically
thin media.  In addition, Hearn's (1972) approximations
resulted in a replacement of the factor $(1+0.75p/E)$ multiplying the
damping and growth rates in equation (\ref{omegalockhydro}) with unity.
This factor becomes very important in gas pressure dominated media.

We note that our result that pure Thomson scattering media exhibit no local
hydrodynamic acoustic wave instability disagrees with the conclusion of
Shaviv (2001), who claimed that his ``Type II'' instability represents just
such a local instability.  In fact, this instability appears to have rather
long vertical wavelengths.  The growth rates actually
depend on the boundary conditions , and also the location of the boundary
(Glatzel 2003, private communication), indicating that it is global in
nature, not local.

\subsection{The Limit of Negligible Gas Pressure}

A number of authors have studied versions of these instabilities in
the limit where gas pressure is completely negligible compared to
radiation pressure.  The gas sound speed then vanishes, and
one then loses the acoustic wave nature of the instability in the
short wavelength limit.

Setting $p=0=\cgas^2$ in the hydrodynamic dispersion relation
(\ref{staticdisphydro}), and then adopting the ansatz that
$\omega\propto k^{1/2}$ as $k\rightarrow\infty$ (Glatzel 1994, Gammie
1998), it is straightforward
to show that there are two modes with
\begin{equation}
\omega^2={i\kappaf\over c}{\bf k}\cdot{\bf F}\thetarho=
-i{\bf k}\cdot{\bf g}\thetarho.
\label{phbubblehydro}
\end{equation}
Assuming $\thetarho>0$, one of these modes corresponds to an upward
propagating unstable wave, while the other corresponds to a downward
propagating wave which is damped.
It is interesting to note that this result agrees with the thermally
locked acoustic wave frequency, equation (\ref{omegalockhydro}), if
one squares that frequency and then takes the limit of negligible gas
pressure.  Note that the damping terms, which set the threshold of
instability, vanish in the limit of negligible gas pressure.

If we first assume negligible thermal coupling between the gas and
the radiation ($\omega_{\rm a}\rightarrow0$), and then consider the zero
gas pressure limit, we get instead two modes that reflect the form
of the driving in the two temperature regime:
\begin{equation}
\omega^2={i\kappaf\over c}{\bf k}\cdot{\bf F}\left[\thetarho+
(\gamma-1)\thetatg\right]=-i{\bf k}\cdot{\bf g}\left[\thetarho+
(\gamma-1)\thetatg\right].
\end{equation}
Once again, the growth rate exhibits a $k^{1/2}$ dependence.  Note that
the damping terms again vanish in this particular limit.

The acoustic wave instabilities we have been discussing throughout this
section are the local, WKB versions of the strange mode instability
discussed in the stellar oscillation literature.  Glatzel (1994) has
discussed a physical origin of strange modes in which he presents a WKB
analysis of the growth rates of purely radial modes (i.e. vertical modes in
our geometry) in the limit of zero gas pressure.  His equation (5.8)
is (nearly\footnote{Glatzel's
equation 5.8 is $\omega^2=\pm i2g\thetarho k$, and we actually only recover
the version of this equation with a plus sign.  In addition, we do not
have his factor of two.  The reason for both of these facts is that his
definition of the wavenumber $k$ differs from ours.  Glatzel performed a WKB
analysis on a variable $\Pi$ which was a transformation of the Eulerian
pressure perturbation (his equation 5.4).  Our analysis only recovers modes
with short wavelengths in the pressure perturbation, and this is true of
only one sign of his dispersion relation.  Moreover, our wavenumber for these
modes is twice his wavenumber.}) identical to our equation
(\ref{phbubblehydro}).  He did not recover the damping terms of
equation (\ref{omegalockhydro}) for two reasons: he assumed negligible
gas pressure, and he also invoked the NAR approximation, a point that we
discuss further below in section 4.1.
Glatzel (1994) also presented a global
analytic solution for unstable strange modes in which he showed that unstable
and damped waves propagate in opposite directions (see discussion after
his equation 5.17), in agreement with our conclusions.

\subsection{Numerical Results}

Table 1 summarizes the basic conclusions of the analytic work on acoustic
wave instabilities presented in the previous two subsections.  We now discuss
numerical solutions to the full dispersion relation (\ref{staticdisphydro})
which show how these instabilities operate in different regimes, illustrating
the basic formulas in Table 1.

At sufficiently short wavelengths, photons diffuse across a wavelength
faster than the wave period, and radiation temperature fluctuations therefore
become small.  Whether or not the gas and radiation temperatures are locked
together is therefore determined by whether absorption and emission are
rapid enough to drive the gas temperature fluctuations to be small as well.
From the perturbed gas energy equation (\ref{ddpgdt}), it is clear that the gas
temperature will be locked to the radiation temperature provided the wave
frequency $\omega$ is much less than a characteristic thermal frequency
\begin{equation}
\omega_{\rm th}\equiv\left[{4(\gamma-1)E\over p}\right]\omegaa.
\label{omegathermal}
\end{equation}

Figure 1 depicts unstable wave growth rates as a function of wavenumber
for a radiation pressure dominated equilibrium, for different, small values of
$\omega_{\rm th}$.  Although such equilibria are
generally dominated by Thomson scattering opacity, for purposes of
illustrating the physics, we have assumed Kramers type values for the flux
mean opacity derivatives: $\thetarho=1$ and $\thetatg=-3.5$.  In any
real application to a radiation pressure dominated equilibrium, these
derivatives will be reduced approximately by the ratio of the absorption
opacity $\kappa_{\rm a}\equiv\omegaa/(\rho c)$ to
$\kappat$, and
this would lead to a reduction in the unstable growth rates, or possibly
stabilization.  We present results appropriate for more realistic
applications later on below.

At the low values of $\omega_{\rm th}$ shown in Figure 1, gas and radiation
temperatures are decoupled at high wavenumber, and the asymptotic growth
rates agree with equation (\ref{omeganocouphydro}).  These unstable
waves propagate downward because $\thetarho+(\gamma-1)\thetatg<0$.
As $\omega_{\rm th}$ increases, their asymptotic
growth rates decrease and eventually damp.  At the same time, a new set of
modes appear at low wavenumber with growth rates that climb with increasing
thermal coupling between the gas and radiation, as shown more clearly
in Figure 1(b).  These modes propagate upward and at higher thermal coupling
turn into the one-temperature acoustic waves described by equation
(\ref{omegalockhydro}) at high wavenumber.

This is shown in Figure 2(a), which depicts the growth rates as a function
of wavenumber for high values of $\omega_{\rm th}$, for the same equilibrium
parameters used in Figure 1.  In contrast to the two-temperature behavior,
the asymptotic growth rates at high wavenumber must eventually fail because
at short enough wavelengths the modes must return to being
adiabatic in the gas alone, and the assumption of thermal locking of the
gas and radiation underlying equation (\ref{omegalockhydro}) must break
down.  Figure 2(b) illustrates that for wavenumbers
$k\gta\omega_{\rm th}/\cgas$, the real part of the phase
velocity returns to the adiabatic gas sound speed.

The one temperature instability growth rate cuts off at even lower wavenumbers
than this, however, and the reason appears to be a breakdown of thermal
locking related to an interplay between the equilibrium radiation flux,
absorption and emission, and fluid inertia.  Numerical
exploration shows that the growth rate cuts off for wavenumbers
$k\gta(\omegath g|1+\thetarho|/c_{\rm i}^3)^{1/2}$ for both gas and radiation
pressure dominated equilibria, and this formula for the maximum cutoff
wavenumber is included in Table 1.

For demonstration purposes, we have been considering equilibria with rather
artificial opacities.
Figure 3 shows the unstable wave growth rate as a function of wavenumber for
a {\it gas}-pressure dominated equilibrium with a realistic Kramers type
opacity law, appropriate for mid- to upper main sequence stars.  For
sufficiently high thermal coupling, as is indeed present in these stars
(see section 7.2 below), unstable upward propagating waves exist over a
broad range of high wavenumbers.  Interestingly, long wavelength instabilities
also appear in downward propagating waves.  These downward wave instabilities
exist because of the temperature dependence of the opacity ($\thetatg=-3.5$).
Because the gas and radiation temperatures are tightly locked,
temperature fluctuations in the gas are smoothed out by radiative
diffusion.  Hence opacity fluctuations caused by $\delta T$ are smaller
than opacity fluctuations caused by $\delta\rho$.  Therefore the temperature
dependence of the opacity does not affect the growth rate at short
wavelengths.  This agrees with equation (\ref{omegalockhydro}).  However,
at long wavelengths where radiative diffusion is slow, significant gas
temperature fluctuations exist in the wave, and the resulting opacity
fluctuations reverse the radiative driving.  Note that these long
wavelength, downward unstable waves do not appear to exist in radiation
pressure dominated equilibria, as shown in Figure 2.

\section{Physics of the Acoustic Wave Instabilities}

The results of the previous section are based on a brute force analytic and
numerical attack on the full dispersion relation.  In this section, we present
a more physically-motivated derivation of some of these results.
(This turns out to be a useful starting point for understanding the
unstable MHD waves that we discuss below in sections 5 and 6.)
It is the coupling of the equilibrium radiation flux to opacity fluctuations
that drives acoustic waves unstable.  To see how this works physically,
consider the forces acting on a perturbed parcel of gas
in the acoustic wave.  [This approach is somewhat analogous to using a
zeroth order eigenfunction to compute damping or driving rates through
a work integral in stellar oscillation theory (Cox 1980).]

The gas and radiation energy equations (\ref{ddpgdt}) and (\ref{eqde2}) may
be written in the WKB limit as
\begin{equation}
\delta p=\cgas^2\delta\rho-{i\over\omega}\delta{\bf v}\cdot\left({\bnabla}p
-\cgas^2{\bnabla}\rho\right)+{4i(\gamma-1)\omegaa E\over\omega}
\left({\delta T_{\rm r}-\delta T_{\rm g}\over T}\right)
\label{ddpgdt2}
\end{equation}
and
\begin{eqnarray}
\left(1+{ick^2\over3\kappaf\rho\omega}\right){\delta E\over3}&=&
\crad^2\delta\rho-{i\rho T\over3\omega}\delta{\bf v}\cdot{\bnabla}S_{\rm r}
-{4i\omegaa E\over3\omega}
\left({\delta T_{\rm r}-\delta T_{\rm g}\over T}\right)\nonumber\\
&-&{{\bf k}\cdot{\bf F}\over3\omega}\left[(1+\thetarho){\delta\rho\over\rho}
+\thetatr{\delta T_{\rm r}\over T}+\thetatg{\delta T_{\rm g}\over T}\right],
\label{eqde3}
\end{eqnarray}
respectively.
The radiation energy equation (\ref{eqde3}) immediately implies that for
short wavelength ($k\rightarrow\infty$) acoustic waves,
$\delta E\sim{\cal O}(k^{-1})\delta\rho$.  Physically, at high $k$ the
radiation diffusion time across a wavelength is shorter than a wave period,
so that radiation temperature fluctuations are smoothed out.  On the other
hand, because acoustic waves in the gas are adiabatic in this limit, the
gas temperature fluctuation is given by
$\delta T_{\rm g}/T=(\gamma-1)\delta\rho/\rho$.  Hence the gas pressure
perturbation is given by
\begin{equation}
\delta p=\cgas^2\delta\rho-{i\over\omega}\delta{\bf v}\cdot\left({\bnabla}p
-\cgas^2{\bnabla}\rho\right)-{4i(\gamma-1)^2\omegaa E\over\omega}\left(
{\delta\rho\over\rho}\right)+{\cal O}(k^{-2}).
\label{dpghighk}
\end{equation}
The first term on the right hand side is the dominant term at short wavelengths,
and represents the acoustic response of the gas to compressive perturbations.
The second term arises from buoyancy.  The last term is a damping term
caused by radiative emission/absorption: temperature fluctuations in
the gas emit and absorb photons to try and come into equilibrium with the
nearly uniform radiation temperature.

To the same order in $k^{-1}$, the radiation pressure perturbation is
\begin{equation}
{\delta E\over3}=-{i\kappaf\over c}\left[
{4E\over3}\left({\omega\over k^2}\right)\right]\delta\rho
+{i\kappaf\over k^2c}\left({\bf k}\cdot{\bf F}\right)\left[1+\Theta_\rho
+(\gamma-1)\thetatg\right]\delta\rho
+{\cal O}(k^{-2}).
\label{dprhighk}
\end{equation}
The first term on the right hand side represents damping by radiative
diffusion: a density fluctuation will try to compress photons, but
these diffuse quickly and transport heat out of the fluctuation.
The last term is the most interesting.
It is an extra radiation pressure force arising from the equilibrium radiation
flux acting on density and opacity fluctuations in the gas.
As we shall see, it is this extra force
that is destabilizing if it can overcome the damping forces.

Using equations (\ref{dpghighk})-(\ref{dprhighk}) to eliminate the total
pressure perturbation,
the gas momentum equation (\ref{dgasmom}) for short wavelength acoustic
perturbations may be written
\begin{eqnarray}
\rho{\partial\delta{\bf v}\over\partial t}
&=&-i{\bf k}\cgas^2\delta\rho
-{{\bf k}\over\omega}\delta{\bf v}\cdot
\left({\bnabla}p-\cgas^2{\bnabla}\rho\right)
-{\kappaf\over c}{\bf k}\left[
{4E\over3}\left({\omega\over k^2}\right)\right]\delta\rho-{\kappaf\over c}
{\bf k}\left[{4Ec\omegaa(\gamma-1)^2\over
\kappaf\rho\omega}\right]\delta\rho\nonumber\\
&+&{\kappaf\over k^2c}{\bf k}\left({\bf k}\cdot{\bf F}\right)\left[1+\Theta_\rho
+(\gamma-1)\thetatg\right]\delta\rho
+{\bf g}\delta\rho+{\cal O}(k^{-1})\delta\rho.
\label{dvdthighk}
\end{eqnarray}
The destabilizing force density is ${\bf k}({\bf k}\cdot{\bf F})\delta\rho$
times a numerical factor depending on opacity derivatives that may be positive
or negative.  For the purposes of discussion, let us assume for the moment that
it is positive.  Figure 4(a) illustrates the geometry of this force acting
on a downward propagating acoustic wave.  In this case
${\bf k}({\bf k}\cdot{\bf F})\delta\rho$ is everywhere opposite to the
velocity perturbation, and the wave is therefore damped.  However, for an
upward propagating wave, shown in Figure 4(b), the force is everywhere in
phase with the velocity perturbation, and therefore amplifies the wave.  If
we had chosen the numerical factor depending on opacity derivatives
to be negative, then it would have been the upward propagating wave that
is damped and the downward propagating wave that is amplified.

It is also possible to consider the effects of the destabilizing force density
in terms of whether it induces a net lag or lead between the total pressure
and density perturbations in the sound wave.  Once again, only one direction
of propagation is destabilizing when viewed in this way.

Examination of the growth rate in equation (\ref{omeganocouphydro}) reveals
that it is just the combination of opacity derivatives
$\thetarho+(\gamma-1)\thetatg$ that determines the action of the force,
not the full factor $[1+\Theta_\rho+(\gamma-1)\thetatg]$ in equation
(\ref{dvdthighk}).  Pure Thomson scattering media are not unstable by
this mechanism.  To see why this is so, it is necessary to examine the
equation of motion (\ref{dvdthighk}) more closely.  Using the perturbed
continuity equation (\ref{dcont}) and regrouping, equation (\ref{dvdthighk})
may be rewritten as
\begin{eqnarray}
\rho{\partial\delta{\bf v}\over\partial t}
&=&-{i{\bf k}\cgas^2\rho\over\omega}\left({\bf k}\cdot\delta{\bf v}\right)
-{\kappaf\over c}{\bf k}\left[
{4E\over3}\left({\omega\over k^2}\right)\right]\delta\rho-{\kappaf\over c}
{\bf k}\left[{4Ec\omegaa(\gamma-1)^2\over
\kappaf\rho\omega}\right]\delta\rho\nonumber\\
&+&{\kappaf\over k^2c}{\bf k}\left({\bf k}\cdot{\bf F}\right)\left[\Theta_\rho
+(\gamma-1)\thetatg\right]\delta\rho\nonumber\\
&+&\left[{\kappaf\over k^2c}{\bf k}\left({\bf k}\cdot{\bf F}\right)
\delta\rho-{{\bf k}\over\omega}\delta{\bf v}\cdot{\bnabla}p+{\bf g}\delta\rho
\right]+{\cal O}(k^{-1})\delta\rho.
\label{dvdthighk2}
\end{eqnarray}
It is convenient to rewrite this equation in terms of the Lagrangian
displacement ${\bxi}$, where
\begin{equation}
\delta{\bf v}={\partial{\bxi}\over\partial t}=-i\omega{\bxi}.
\end{equation}
Using the fact that to lowest order in $k^{-1}$,
$\delta\rho=-i\rho{\bf k}\cdot{\bxi}$ from the continuity equation
(\ref{dcont}), we obtain
\begin{eqnarray}
{\partial^2{\bxi}\over\partial t^2}
&=&-{\bf k}\cgas^2\left({\bf k}\cdot{\bxi}\right)\nonumber\\
&+&i({\bf k}\cdot{\bxi}){\bf k}{\kappaf\over c}\left\{
{4E\over3}\left({\omega\over k^2}\right)+{4Ec\omegaa(\gamma-1)^2\over\kappaf
\rho\omega}-{1\over k^2}({\bf k}\cdot{\bf F})\left[\thetarho+(\gamma-1)
\thetatg\right]\right\}\nonumber\\
&-&i\left[{\kappaf\over k^2c}{\bf k}\left({\bf k}\cdot{\bf F}\right)\left(
{\bf k}\cdot{\bxi}\right)-{1\over\rho}{\bf k}{\bxi}\cdot{\bnabla}p+
{\bf g}\left({\bf k}\cdot{\bxi}\right)\right]+
{\cal O}(k^0)|{\bxi}|.
\label{dxidthighk}
\end{eqnarray}

The destabilizing ${\bf k}\cdot{\bf F}$ term that would be present in
Thomson scattering atmospheres competes with the equilibrium gravitational
acceleration and gas pressure gradient in the last term in square brackets
in equation (\ref{dxidthighk}).  Figure 4 illustrates this by showing that
the projection of ${\bf g}\delta\rho$ along the wave acts in the opposite
direction to ${\bf k}\cdot{\bf F}{\bf k}\delta\rho$.
Eliminating the gas pressure gradient using the hydrostatic equilibrium
equation (\ref{mechequil}), the last term in square brackets in equation
(\ref{dxidthighk}) may be written
\begin{equation}
{\kappaf\over k^2c}{\bf k}\left({\bf k}\cdot{\bf F}\right)\left(
{\bf k}\cdot{\bxi}\right)-{\bf k}{\bxi}\cdot{\bnabla}p+{\bf g}\left(
{\bf k}\cdot{\bxi}\right)=
-{\kappaf\over c}{\bf k}(\hat{\bf k}\times{\bf F})\cdot
(\hat{\bf k}\times{\bxi})-{\bxi}\times({\bf k}\times{\bf g}).
\end{equation}
Because hydrodynamic acoustic waves are longitudinal to lowest order in
$k^{-1}$, ${\bxi}\propto{\bf k}$, and the term involving the flux vanishes.
In addition, the second term does no work as it is perpendicular to the
displacement.  Hence these terms contribute nothing to the growth rate - the
destabilizing radiation pressure force that would be present in a Thomson
scattering medium is exactly cancelled due to the hydrostatic equilibrium
of the unperturbed flow, at least for longitudinal waves.
In the high-$k$ limit, the modes with density perturbations
that can couple to the equilibrium radiation force are
{\it necessarily} longitudinal.  This is why gravity waves do not experience
radiative driving in the short wavelength limit.

Taking the scalar product of equation (\ref{dxidthighk}) with the wave vector
${\bf k}$, we obtain
\begin{eqnarray}
{\partial^2\over\partial t^2}({\bf k}\cdot{\bxi})
&=&-k^2\cgas^2\left({\bf k}\cdot{\bxi}\right)
+ik^2({\bf k}\cdot{\bxi}){\kappaf\over c}\Biggl\{
{4E\over3}\left({\omega\over k^2}\right)+{4Ec\omegaa(\gamma-1)^2\over\kappaf
\rho\omega}\nonumber\\
&-&{1\over k^2}({\bf k}\cdot{\bf F})\left[\thetarho+(\gamma-1)
\thetatg\right]\Biggr\}+{\cal O}(k)|{\bxi}|.
\label{dxidthighk2}
\end{eqnarray}
Replacing $\partial/\partial t$ with $-i\omega$, we immediately
recover equation (\ref{omeganocouphydro}) for the mode frequency
and growth rate.

We may repeat this procedure to better understand the instability in the limit
where the gas and radiation exchange heat extremely rapidly.  The
gas and radiation temperatures are then locked together, so that
\begin{equation}
\delta T_{\rm g}=\delta T_{\rm r}\equiv \delta T={T\over4E}\delta E=T\left(
{\delta p\over p}-{\delta\rho\over\rho}\right).
\label{eqdt}
\end{equation}
In this regime
we must replace the separate gas and radiation energy equations with a
single, combined energy equation which, when perturbed, gives
\begin{eqnarray}
-i\omega\left({\delta p\over\gamma-1}+\delta E\right)&=&
-{k^2c\over3\kappaf\rho}\delta E
-i\omega\left({\cgas^2\over\gamma-1}+3\crad^2\right)\delta\rho
-\rho T\delta{\bf v}\cdot{\bnabla}(S_{\rm g}+S_{\rm r})\nonumber\\
&+&i{\bf k}\cdot{\bf F}\left[(1+\thetarho){\delta\rho\over\rho}+
(\thetatg+\thetatr){\delta T\over T}\right].
\label{eqdelock}
\end{eqnarray}
Once again, equation (\ref{eqdelock}) immediately implies that
$\delta E\sim{\cal O}(k^{-1})\delta\rho$ for short wavelength acoustic
waves.  Rapid radiative diffusion smooths out both gas and radiation
temperature fluctuations in this case.  Equation (\ref{eqdt}) then
implies that $\delta p=c_{\rm i}^2\delta\rho$ to lowest order, i.e.
that the acoustic waves are isothermal in this limit.  Equation
(\ref{eqdelock}) can then be solved for the radiation pressure perturbation,
\begin{equation}
{\delta E\over3}=-{i\kappaf\over c}\left(p+{4E\over3}\right)
\left({\omega\over k^2}\right)\delta\rho+{i\kappaf\over k^2c}
({\bf k}\cdot{\bf F})(1+\thetarho)\delta\rho+{\cal O}(k^{-2})\delta\rho.
\label{dprhighklock}
\end{equation}
From equation (\ref{eqdt}), the gas pressure perturbation is given to
the same order by
\begin{equation}
\delta p=c_{\rm i}^2\delta\rho
-{i\kappaf\over c}\left(1+{3p\over4E}\right)p
\left({\omega\over k^2}\right)\delta\rho+{i\kappaf\over k^2c}
\left({3p\over4E}\right)
({\bf k}\cdot{\bf F})(1+\thetarho)\delta\rho+{\cal O}(k^{-2})\delta\rho.
\label{dpghighklock}
\end{equation}

The first term on the right hand side of equation (\ref{dprhighklock})
and the second term on the right hand side of equation (\ref{dpghighklock})
are the damping terms due to radiative diffusion.  The last terms in both
equations are the potentially destabilizing terms from the equilibrium radiation
flux acting on density and opacity fluctuations.  Note that the opacity
derivative with respect to temperature, $\thetatg$, has been lost because
the gas temperature is locked to the uniform radiation temperature.  There
is also no buoyancy response in the pressure perturbations to this order,
as expected because of the suppression of gravity waves in this limit.

Using equations (\ref{dprhighklock})-(\ref{dpghighklock}),
the gas momentum equation (\ref{dgasmom}) becomes
\begin{eqnarray}
\rho{\partial\delta{\bf v}\over\partial t}
&=&-i{\bf k}c_{\rm i}^2\delta\rho
-{\kappaf\over c}{\bf k}\left(1+{3p\over4E}\right)\left({4E\over3}+p\right)
\left({\omega\over k^2}\right)\delta\rho\nonumber\\
&+&{\kappaf\over k^2c}{\bf k}\left({\bf k}\cdot{\bf F}\right)
\left(1+{3p\over4E}\right)\left(1+\Theta_\rho\right)\delta\rho
+{\bf g}\delta\rho+{\cal O}(k^{-1})\delta\rho.
\label{dvdthighklocked}
\end{eqnarray}
Once again, the destabilizing force density term is proportional to
${\bf k}({\bf k}\cdot{\bf F})\delta\rho$, and the geometry of Figure
4 still applies.  Repeating the same steps as before, we obtain
\begin{eqnarray}
{\partial^2{\bxi}\over\partial t^2}&=&-{\bf k}c_{\rm i}^2({\bf k}\cdot{\bxi})
\nonumber\\
&+&i({\bf k}\cdot{\bxi}){\bf k}{\kappaf\over c}\left(1+{3p\over4E}\right)
\left[\left({4E\over3}+p\right){\omega\over k^2}-{1\over k^2}({\bf k}\cdot
{\bf F})\thetarho\right]\nonumber\\
&-&i\left[\left(1+{3p\over4E}\right){\kappaf\over k^2c}{\bf k}({\bf k}\cdot
{\bf F})({\bf k}\cdot{\bxi})-{\bf k}c_{\rm i}^2{\bxi}\cdot{\bnabla}\ln\rho
+{\bf g}({\bf k}\cdot{\bxi})\right]+{\cal O}(k^0)|{\bxi}|,
\label{dxidthighklock}
\end{eqnarray}
which should be compared with equation (\ref{dxidthighk}).  The destabilizing
${\bf k}\cdot{\bf F}$ term that would remain in a pure Thomson scattering
atmosphere again competes with other forces in the last term in square
brackets.  Using the equilibrium LTE condition (\ref{eqlte}), the hydrostatic
equilibrium equation (\ref{mechequil}), and the radiative diffusion equation
(\ref{raddiff}), the equilibrium density gradient is given by
\begin{equation}
c_{\rm i}^2{\bnabla}\ln\rho={\bf g}+\left(1+{3p\over4E}\right)
{\kappaf\over c}{\bf F}.
\end{equation}
The last term in square brackets in equation (\ref{dxidthighklock}) is
therefore
\begin{equation}
\left(1+{3p\over4E}\right){\kappaf\over k^2c}{\bf k}({\bf k}\cdot
{\bf F})({\bf k}\cdot{\bxi})-{\bf k}c_{\rm i}^2{\bxi}\cdot{\bnabla}\ln\rho
+{\bf g}({\bf k}\cdot{\bxi})=-\left(1+{3p\over4E}\right){\kappaf\over c}{\bf k}
(\hat{\bf k}\times{\bf F})\cdot(\hat{\bf k}\times{\bxi})-{\bxi}\times({\bf k}
\times{\bf g}).
\end{equation}
Once again, these terms do not affect the growth rate for longitudinal
waves.  After taking the scalar product of equation (\ref{dxidthighklock})
with ${\bf k}$, we finally obtain
\begin{eqnarray}
{\partial^2\over\partial t^2}\left({\bf k}\cdot{\bxi}\right)
&=&-k^2c_{\rm i}^2\left({\bf k}\cdot{\bxi}\right)
+ik^2({\bf k}\cdot{\bxi}){\kappaf\over c}\left(1+{3p\over4E}\right)
\Biggl[\left({4E\over3}+p\right){\omega\over k^2}\nonumber\\
&-&{1\over k^2}\left({\bf k}\cdot{\bf F}\right)\thetarho\Biggr]+
{\cal O}(k)|{\bxi}|,
\label{dxidthighklock2}
\end{eqnarray}
from which we recover the growth rate (\ref{omegalockhydro}).

\subsection{Relation to the NAR Approximation}

Gautschy \& Glatzel (1990) have shown that strange modes are well-described
by a so-called non-adiabatic reversible (NAR) approximation.
Mathematically, this approximation reduces to taking the scalar product
of the perturbed radiative diffusion equation (\ref{eqdeltaf}) with ${\bf k}$,
and then setting ${\bf k}\cdot\delta{\bf F}=0$.  One also ignores the
radiation energy equation (\ref{eqde}).  In addition,
if the gas and radiation temperatures are locked together by rapid
emission and absorption of photons, the gas energy equation
(\ref{eqdu}) is also ignored.  The perturbed radiation pressure
is then given by
\begin{equation}
-ik^2{\delta E\over3}={\kappaf\rho\over c}{\bf k}\cdot{\bf F}
\left({\delta\rho\over\rho}+{\delta\kappaf\over\kappaf}\right).
\end{equation}
This equation recovers our high $k$ expressions (\ref{dprhighk}) and
(\ref{dprhighklock}) for the
perturbed radiation pressure, except that the NAR approximation cannot
reproduce the radiation diffusion damping terms (as it neglects the
radiation energy equation).  Indeed, if we use equations (\ref{eqdeltaf}),
(\ref{dprhighk}) and (\ref{dprhighklock}) to derive the perturbed radiative
flux, we find
\begin{equation}
\delta{\bf F}=
\cases{
-3\crad^2\left({\omega\over k^2}\right){\bf k}\delta\rho
+\left[{{\bf k}({\bf k}\cdot{\bf F})\over k^2}-{\bf F}\right]\left[
1+\thetarho+(\gamma-1)\thetatg\right]{\delta\rho\over\rho}+{\cal O}(k^{-1})
{\delta\rho}
& if $\delta T_{\rm r}\ne\delta T_{\rm g}$,\cr
-(c_{\rm i}^2+3\crad^2)\left({\omega\over k^2}\right){\bf k}\delta\rho
+\left[{{\bf k}({\bf k}\cdot{\bf F})\over k^2}-{\bf F}\right]
(1+\thetarho){\delta\rho\over\rho}+{\cal O}(k^{-1}){\delta\rho}
& if $\delta T_{\rm r}=\delta T_{\rm g}$.
}
\end{equation}
When unstable driving (the second term on the right hand sides) strongly
dominates
radiative diffusion damping (the first term on the right hand sides),
the NAR approximation that ${\bf k}\cdot\delta{\bf F}=0$ is excellent.

By not invoking the NAR approximation, however, we are able to deduce the
threshold criteria listed in Table 1 required for the unstable driving
to exceed the damping by radiative diffusion.  At the same time, we
can also calculate the actual luminosity perturbation associated with
the unstable wave.

\section{MHD Instabilities in a Static Medium}

We now extend the analysis of the previous section by including
magnetic stresses.  It turns out that magnetic fields widen the
domain of the acoustic wave instabilities to include even pure Thomson
scattering media.  This is because magnetic tension endows acoustic waves
with a mixed longitudinal/transverse character, except in the special
cases of propagation purely along or perpendicular to the magnetic field.

\subsection{Short Wavelength Limit}

The dispersion relation (\ref{staticdisp}) again factors easily in the
short wavelength limit:
\begin{eqnarray}
0&=&\left[\omega^2-({\bf k}\cdot{\bf v}_{\rm A})^2\right]
\left[\omega^4-\omega^2k^2(v_{\rm A}^2+\cgas^2)+k^2\cgas^2({\bf k}
\cdot{\bf v}_{\rm A})^2\right]\nonumber\\
&\times&\left(\omega+{ick^2\over3\kappaf\rho}\right)
\left[\omega+{4i\omegaa E(\gamma-1)\over\gamma p}\right].
\label{highkmhd}
\end{eqnarray}
From left to right, the factors correspond to Alfv\'en waves, fast and
slow magnetosonic waves, the damped radiative diffusion mode, and a damped
$\delta T_{\rm g}\ne\delta T_{\rm r}$ mode.\footnote{In our previous paper
(BS01), we
found only seven nonzero frequency modes in our analysis of the perturbed
radiation MHD equations.  This is because we did not include heat exchange
between the gas and radiation through absorption and emission, a process
which turns out to be important in most applications (see section 7).
Setting $\omegaa=0$ in equation (\ref{highkmhd}) produces an additional zero
frequency mode.}
The gravity waves of equation
(\ref{highkhydro}) have been lost because they are dominated by magnetic
fields in this short wavelength limit, at least when
${\bf k}\cdot{\bf v}_{\rm A}\ne0$, i.e. for wave vectors that are not
orthogonal to the equilibrium magnetic field.

The first order corrections to the Alfv\'en modes vanish, so that
$\omega=\pm{\bf k}\cdot{\bf v}_{\rm A}+{\cal O}(k^{-1})$.  Because Alfv\'en
waves are purely transverse and thus do not involve density fluctuations
in this short wavelength limit, they do not couple to the radiation physics.

To the same order, the magnetosonic wave frequencies are given by
\begin{eqnarray}
\omega&=&\pm k\vph-{i\kappaf\over2c\vph}\left[{\vph^2-(\hat{\bf k}\cdot
{\bf v}_A)^2\over2\vph^2-v_{\rm A}^2-\cgas^2}\right]\left\{{4E\vph\over3}+
{4Ec\omega_{\rm a}(\gamma-1)^2\over\kappaf\rho\vph}
\mp\left(\hat{\bf k}\cdot{\bf F}\right)\left[
\Theta_\rho+(\gamma-1)\Theta_{T{\rm g}}\right]\right\}\nonumber\\
& &\pm{i\kappaf\over2c\vph(2\vph^2-v_{\rm A}^2-\cgas^2)}
(\hat{\bf k}\cdot{\bf v}_{\rm A})(\hat{\bf k}\times{\bf v}_{\rm A})
\cdot(\hat{\bf k}\times{\bf F})+{\cal O}(k^{-1}),
\label{omeganocoupmhd}
\end{eqnarray}
where $\vph>0$ is the phase speed of the fast and slow magnetosonic waves,
given by
\begin{equation}
\vph^2\equiv{1\over2}\left\{v_{\rm A}^2+\cgas^2\pm\left[(v_A^2+\cgas^2)^2-
4(\hat{\bf k}\cdot{\bf v}_{\rm A})^2\cgas^2\right]^{1/2}\right\}.
\label{magnetosonic}
\end{equation}
Equation (\ref{omeganocoupmhd}) bears a strong resemblance to its
hydrodynamic counterpart (\ref{omeganocouphydro}).  The gas sound speed
$\cgas$ has been replaced by the phase velocity of the relevant magnetosonic
wave $\vph$, and the ``hydrodynamic'' corrections to the mode frequency have
been multiplied by a factor
\begin{equation}
\left({\vph^2-(\hat{\bf k}\cdot
{\bf v}_A)^2\over2\vph^2-v_{\rm A}^2-\cgas^2}\right)\ge0.
\end{equation}
The last term in equation (\ref{omeganocoupmhd}) is a new, destabilizing term
that exists only in the presence of magnetic fields.  As we discuss in
more detail in section 6 below, its physical origin
ultimately arises from the ability of magnetic stresses to support velocity
perturbations that are not purely longitudinal, i.e. that are not parallel
or antiparallel to the wave vector ${\bf k}$.  Note that this term vanishes
for wave vectors that are either parallel or perpendicular to the equilibrium
magnetic field.  In each of these cases the compressive magnetosonic wave is
longitudinal.

In the rapid heat exchange ($\omega_{\rm a}\rightarrow\infty$) limit, one
of the eight modes is lost for the same reason as in the hydrodynamic case,
and the remaining seven modes factor in the short wavelength limit as
\begin{eqnarray}
0&=&\left[\omega^2-({\bf k}\cdot{\bf v}_{\rm A})^2\right]
\left[\omega^4-\omega^2k^2(v_{\rm A}^2+c_{\rm i}^2)+k^2c_{\rm i}^2({\bf k}
\cdot{\bf v}_{\rm A})^2\right]\nonumber\\
&\times&\left[\omega+{ick^2\over3\kappaf\rho}
\left({4(\gamma-1)E\over p+4(\gamma-1)E}\right)\right].
\label{highkmhdlocked}
\end{eqnarray}
Once again, instabilities only occur in the fast and slow magnetosonic
waves, whose frequencies are given to first order by
\begin{eqnarray}
\omega&=&\pm k\vph-{i\kappaf\over2c\vph}\left[{\vph^2-(\hat{\bf k}\cdot
{\bf v}_A)^2\over2\vph^2-v_{\rm A}^2-c_{\rm i}^2}\right]
\left(1+{3p\over4E}\right)\left\{\left({4E\over3}+p\right)\vph
\mp\left(\hat{\bf k}\cdot{\bf F}\right)\Theta_\rho
\right\}\nonumber\\
& &\pm{i\kappaf\over2c\vph(2\vph^2-v_{\rm A}^2-c_{\rm i}^2)}
\left(1+{3p\over4E}\right)
(\hat{\bf k}\cdot{\bf v}_{\rm A})(\hat{\bf k}\times{\bf v}_{\rm A})
\cdot(\hat{\bf k}\times{\bf F})+{\cal O}(k^{-1}),
\label{omegalockmhd}
\end{eqnarray}
where $\vph$ is given by equation (\ref{magnetosonic}) with $\cgas$ replaced
by the isothermal sound speed $c_{\rm i}$.

Equation (\ref{magnetosonic}) implies that $(2\vph^2-v_{\rm A}^2-\cgas^2)$
is positive for fast modes and negative for slow modes.  Hence if the last
``magnetic'' destabilizing term in equation (\ref{omeganocoupmhd}), or
(\ref{omegalockmhd}), dominates the hydrodynamic terms, the unstable fast
and slow waves will have {\it opposite} vector phase velocities.  It is
also easy to see that in this case the slow mode will always grow faster
than the fast mode.  However, if hydrodynamic driving (the opacity derivative
terms in eqs. [\ref{omeganocoupmhd}] and [\ref{omegalockmhd}]) dominates
magnetic driving, then whichever of the two modes has the larger density
perturbation for a given velocity perturbation will be the one that grows
fastest.  If
$v_{\rm A}^2\gg\cgas^2$, the slow mode grows faster, whereas if
$v_{\rm A}^2\ll\cgas^2$, the fast mode grows faster.

\subsection{Short Wavelength Limit With
${\bf k}\cdot{\bf v}_{\rm A}=0$:  Recovery of Gravity Waves}

If we consider wave vectors that are orthogonal to the equilibrium magnetic
field, then the short wavelength limit recovers the fast magnetosonic modes,
the radiative diffusion mode, and gives us a cubic equation describing
coupled gravity and $\delta T_{\rm g}\ne\delta T_{\rm r}$ modes:
\begin{eqnarray}
0&=&(v_{\rm A}^2+\cgas^2)\omega^3+{4i(\gamma-1)\omegaa E\over p}
(v_{\rm A}^2+c_{\rm i}^2)\omega^2
-\left[1-(\hat{\bf k}\cdot\hat{\bf z})^2\right]
\left(\cgas^2N_{\rm g}^2+v_{\rm A}^2{\bf g}\cdot{\bnabla}\ln\rho\right)\omega
\nonumber\\
&-&\left[1-(\hat{\bf k}\cdot\hat{\bf z})^2\right]
{4i(\gamma-1)\omegaa E\over p}
v_{\rm A}^2{\bf g}\cdot{\bnabla}\ln\rho.
\end{eqnarray}
The reason we recover gravity waves in this limit is that magnetic tension
does not exist when ${\bf k}\cdot{\bf v}_{\rm A}=0$, and therefore no
longer dominates buoyancy at short wavelengths.
For weak thermal coupling between the gas and radiation, $\omegaa\rightarrow0$,
the roots of this dispersion relation correspond to magnetically altered
gravity waves in the gas,
\begin{equation}
\omega^2={1-(\hat{\bf k}\cdot\hat{\bf z})^2\over v_{\rm A}^2+\cgas^2}
\left(\cgas^2N_{\rm g}^2+v_{\rm A}^2{\bf g}\cdot{\bnabla}\ln\rho\right),
\end{equation}
and a mode
\begin{equation}
\omega\simeq-{4i(\gamma-1)\omegaa E\over p}\left({v_{\rm A}^2
{\bf g}\cdot{\bnabla}\ln\rho\over
\cgas^2N_{\rm g}^2+v_{\rm A}^2{\bf g}\cdot{\bnabla}\ln\rho}\right).
\end{equation}
In the rapid heat exchange limit, $\omegaa\rightarrow\infty$, the last mode
is lost and the gravity waves depend only on the vertical density profile,
\begin{equation}
\omega^2={1-(\hat{\bf k}\cdot\hat{\bf z})^2\over v_{\rm A}^2+c_{\rm i}^2}
v_{\rm A}^2{\bf g}\cdot{\bnabla}\ln\rho.
\end{equation}

\subsection{The Limit of Zero Gas Pressure: Photon Bubbles}

Just as in the hydrodynamic case, a number of authors (Arons 1992, Gammie 1998)
have investigated the slow mode instability in the limit of zero
gas pressure.  Because the phase speed of the slow mode then vanishes,
the instability then loses its obvious connection to this mode.

Setting $p=0=\cgas^2$ in the dispersion relation (\ref{staticdisp}), and
then assuming $\omega\propto k^{1/2}$ as $k\rightarrow\infty$, we find two
modes with
\begin{eqnarray}
\omega^2&=&{i\kappaf\over cv_{\rm A}^2}{\bf k}\cdot{\bf v}_{\rm A}
\left[(\hat{\bf k}\cdot{\bf v}_{\rm A})(\hat{\bf k}\cdot{\bf F})\thetarho
-(\hat{\bf k}\times{\bf v}_{\rm A})\cdot(\hat{\bf k}\times{\bf F})\right]
\nonumber\\
&=&-{i\over v_{\rm A}^2}{\bf k}\cdot{\bf v}_{\rm A}
\left[(\hat{\bf k}\cdot{\bf v}_{\rm A})(\hat{\bf k}\cdot{\bf g})\thetarho
-(\hat{\bf k}\times{\bf v}_{\rm A})\cdot(\hat{\bf k}\times{\bf g})\right].
\label{phbubblemhd}
\end{eqnarray}
Just as in the hydrodynamic case [eq. (\ref{phbubblehydro})], equation
(\ref{phbubblemhd}) can be obtained from the thermally locked slow mode
frequency in equation (\ref{omegalockmhd}) by first squaring the
frequency and then taking the limit of zero gas pressure.  Again, we
lose the damping terms in this limit, which determine the instability
threshold for short wavelength modes.

If we first assume negligible thermal coupling between the gas and radiation,
and then take the limit of zero gas pressure, we obtain instead
\begin{eqnarray}
\omega^2&=&{i\kappaf\over cv_{\rm A}^2}{\bf k}\cdot{\bf v}_{\rm A}
\left\{(\hat{\bf k}\cdot{\bf v}_{\rm A})(\hat{\bf k}\cdot{\bf F})[\thetarho
+(\gamma-1)\thetatg]
-(\hat{\bf k}\times{\bf v}_{\rm A})\cdot(\hat{\bf k}\times{\bf F})\right\}
\nonumber\\
&=&-{i\over v_{\rm A}^2}{\bf k}\cdot{\bf v}_{\rm A}
\left\{(\hat{\bf k}\cdot{\bf v}_{\rm A})(\hat{\bf k}\cdot{\bf g})[\thetarho
+(\gamma-1)\thetatg]
-(\hat{\bf k}\times{\bf v}_{\rm A})\cdot(\hat{\bf k}\times{\bf g})\right\}.
\label{phbubblemhdnocoup}
\end{eqnarray}

In a Thomson scattering medium, $\thetarho=\thetatg=0$, and equations
(\ref{phbubblemhd})-(\ref{phbubblemhdnocoup}) are
identical to the photon bubble dispersion relation derived by
Gammie [1998, his eq. (41)].  As noted both by Gammie (1998) and BS01,
this growth rate only applies for wavelengths longer than the gas pressure
scale height $\sim\cgas^2/g$.  Unless this wavelength scale is optically
thin, in which case our diffusion-based analysis breaks down, then the
maximal growth rate occurs at wavelengths shorter than the gas pressure
scale height, and is given by equations (\ref{omeganocoupmhd})
or (\ref{omegalockmhd}).

There has been some confusion as to how Gammie's (1998) work on photon
bubbles relates to the original analysis performed by Arons (1992), who
also assumed a pure Thomson scattering medium with zero gas pressure.
Gammie's (1998) result can in fact be obtained as the short wavelength
limit of Arons' (1992) dispersion relation [his eq. (33)], at least within
the latter's assumed vertical magnetic field geometry.  Arons' (1992)
numerical growth rates all climb toward higher wavenumber, and must
therefore eventually recover the $k^{1/2}$ dependence of equation
(\ref{phbubblemhd}), provided the medium remains optically thick at these
short wavelengths.  Arons (1992) was mainly interested in long wavelengths
where radiation diffusion was slow compared to the (radiation) acoustic wave
period.  Gammie (1998), on the other hand, was working in the rapid diffusion
regime at shorter wavelengths.

\subsection{Numerical Results}

Table 2 summarizes the instability thresholds, growth rates, and characteristic
wavenumbers for the magnetoacoustic wave instabilities in media with a variety
of ratios of magnetic, gas thermal, and radiation energy densities.
We have assumed a pure Thomson scattering equilibrium throughout this
table, so that the hydrodynamic driving forces that we addressed in sections
3 and 4 vanish, and only the magnetically-based driving remains.

Figure 5 illustrates growth rates of unstable slow waves for small values of
the thermal coupling frequency $\omega_{\rm th}$ for a radiation pressure
dominated equilibrium, for which the asymptotic high wavenumber growth
rate is accurately given by equation (\ref{omeganocoupmhd}).  Just as
in the hydrodynamic case, the short wavelength growth rate declines with
increasing thermal coupling.  This continues until $\omega_{\rm th}$ exceeds
$g/\cgas$, beyond which the gas and radiation temperatures are effectively
locked and the waves reach unstable asymptotic growth rates given by
equation (\ref{omegalockmhd}), as shown in Figure 6.

Just as in the hydrodynamic case, the unstable growth rate for high thermal
coupling only extends over a finite range of wavenumbers.  Above a critical
cutoff wavenumber, the waves again become damped.  After numerical
experimentation, we deduced the physical dependencies of these cutoff
wavenumbers, and they are listed in Table 2.  Figure 7 illustrates the
accuracy of these formulas in two specific cases, one with $v_{\rm A}>\cgas$
and one with $v_{\rm A}<\cgas$.

\section{Physics of the MHD Wave Instabilities}

The magnetoacoustic wave instabilities are much more robust than their
hydrodynamic counterparts, and in particular they survive unscathed even
in pure Thomson scattering media.  In order to see how this arises physically,
we repeat the analysis of section 4 here to re-examine the competing forces
acting on a parcel of gas in the wave.  There are now two relevant waves
that possess density fluctuations: the fast and slow magnetosonic waves.
For reference, their polarizations are shown in Figure 8.

Using equation (\ref{dfluxfreezing}) to eliminate the magnetic field
perturbation, and equations (\ref{dpghighk})-(\ref{dprhighk})
to eliminate the total pressure
perturbation, the perturbed gas momentum equation (\ref{dgasmom}) may
be written in the high wavenumber limit as
\begin{eqnarray}
{\partial^2{\bxi}\over\partial t^2}&=&-({\bf k}\cdot{\bf v}_{\rm A})^2{\bxi}
+({\bf k}\cdot{\bxi})({\bf k}\cdot{\bf v}_{\rm A}){\bf v}_{\rm A}
+({\bf k}\times{\bf v}_{\rm A})\cdot({\bf v}_{\rm A}\times{\bxi}){\bf k}
-{\bf k}\cgas^2({\bf k}\cdot{\bxi})\nonumber\\
&+&i({\bf k}\cdot{\bxi}){\bf k}{\kappaf\over c}\left\{
{4E\over3}\left({\omega\over k^2}\right)+{4Ec\omegaa(\gamma-1)^2\over\kappaf
\rho\omega}-{1\over k^2}({\bf k}\cdot{\bf F})\left[\thetarho+(\gamma-1)
\thetatg\right]\right\}\nonumber\\
&+&i\left[{\kappaf\over c}{\bf k}(\hat{\bf k}\times{\bf F})\cdot
(\hat{\bf k}\times{\bxi})+{\bxi}\times({\bf k}\times{\bf g})\right]+
{\cal O}(k^0)|{\bxi}|.
\label{mhdmomhighk}
\end{eqnarray}

Apart from the additional magnetic terms, this equation is of course identical
to its hydrodynamic counterpart, equation (\ref{dxidthighk}), which we
discussed above in section 4.  There we found that the last term in square
brackets did not affect the growth rate.  This continues to be true for
the piece of this term that involves the gravitational acceleration ${\bf g}$,
as this piece does no work.  However the first piece that involves the
radiative flux now can play a role because magnetosonic waves are, in
general, {\it not} longitudinal due to the effects of magnetic tension.

In the short wavelength limit, the fast and slow magnetosonic
modes described by equation (\ref{mhdmomhighk})
are polarized in the plane of ${\bf k}$ and ${\bf v}_{\rm A}$, and their
Lagrangian displacements may be written as
\begin{equation}
{\bxi}=[v_{\rm ph}^2{\bf k}-({\bf k}\cdot{\bf v}_{\rm A}){\bf v}_{\rm A}]\psi
\equiv{\bepsilon}\psi,
\end{equation}
where $\psi$ is some complex scalar amplitude.
Taking the scalar product of equation (\ref{mhdmomhighk}) with the
polarization vector ${\bepsilon}$, we obtain
\begin{eqnarray}
{\partial^2\over\partial t^2}({\bepsilon\cdot\bxi})&=&-k^2v_{\rm ph}^2
({\bepsilon\cdot\bxi})\nonumber\\
&+&ik^2[v_{\rm ph}^2-(\hat{\bf k}\cdot{\bf v}_{\rm A})^2]({\bf k}\cdot{\bxi})
{\kappaf\over c}\Biggl\{{4E\over3}\left({\omega\over k^2}\right)+
{4Ec\omegaa(\gamma-1)^2\over\kappaf\rho\omega}\nonumber\\
&-&{1\over k^2}({\bf k}\cdot{\bf F})\left[\thetarho+(\gamma-1)
\thetatg\right]\Biggr\}
-i{\kappaf\over c}({\bf k}\cdot{\bf v}_{\rm A})({\bf k}\times{\bf v}_{\rm A})
\cdot({\bf k}\times{\bf F})({\bf k}\cdot{\bxi}),
\label{mhdmomepshighk}
\end{eqnarray}
where, in the last term, we have used the fact that
$({\bf k}\cdot{\bepsilon})({\bf k}\times{\bxi})=({\bf k}\cdot{\bxi})
({\bf k}\times{\bepsilon})$.  The expression for the growth rate, equation
(\ref{omeganocoupmhd}), now follows immediately from equation
(\ref{mhdmomepshighk}) and the fact that, to lowest order in $k^{-1}$,
\begin{equation}
{{\bf k}\cdot{\bxi}\over{\bepsilon}\cdot{\bxi}}={1\over2\vph^2-\cgas^2-
v_{\rm A}^2}.
\label{kxikeps}
\end{equation}
As illustrated in Figure 8, the polarizations of the fast and slow waves
for fixed ${\bf k}$ are orthogonal, and this normalized projection of ${\bxi}$
onto ${\bf k}$ therefore always has opposite signs for the two modes.  Hence the
unstable fast and slow modes propagate in different
directions if the magnetic driving term dominates over the hydrodynamic
terms.

To understand the physics of the MHD driving further, consider again the
relevant force density ($\rho$ times the second to last term in equation
[\ref{mhdmomhighk}]), which can be written as
\begin{equation}
i{\kappaf\rho\over c}{\bf k}(\hat{\bf k}\times{\bf F})\cdot(\hat{\bf k}\times
{\bxi})=-i{\kappaf\rho\over c}{\bf k}\left[(\hat{\bf k}
\cdot{\bxi})(\hat{\bf k}\cdot{\bf F})-({\bxi}\cdot{\bf F})\right].
\label{mhddriving}
\end{equation}
MHD driving results fundamentally from a breakdown in cancellation of the two
force densities on the right hand side of this equation, whenever the waves
are not longitudinal.
The first, $(\kappaf/c)\hat{\bf k}(\hat{\bf k}\cdot {\bf F})\delta\rho$,
is the driving force based on density, not opacity, fluctuations
that we illustrated in Figure 4.  The second arises from a
change in the radiation pressure along a fluid displacement, and
is $i(\kappaf\rho/c){\bf k}({\bxi}\cdot{\bf F})$.  If we ignore the
radiation diffusion damping and opacity fluctuations in equation
(\ref{dprhighk}), these two force densities simply combine
to give the negative gradient of the Lagrangian perturbation in radiation
pressure $\Delta E/3$,
\begin{equation}
i{\kappaf\rho\over c}{\bf k}(\hat{\bf k}\times{\bf F})\cdot(\hat{\bf k}\times
{\bxi})=-{\bnabla}\left({1\over3}\Delta E\right)=
-{\bnabla}\left({\delta E\over3}
+{{\bxi}\cdot{\bnabla}E\over3}\right).
\label{pradlagrange}
\end{equation}
For any purely longitudinal wave (${\bxi}$ and ${\bf k}$ parallel),
the two force densities exactly
cancel, and the Lagrangian radiation pressure perturbation then vanishes
(again ignoring radiation diffusion damping and opacity fluctuations).
For an MHD wave with fixed
density perturbation $\delta\rho$ and fixed ${\bf k}$, magnetic
tension can alter the direction of ${\bxi}$ off of ${\bf k}$,
provided propagation is neither along or perpendicular to the
field.  Spatial gradients in the flow (which determine the
radiative heat flow) are no longer parallel to fluid displacements.
The second force density on the right hand side of equation (\ref{mhddriving})
will now be greater or less than the first, depending on the type of MHD wave
(slow or fast) and the propagation direction.  Note that the two force
densities also cancel for vertical propagation.

The physics is very similar in the case where gas and radiation exchange
heat rapidly.  The perturbed gas momentum equation is then
\begin{eqnarray}
{\partial^2{\bxi}\over\partial t^2}&=&-({\bf k}\cdot{\bf v}_{\rm A})^2{\bxi}
+({\bf k}\cdot{\bxi})({\bf k}\cdot{\bf v}_{\rm A}){\bf v}_{\rm A}
+({\bf k}\times{\bf v}_{\rm A})\cdot({\bf v}_{\rm A}\times{\bxi}){\bf k}
-{\bf k}c_{\rm i}^2({\bf k}\cdot{\bxi})\nonumber\\
&+&i({\bf k}\cdot{\bxi}){\bf k}{\kappaf\over c}\left(1+{3p\over4E}\right)
\left[\left({4E\over3}+p\right){\omega\over k^2}-{1\over k^2}
({\bf k}\cdot{\bf F})\thetarho\right]\nonumber\\
&+&i\left[\left(1+{3p\over4E}\right)
{\kappaf\over c}{\bf k}(\hat{\bf k}\times{\bf F})\cdot
(\hat{\bf k}\times{\bxi})+{\bxi}\times({\bf k}\times{\bf g})\right]+
{\cal O}(k^0)|{\bxi}|.
\label{mhdmomhighklock}
\end{eqnarray}
Taking the scalar product of this equation with ${\bepsilon}$, we obtain
\begin{eqnarray}
{\partial^2\over\partial t^2}({\bepsilon\cdot\bxi})&=&-k^2v_{\rm ph}^2
({\bepsilon\cdot\bxi})\nonumber\\
&+&ik^2[v_{\rm ph}^2-(\hat{\bf k}\cdot{\bf v}_{\rm A})^2]({\bf k}\cdot{\bxi})
{\kappaf\over c}\left(1+{3p\over4E}\right)\Biggl[\left({4E\over3}+p\right)
{\omega\over k^2}\nonumber\\
&-&{1\over k^2}({\bf k}\cdot{\bf F})\thetarho\Biggr]
-i\left(1+{3p\over4E}\right){\kappaf\over c}
({\bf k}\cdot{\bf v}_{\rm A})({\bf k}\times{\bf v}_{\rm A})
\cdot({\bf k}\times{\bf F})({\bf k}\cdot{\bxi}).
\label{mhdmomepshighklock}
\end{eqnarray}
With the help of equation (\ref{kxikeps}) with $\cgas$ replaced by the
isothermal sound speed $c_{\rm i}$, this now gives us the growth
rate of equation (\ref{omegalockmhd}).

We have extended Hearn's (1972) hydrodynamic acoustic wave analysis of
optically {\it thin} media to the magnetosonic
wave instabilities, and have found that the extra magnetic piece of the
instability vanishes completely in this regime.
Physically, this is easy to understand.
In the limit of vanishing optical depth, the dynamics of the radiation
field can be ignored, and the equilibrium flux ${\bf F}$ merely acts to globally
reduce the equilibrium gravitational acceleration ${\bf g}$ to
${\bf g}_{\rm eff}={\bf g}-\kappaf{\bf F}/c$.  Hence, apart from the
hydrodynamic driving terms that are proportional to logarithmic derivatives
of $\kappaf$, the
radiation flux enters the gas momentum equation only through a
${\bxi}\times({\bf k}\times{\bf g}_{\rm eff})$ term [the last term in
equations (\ref{mhdmomhighk}) and (\ref{mhdmomhighklock})].  This term
performs no work.  Presumably, there will still be some level of
instability in media with optical depths $\lta 1$, but an analysis
of this situation is beyond the scope of the present paper.

\section{Astrophysical Applications}

We intend to investigate applications of the theory presented in this paper
to particular astrophysical phenomena in future work.  For now,
we limit ourselves to a brief discussion of how these instabilities
might manifest themselves in accretion disks and stars.

\subsection{Accretion Disks}

A number of authors have considered applications of strange modes
(Glatzel \& Mehren 1996, Mehren-Baehr \& Glatzel 1999) and magnetic photon
bubbles or magnetoacoustic waves (Gammie 1998; BS01; Begelman 2001, 2002)
to radiation pressure dominated accretion disks around black holes.

The gas temperature is probably reasonably well-locked to the radiation
temperature in standard accretion disk models.  For stellar mass black holes,
the most important atomic absorption opacity is bremsstrahlung, for which the
Planck mean opacity is given in cgs units by
\begin{equation}
\kappap\simeq2.2\times10^{24}\bar{g}_{\rm P}\rho T_{\rm g}^{-7/2},
\label{kpbrems}
\end{equation}
where
\begin{equation}
\bar{g}_{\rm P}\equiv{h\over\kb T_{\rm g}}\int_0^\infty d\nu\bar{g}_{\rm ff}
e^{-h\nu/\kb T_{\rm g}}
\end{equation}
is an appropriate frequency-averaged Gaunt factor (Rybicki \& Lightman
1979).  Equation (\ref{kpbrems}) gives a lower limit to the Planck mean
opacity, which should also include bound-free and bound-bound contributions.
These contributions are generally more important for the cooler disks thought
to exist around supermassive black holes in active galactic nuclei.

By approximating $\bar{g}_{\rm ff}$ to be nearly constant inside the
frequency integrals, it is also straightforward to show that for
bremsstrahlung,
\begin{equation}
1+{1\over4}{\partial\ln\kappaj\over\partial\ln T_{\rm r}}\simeq{\pi^2\over6}
\simeq 1.6.
\end{equation}

For concreteness, consider a standard Shakura \& Sunyaev (1973) disk model
in which the anomalous stress is $\alpha$ times the total pressure.  Using
equations (\ref{omegaabs}) and (\ref{omegathermal}), the thermal coupling
frequency due to bremsstrahlung is
\begin{eqnarray}
\omega_{\rm th, bremss}&\equiv&\kappap\rho c
\left(1+{1\over4}{\partial\ln\kappaj\over\partial\ln T_{\rm r}}\right)
\left[{4(\gamma-1)E\over p}\right]\nonumber\\
&\simeq&
3\times10^4{\rm rad\, s}^{-1}\bar{g}_{\rm P}\eta^2\alpha^{-7/8}
\left({M\over\msun}\right)^{-7/8}\left({L\over L_{\rm Edd}}\right)^{-2}
\left({r\over r_{\rm g}}\right)^{27/16}.
\label{omegathbremss}
\end{eqnarray}
Here $L$ is the disk luminosity, $L_{\rm Edd}$ is the (Thomson scattering)
Eddington luminosity, $M$ is the black hole mass, $\eta\equiv L/(\dot{M}c^2)$
is the disk radiative efficiency, $r$ is the radius of the particular point
of interest in the disk, and $r_{\rm g}\equiv GM/c^2$ is the gravitational
radius of the black hole.  We have neglected all relativistic correction
factors in the disk model.  The particular numerical value of the
thermal coupling frequency in equation (\ref{omegathbremss}) assumes
vertically averaged, disk interior quantities.

The corresponding thermal coupling frequency due to Compton scattering is
\begin{equation}
\omega_{\rm th, Comp}\equiv\kappat\rho c
\left({k_{\rm B}T\over m_{\rm e}c^2}\right)
\left[{4(\gamma-1)E\over p}\right]\simeq2\times10^9
{\rm rad\, s}^{-1}\alpha^{-1}
\left({M\over\msun}\right)^{-1}\left({r\over r_{\rm g}}\right)^{-3/2}.
\label{omegathcomp}
\end{equation}

Comparing equations (\ref{omegathbremss}) and (\ref{omegathcomp}), we see
that Compton scattering generally dominates the thermal coupling of
gas and radiation in the inner regions of the disk, while bremsstrahlung
becomes more important further out, at least in the disk interior for
this class of models.  For hydrodynamic acoustic waves,
the requirement that the gas and radiation temperatures
be locked in the wave is that $\omega_{\rm th}$ exceed the sound
wave frequency on scales of order the gas scale height $\cgas^2/g$,
i.e. that $\omega_{\rm th}\cgas/g\gta1$.  For standard Shakura \&
Sunyaev (1973) models,
\begin{equation}
\omega_{\rm th, bremss}{\cgas\over g}\sim4\times10^{-4}\bar{g}_{\rm P}\eta^3
\alpha^{-1}\left({L\over L_{\rm Edd}}\right)^{-3}
\left({r\over r_{\rm g}}\right)^{9/2}
\label{omegathbremsscg}
\end{equation}
and
\begin{equation}
\omega_{\rm th, Comp}{\cgas\over g}\sim3\times10^1\eta
\alpha^{-9/8}\left({M\over\msun}\right)^{-1/8}
\left({L\over L_{\rm Edd}}\right)^{-1}
\left({r\over r_{\rm g}}\right)^{21/16}.
\label{omegathcompcg}
\end{equation}
Hence while bremsstrahlung may be unable to ensure tight thermal coupling
between the gas and the radiation in the innermost radii, Compton scattering
is easily adequate.  We stress, however, that these estimates are very
uncertain as they do not properly take into account the vertical stratification
in the disk.  In particular, we have assumed a gravitational acceleration
appropriate at the disk scale height but a temperature appropriate to
the disk midplane.  Because $\omega_{\rm th, Comp}\cgas/g\propto T^{3/2}$,
it could be that in regions near the disk photosphere, the
gas temperature may not be able to track the radiation temperature.

If the disk is treated hydrodynamically, then unstable acoustic wave
growth rates are likely to be much slower than the dynamical frequency
in the innermost parts of the disk because Thomson scattering dominates
the Rosseland mean opacity.
From Table 1, the maximum growth rate for hydrodynamic instabilities is roughly
given by
\begin{eqnarray}
\thetarho{g\over \cgas}&\sim&\frac{g}{\crad}\frac{\crad}{\cgas}\frac
{\kappa_{\rm a}}{\kappat}\sim\Omega\frac{\crad}{\cgas}
\frac{\kappa_{\rm a}}{\kappat}\nonumber\\
&\sim&3\times10^{-5}\Omega\bar{g}_{\rm P}\eta
\left({L\over L_{\rm Edd}}\right)^{-1}\left({r\over r_{\rm g}}\right)^{3/2}
\label{grestimate}
\end{eqnarray}
for a radiation pressure supported disk with $\kappa_{a}\sim\kappap$.

For example, Glatzel \& Mehren
(1996) found that a class of global strange mode instabilities
was present at 25~km radii in accretion disk models around
1~$\msun$ black holes with
accretion rates of $10^{-9}$~$\msun$~yr$^{-1}$.  Our estimate (\ref{grestimate})
then gives a growth rate $\sim5\times10^{-3}\Omega$, in remarkable agreement
with their calculated value $\sim3\times10^{-3}\Omega$.
Mehren-Baehr \& Glatzel (1999) also
note that this class of modes has growth rates that decline as the
ratio of gas pressure to radiation pressure increases, which is also
in agreement with our estimate (\ref{grestimate}).\footnote{Mehren-Baehr
\& Glatzel (1999) actually state that the growth rates are inversely
proportional to $\beta\equiv3p/E$, but they do not present direct
quantitative evidence for a dependence on $\crad^2/\cgas^2$ rather
than $\crad/\cgas$.}  In addition to these ``hot'' instabilities, Glatzel \&
Mehren (1996) and Mehren-Baehr \& Glatzel (1999) found a class of unstable
modes in colder regions of disk models where gas pressure was substantial.
They found that these modes required the presence of a vertical opacity peak
due to helium or hydrogen ionization, but that these modes also satisfied
the NAR approximation and were therefore likely to be related to strange
modes.  We speculate that these modes are related to
the local, WKB driving we find in the {\it gas}-pressure dominated
case.  Radiation pressure support is not essential for strange modes
provided the gas and radiation temperatures are locked, as these authors
implicitly assumed in their analysis.

It is almost certainly true, however, that the presence of magnetic fields
cannot be neglected in these flows, and these allow in principle much
faster growth rates through the slow mode instability (Gammie 1998, BS01).
From Table 2, the maximum growth rate in this case is
$\sim\Omega(\crad/\cgas){\rm min} [1,(v_{\rm A}/\cgas)]$, which can be
much faster than the orbital frequency in the radiation pressure dominated
inner regions.  The characteristic length scale of the instability is
given by the reciprocal of the turnover wavenumber, which is the gas scale
height.  Again in the radiation pressure dominated inner regions, this is
smaller than the disk scale height by the ratio of the gas
to radiation pressure.  It is far from clear how this instability might,
if at all, manifest itself in the presence of MRI
turbulence.  The MRI acts on longer time scales (the orbital time), and
possibly length scales as well, and this separation of scales between
the two instabilities suggests that they might both be simultaneously
present.  Numerical simulations of vertically stratified, radiating shear
flows will be required to address this question.

\subsection{Stellar Envelopes on the Main Sequence}

Except possibly for short wavelength acoustic waves near the stellar
photosphere, the gas and radiation temperatures are tightly coupled together
in main sequence stellar envelopes.  Thermal locking at the turnover
wavenumber $\sim g/c_{\rm i}^2$ requires $\omega_{\rm th}\gta g/c_{\rm i}$.
For gas pressure supported envelopes, for example, this translates to the
requirement that
\be
\left(\frac{L}{\lsun}\right)\left(\frac{M}{\msun}\right)^{-1}
\left(\frac{\kappa_F\kappa_{\rm a}}{\kappa_T^2}\right)
\left(\frac{\rho}{10^{-7}{\rm g\,cm}^{-3}}\right)
\left(\frac{2\pi c_{\rm i}/g}{3 {\rm minutes}}\right)^{-1}\gta1,
\ee
which, given realistic envelope opacities, is usually satisfied.

From Table 1, if the gas and radiation temperatures are locked together, then
the vertical radiative flux will destabilize short wavelength
hydrodynamic
acoustic waves in a radiation pressure dominated stellar envelope if
$F\Theta_{\rho} \gta Ec_i$.  It is convenient to express this in terms
of the flux mean (Rosseland) optical depth $\tau_{\rm F}$ of the particular
envelope layer in question.  Radiative diffusion implies that
$F\sim Ec/\tau_{\rm F}$, so that short wavelength acoustic waves will be
unstable in a radiation pressure dominated stellar envelope for all layers
with optical depths satisfying
\begin{equation}
\tau_{\rm F}\lta\left({c\over c_{\rm i}}\right)\thetarho.
\end{equation}
This can also be written as
\begin{equation}
\thetarho\gta3\times10^{-5}\left({c_{\rm i}\over10{\rm km\,s}^{-1}}\right)
\tau_{\rm F}.
\end{equation}

In a gas pressure supported stellar envelope, the instability criterion in
Table 1 can be written in a similar fashion,
\begin{equation}
\tau_{\rm F}\lta\left({c\over c_{\rm i}}\right)\left({E\over p}\right)\thetarho.
\label{stargascrit}
\end{equation}
Now, $p/E\sim L_{\rm Edd}/L$ if gas pressure dominates radiation
pressure, where $L_{\rm Edd}$ is the Eddington luminosity appropriate for
the Rosseland mean opacity $\kappaf$.  Hence the instability criterion
(\ref{stargascrit}) may be rewritten as
\begin{equation}
\thetarho\gta3\times 10^{-5}\frac
{L_{\rm Edd}}{L}\left(\frac{c_i}{10\,{\rm km\,s^{-1}} }\right)\tau_{\rm F}.
\end{equation}
This equation strongly suggests that even in gas pressure supported envelopes
on the mid- to upper main sequence, acoustic waves may be unstable and
strange modes might exist.

%Recall that the mass-luminosity relation for main sequence stars is given by
%$L=L_{\odot}\left(M/M_{\odot}\right)^3$.  This, along with the expression
%$L_{\rm edd}$ tells us that
%$L/L_{\rm edd}\sim 3\times 10^{-5}\left(M/M_{\odot}\right)^2$.
%Using this allows to write the instability criteria in terms of the star's
%mass
%\be
%\frac{\partial{\rm ln}\kappa}{\partial{\rm ln}\rho} > \left(\frac{M_{\odot}}{M}
%\right)^2\left(\frac{c_i }{10^6\,{\rm cm\,s^{-1}} }\right)\tau.
%\ee

We stress that we have only performed a local analysis of the stability
of {\it propagating} waves in this paper.  A demonstration that such waves are
amplified by the background radiative flux does not necessarily imply that
global compressive modes of the star will be unstable.  This is because such
modes can in general be viewed as superpositions of WKB waves propagating in
different directions and, as we have seen, while one propagation direction may
be driven, the other may very well be damped.  Our instability criteria are
therefore necessary, but not sufficient when it comes to global standing waves
of the star.  Note that the radiative driving becomes most important at lower
optical depths, and it is also important to point out that we have not
included the effects of turbulent damping associated with outer convection
zones, when they exist.

Most main sequence stellar envelopes probably contain rather weak magnetic
fields, and therefore the hydrodynamic driving that we have discussed so
far is probably of greatest importance.  There
are exceptions, however, where MHD effects might play a role.
In some chemically peculiar A and B-stars,
for example, highly ordered, roughly dipolar magnetic fields of order
1~kG have been measured (e.g. Landstreet 1992).  The MHD driving that we
discussed in sections 5 and 6 above could produce instabilities in the
upper layers of such
stars, and magnetic effects will almost certainly alter the hydrodynamic
driving.  A particular case in point is that of roAp stars, which exhibit
high order $p$-mode oscillations with low angular momentum quantum number
$\ell$, largely confined to the poles
of a dipole magnetic field (e.g. Kurtz 1990).  Consider the possibility
that hydrodynamic driving due to a large opacity variation dominates magnetic
driving, but in the surface layers where the magnetic energy density dominates
the gas pressure.  From equation (\ref{magnetosonic}), the phase speed
of slow modes in this case is
$\simeq|\hat{\bf k}\cdot{\bf v}_{\rm A}/v_{\rm A}|c_{\rm i}$, assuming
the gas and radiation temperatures are locked together.  Equation
(\ref{omegalockmhd}) then implies a growth rate due to hydrodynamic
driving which is proportional to
$|(\hat{\bf k}\cdot{\bf v}_{\rm A})(\hat{\bf k}\cdot{\bf F})\thetarho|$.
In other words, slow waves will have maximal driving if they are
vertical (i.e. radial) in the regions of the magnetic poles, where the
field is also vertical.  This might therefore be the basic driving mechanism
behind roAp oscillations.  We intend to explore this further in future work.

\section{Conclusions}

We have examined the conditions for local driving of a broad class of
instabilities of propagating waves in optically thick media.  These waves
can be driven unstable by a sufficiently strong equilibrium radiative
heat flux acting on density fluctuations in the wave.  The central
mathematical results of this paper are the local dispersion relations:
equation (\ref{staticdisp}) for the full MHD case and equation
(\ref{staticdisphydro}) for hydrodynamics.  Short wavelength expressions
for the hydrodynamic damping and driving of acoustic waves are given
in equations (\ref{omeganocouphydro}) and (\ref{omegalockhydro}).
Generalization of these for the full MHD case are given in equations
(\ref{omeganocoupmhd}) and (\ref{omegalockmhd}).  In addition to the
acoustic waves, we have also briefly discussed short wavelength gravity waves
in sections 3 and 5.

There are essentially two types of local radiative driving of the acoustic
wave instabilities.  Hydrodynamic driving, which occurs even in the absence
of magnetic fields, requires that the wave possess fluctuations in the
flux mean (i.e. Rosseland mean) opacity.
A medium with pure Thomson scattering opacity possesses no such local driving.
The fastest growth rates occur for waves with vector phase velocities parallel
(or possibly antiparallel) to the equilibrium flux ${\bf F}$.  Table 1
summarizes the properties of acoustic wave instabilities subject to
hydrodynamic driving.  It is this driving that is responsible for the
existence of global strange mode oscillations in stars, which have
normally been found in radiation pressure dominated stellar envelopes.
We find that hydrodynamic driving can also produce fast growth rates
in propagating acoustic waves in gas pressure dominated envelopes, at
least for sufficiently large radiative heat fluxes and provided that
the thermal emission and absorption effectively lock the gas and radiation
temperatures together.

The presence of a large scale, equilibrium magnetic field expands the ability
of an equilibrium radiative flux to drive acoustic waves unstable.  In
particular, MHD driving requires only that the wave possess density
fluctuations, and even a medium with pure Thomson scattering opacity
can have unstable magnetosonic waves.  If MHD driving dominates over
hydrodynamic driving, then the fastest growth rates occur for slow
magnetosonic waves, and it is straightforward to show that these growth
rates occur for phase velocities that are in the plane of the equilibrium
flux ${\bf F}$ and magnetic field ${\bf B}$.  In general, the fastest
growing waves have phase velocities that are at some angle to both
${\bf F}$ and ${\bf B}$, and the MHD driving in fact vanishes for
propagation along ${\bf F}$, along ${\bf B}$, or perpendicular to ${\bf B}$.
The characteristics of MHD driving in various limits are summarized in
Table 2.  MHD driving is responsible for ``photon bubble'' modes in
accreting X-ray pulsars and accretion disks around black holes and
neutron stars.  It appears to require diffusive radiation transport:
MHD driving vanishes in the optically thin limit, in contrast to
hydrodynamic driving which survives unscathed (Hearn 1972).

It it important to note that magnetic fields can also
alter the behavior of the hydrodynamic driving that exists in the presence
of opacity fluctuations, and this is summarized in equations
(\ref{omeganocoupmhd}) and (\ref{omegalockmhd}).  We suggest that this
magnetically modified hydrodynamic driving may play a role in the
excitation of observed p-mode oscillations in roAp stars.

The physics of radiative driving may be more generic than just photon-matter
interactions.  In a coupled two-fluid system, provided one of the fluids
dominates the inertia and the other provides rapid diffusive heat transport
through which momentum exchange can occur, then unstable driving of density
fluctuations of the sort we have discussed here may occur.  One example
might be diffusive neutrino transport in proto-neutron stars or hyper-Eddington
accretion flows.  Rapid streaming of cosmic rays in the interstellar medium
can also drive acoustic waves unstable (Begelman \& Zweibel 1994), though
it is unclear whether and how this might be related to the instabilities
we have considered here.  We hope to explore more detailed applications of
radiatively driven instabilities to particular astrophysical phenomena
in future work.

\acknowledgments{We are grateful to J. Arons, P. Arras, L. Bildsten, P.
Chang, S. Davis, C. Fryer, C. Gammie, W. Glatzel, P. Goldreich,
and N. Turner for very useful discussions.  We also thank the anonymous
referee for constructive criticism which greatly improved this paper.
This work was supported by NSF grant AST-9970827 and
UCSB/LANL CARE grant SBB-001A.  OB also thanks the generous hospitality of
the Caltech Tapir group, where much of this work was carried out.}

\appendix

\section{Appendix:  Additional Mathematical Justification for the
WKB Damping and Driving Terms}

Our WKB treatment presented in sections 3-6 is not rigorous, particularly
as the interesting physics (the damping and driving effects on the
underlying acoustic and magnetoacoustic waves) lies in a first order
correction to the infinitely short wavelength limit.  Because we invoked
the WKB approximation on our separate perturbation equations before attempting
to combine them into one, these corrections are vulnerable to alteration
by other equilibrium vertical gradients.

A more rigorous approach would be to combine all nine of our original
linear perturbation equations together into one single ordinary differential
equation in $z$, and then apply WKB techniques on that equation alone.
Unfortunately, we have not discovered a way of doing this in general.  However,
there are two particular cases where this can be done, and we present them
here.  The results of these cases fully agree with the asymptotic WKB
results that we obtained in sections 3-6.

\subsection{Case 1: Vertically Propagating Hydrodynamic Acoustic Waves With
Equal Gas and Radiation Temperatures}

Assume that the gas and radiation exchange energy quickly enough to
guarantee that $T_{\rm r}=T_{\rm g}\equiv T$, and neglect the effects
of magnetic fields.  Consider perturbations that have no variation
in the horizontal direction, i.e. waves which propagate (up or down)
in the vertical ($z$) direction.  The linear perturbation equations for these
waves are
\begin{equation}
{\partial\delta\rho\over\partial t}+\delta v{d\rho\over dz}+
\rho{\partial\delta v\over\partial z}=0,
\label{a1cont}
\end{equation}
\begin{equation}
\rho{\partial\delta v\over\partial t}=-{\partial\delta p\over\partial z}
-g\delta\rho-{1\over3}{\partial\delta E\over\partial z},
\label{a1mom}
\end{equation}
\begin{equation}
{\partial\over\partial t}(\delta u+\delta E)+\delta v\left({du\over dz}+
{dE\over dz}\right)+\left(\gamma u+{4\over3}E\right){\partial\delta v\over
\partial z}=-{\partial\delta F\over\partial z},
\label{a1energy}
\end{equation}
\begin{equation}
{1\over3}{\partial\delta E\over\partial z}=-{\kappaf\rho\over c}\delta F
-{\kappaf F\over c}(1+\thetarho)\delta\rho-{\kappaf F\over c}\thetat
{\delta T\over T},
\label{a1diff}
\end{equation}
together with $\delta u=\delta p/(\gamma-1)$, $\delta p/p=\delta\rho/\rho+
\delta T/T$, and $\delta E/E=4\delta T/T$.  Note that in this case of
vertical propagation, $\delta {\bf v}$ and $\delta{\bf F}$ only have
nonzero components in the vertical direction.

These equations are still too complicated to combine into a single wave
equation, and we must approximate the thermodynamics somewhat.  The
perturbed energy equation (\ref{a1energy}) can be written just in terms
of $\delta T$, $\delta F$, and $\delta v$ as
\begin{equation}
{1\over T}\left({p\over\gamma-1}+4E\right)
\left({\partial\delta T\over\partial t}+\delta v{dT\over dz}\right)
+\left(p+{4\over3}E\right){\partial\delta v\over\partial z}=
-{\partial\delta F\over\partial z}.
\label{a1energy1}
\end{equation}
Now, at very short wavelengths ($k\rightarrow\infty$), we expect rapid
radiative diffusion to smooth out temperature fluctuations, so that
$\delta T={\cal O}(k^{-1})\delta v$ is very small.
Hence the dominant terms in this
equation at short wavelengths are the last term on the left hand side and the
term on the
right hand side.\footnote{This approximation was also used by Begelman (2001)
in his treatment of nonlinear, radiatively driven MHD waves, cf. his equation
(7).}  At the same level of approximation, we may therefore rewrite this
equation as
\begin{equation}
{\partial\over\partial z}\left[\left(p+{4\over3}E\right)\delta v+\delta F
\right]=0,
\end{equation}
which can be immediately integrated to give
\begin{equation}
\left(p+{4\over3}E\right)\delta v+\delta F=0.
\label{a1energy2}
\end{equation}
Similarly, we also neglect all $\delta T$ terms in the momentum and
diffusion equations (\ref{a1mom}) and (\ref{a1diff}) that are not
gradients to give
\begin{equation}
\rho{\partial\delta v\over\partial t}=-{p\over\rho T}{dT\over dz}\delta\rho
-{p\over\rho}{\partial\delta\rho\over\partial z}-{p\over T}{\partial
\delta T\over\partial z}-g\delta\rho-
{4E\over3T}{\partial\delta T\over\partial z},
\label{a1mom2}
\end{equation}
and
\begin{equation}
{4E\over3T}{\partial\delta T\over\partial z}=-{\kappaf\rho\over c}\delta F
-{\kappaf F\over c}(1+\thetarho)\delta\rho,
\label{a1diff1}
\end{equation}
respectively.

With these approximations, equations (\ref{a1cont}),
(\ref{a1energy2}), (\ref{a1mom2}), and (\ref{a1diff1})
can now be combined into a single
equation.  Without further loss of generality, we assume all perturbation
variables have a time-dependence of the form $\exp(-i\omega t)$.  Then after
some algebra, we obtain a single ordinary differential equation in $\delta v$:
\begin{eqnarray}
0&=&p{d^2\delta v\over dz^2}+\left[{dp\over dz}-\left(1+{3p\over4E}\right)
\thetarho{\kappaf\rho F\over c}\right]{d\delta v\over dz}
+\Biggl[\omega^2\rho+{i\omega\kappaf\rho\over c}\left(p+{4E\over3}\right)
\left(1+{3p\over4E}\right)\nonumber\\
& &-{3p\over4E}{\kappaf F\over c}{d\rho\over dz}-{\kappaf F\over c}\thetarho
\left(1+{3p\over4E}\right){d\rho\over dz}-{p\over\rho^2}
\left({d\rho\over dz}\right)^2+{p\over\rho}{d^2\rho\over dz^2}\Biggr]\delta v.
\label{a1dvode}
\end{eqnarray}
It is convenient to define a new perturbation variable $\delta\psi$ by
\begin{equation}
\delta\psi=p^{1/2}\delta v.
\label{a1psi}
\end{equation}
Then equation (\ref{a1dvode}) becomes
\begin{eqnarray}
0&=&p{d^2\delta\psi\over dz^2}-\left(1+{3p\over4E}\right)
\thetarho{\kappaf\rho F\over c}{d\delta\psi\over dz}\nonumber\\
&+&\Biggl[\omega^2\rho+{i\omega\kappaf\rho\over c}\left(p+{4E\over3}\right)
\left(1+{3p\over4E}\right)+{1\over2p}{dp\over dz}\thetarho
{\kappaf\rho F\over c}\left(1+{3p\over4E}\right)\nonumber\\
& &+{1\over4p}\left({dp\over dz}\right)^2-{1\over2}{d^2p\over dz^2}
-{3p\over4E}{\kappaf F\over c}{d\rho\over dz}-{\kappaf F\over c}\thetarho
\left(1+{3p\over4E}\right){d\rho\over dz}\nonumber\\
& &-{p\over\rho^2}
\left({d\rho\over dz}\right)^2+{p\over\rho}{d^2\rho\over dz^2}\Biggr]
\delta\psi.
\label{a1dpsiode}
\end{eqnarray}
If we now employ the WKB ansatz that $\delta\psi\propto\exp(ikz)$, this
equation gives the dispersion relation
\begin{equation}
\omega^2=k^2c_{\rm i}^2+i{\kappaf\over c}\left(1+{3p\over4E}\right)
\left[\thetarho kF-\omega\left(p+{4E\over3}\right)\right]+{\cal O}(k^0),
\end{equation}
or
\begin{equation}
\omega=\pm kc_{\rm i}-i{\kappaf\over2cc_{\rm i}}\left(1+{3p\over4E}\right)
\left[\left(p+{4E\over3}\right)c_{\rm i}\mp\thetarho F\right]+{\cal O}(k^{-1}),
\label{a1dispk}
\end{equation}
in exact agreement with equation (\ref{omegalockhydro}) for vertically
propagating waves.

An astute reader familiar with WKB techniques may notice that we cheated
slightly.
The formal WKB treatment first involves eliminating the first order derivative
term, thereby forcing the differential equation to resemble the harmonic
oscillator equation.  We can do this by transforming from $\delta v$ to
$\tilde{\delta\psi}$, defined by
\begin{equation}
\tilde{\delta\psi}=
p^{1/2}\delta v\exp\left[-\int\thetarho{\kappaf\rho F\over2cp}
\left(1+{3p\over4E}\right)dz\right].
\label{a1psit}
\end{equation}
When we employ the WKB ansatz to the differential equation that then results,
$\tilde{\delta\psi}\propto\exp(i\tilde{k}z)$, we obtain
\begin{equation}
\omega=\pm \tilde{k}c_{\rm i}-i{\kappaf\over2c}\left(1+{3p\over4E}
\right)\left(p+{4E\over3}\right)+{\cal O}(\tilde{k}^{-1}).
\label{a1dispkt}
\end{equation}
This appears to be an acoustic wave which is merely damped by radiative
diffusion, and we have apparently lost our radiative driving term.  However,
one must be very careful in the physical interpretation of the wave
vector $\tilde k$.  It is related to the wave vector $k$ we employed above
by
\begin{equation}
\tilde{k}=k+i\thetarho{\kappaf\rho F\over2cp}\left(1+{3p\over4E}\right).
\end{equation}
Substituting this expression into equation (\ref{a1dispkt}), we recover the
dispersion relation (\ref{a1dispk}), including the radiative driving term.
In order to interpret this ambiguous behavior, note that the velocity
perturbation is
\begin{equation}
\delta v\propto p^{-1/2}\exp(-i\omega t+ikz)\propto p^{-1/2}\exp\left[
-i\omega t+i\tilde{k}z+\int\thetarho{\kappaf\rho F\over2cp}
\left(1+{3p\over4E}\right)dz\right].
\label{a1dv}
\end{equation}
For concreteness, assume $\thetarho>0$ and consider upward propagating waves.
We may choose $k$ to be real and large (the interpretation we have adopted
throughout this paper), in which case equation (\ref{a1dispk}) is the
correct dispersion relation.  If radiative driving dominates damping, then
the wave will grow in time.  Alternatively, we may choose $\tilde{k}$ to
be real and large, in which case equation (\ref{a1dispkt}) is the
relevant dispersion relation.  The wave will tend to damp with time, but
if the same quantitative requirement that the driving term dominates the
damping term in equation (\ref{a1dispk}) still holds, then equation
(\ref{a1dv}) implies that it will grow
exponentially as it propagates spatially.  Either way, we have instability:
the wave is driven to larger and larger amplitude by the background radiative
flux.

Finally, we note that the original {\it partial} differential
equations (\ref{a1cont}), (\ref{a1energy2}),
(\ref{a1mom2}), and (\ref{a1diff1})
can be combined into the following energy conservation
equation for the waves,
\begin{equation}
{\partial\over\partial t}\left({1\over2}\rho\delta v^2+
{p\over2\rho^2}\delta\rho^2\right)+{\partial\over\partial z}\left({p\over\rho}
\delta\rho\delta v\right)=\left(1+{3p\over4E}\right)\left[{\kappaf F\over c}
\thetarho\delta\rho\delta v-\left(p+{4E\over3}\right)
{\kappaf\rho\over c}\delta v^2\right].
\label{a1ewave}
\end{equation}
This equation leaves no ambiguity in interpretation.  The two terms on the
right hand side represent radiative driving and damping by diffusion,
respectively, and are consistent with the dispersion relation (\ref{a1dispk}).
Note that for $\thetarho>0$, radiative driving requires
$\delta\rho\delta v>0$, i.e. the density and velocity perturbations
must be in phase, which occurs only for upward propagating waves.  This
is also fully consistent with the physics we described in section 4 and
illustrated in figure 4.

\subsection{Case 2: Infinitely Strong Horizontal Field, Equal Gas and
Radiation Temperatures}

Assume again that the gas and radiation exchange energy quickly so that
they have the same temperature, and consider a case
where there is a horizontal equilibrium magnetic field
that is so strong $(v_{\rm A}/\cgas\rightarrow\infty)$ that fluid elements
are constrained to move horizontally.  In a Cartesian coordinate system
$(x,y,z)$, with $x$ along the field and $z$ in the upward vertical direction,
the linear perturbation equations in this case may be written as
\begin{equation}
{\partial\delta\rho\over\partial t}+\rho{\partial\delta v\over\partial x}=0,
\end{equation}
\begin{equation}
\rho{\partial\delta v\over\partial t}=-{\partial\delta p\over\partial x}+
{\kappaf\rho\over c}\delta F_x,
\end{equation}
\begin{equation}
{\partial\over\partial t}(\delta u+\delta E)+\left(\gamma u+{4\over3}E\right)
{\partial\delta v\over\partial x}=-{\bnabla}\cdot\delta{\bf F},
\end{equation}
\begin{equation}
{1\over3}{\partial\delta E\over\partial x}=-{\kappaf\rho\over c}\delta F_x,
\end{equation}
\begin{equation}
{1\over3}{\partial\delta E\over\partial y}=-{\kappaf\rho\over c}\delta F_y,
\end{equation}
\begin{equation}
{1\over3}{\partial\delta E\over\partial z}=-{\kappaf\rho\over c}\delta F_z
-{\kappaf F\over c}(1+\thetarho)\delta\rho-{\kappaf\rho F\over c}\thetat
{\delta T\over T},
\end{equation}
together with $\delta u=\delta p/(\gamma-1)$, $\delta p/p=\delta\rho/\rho+
\delta T/T$, and $\delta E/E=4\delta T/T$.  The magnetic field has vanished
entirely from these equations, as it is considered to be so strong that
fluid motions do not bend the field lines.  As a result, Alfv\'en and
fast magnetosonic modes have been eliminated from this system of equations.
Note that, in contrast to the previous case we considered, we have not
made any approximations concerning the rate of radiative diffusion.

Without loss of generality,
we assume that all perturbation variables have an $(x,y,t)$ dependence
proportional to $\exp[i(k_x x+k_y y-\omega t)]$.  Then with some algebraic
work, all perturbation variables can be eliminated in favor of $\delta T$,
giving us a single, second order ordinary differential equation in $z$, viz.
\begin{eqnarray}
0&=&{4Ec\over3\kappaf\rho}{d^2\over dz^2}\left({\delta T\over T}
\right)+\left[{k_x^2F(1+\thetarho)(p+4E/3)\over\rho\omega^2-pk_x^2}+
F(\thetat-4)+{d\over dz}\left({4Ec\over3\kappaf\rho}\right)\right]{d\over dz}
\left({\delta T\over T}\right)\nonumber\\
&+&\Biggl[i\omega\left({p\over\gamma-1}+4E\right)-{ik_x^2\omega(p+4E/3)^2\over
(\rho\omega^2-pk_x^2)}-{4Ec\over3\kappaf\rho}(k_x^2+k_y^2)\nonumber\\
& &+k_x^2F(1+\thetarho)
{d\over dz}\left({p+4E/3\over\rho\omega^2-pk_x^2}\right)\Biggr]
{\delta T\over T}.
\label{a2dtode}
\end{eqnarray}
(We have assumed for simplicity that $\thetarho$ and $\thetat$ are constants.)
Now, we expect that slow magnetosonic modes will have a dispersion relation
of the form $\omega^2=k_x^2c_{\rm i}^2=k_x^2p/\rho$
plus higher order corrections.  Hence
adopt the WKB ansatz that $\delta T/T\propto\exp(ik_z z)$ with large
$k_x$, $k_y$, and $k_z$; $\omega\propto k$; and $\omega^2-k_x^2p/\rho\propto k$.
Then equation (\ref{a2dtode}) gives the dispersion relation
\begin{equation}
\omega^2=k_x^2c_{\rm i}^2+i{k_x^2\over k^2}{\kappaf\rho\over c}\left(1+
{3p\over4E}\right)\left[{k_zF(1+\thetarho)\over\rho}-\left(p+{4E\over3}\right)
{\omega\over\rho}\right]+{\cal O}(k^0),
\end{equation}
or
\begin{equation}
\omega=\pm|k_x|c_{\rm i}=-i{\kappaf\over2cc_{\rm i}}\left(1+{3p\over4E}\right)
\left[\left({k_x^2\over k^2}\right)\left(p+{4E\over3}\right)c_{\rm i}\mp
{|k_x|k_z\over k^2}F(1+\thetarho)\right]+{\cal O}(k^{-1}),
\end{equation}
which agrees completely with equation (\ref{omegalockmhd}) in the
$v_{\rm A}\rightarrow\infty$ limit.  This confirms
that our crude WKB analysis presented in the main body of the paper also
appears to work in the MHD case, and is robust with regard to the radiation
physics.

\section{Appendix: Radiative Diffusion and Gas/Radiation Heat Exchange
Effects on the Magnetorotational Instability}

In our first paper (BS01), in addition to pursuing an initial exploration
of radiatively driven instabilities in magnetoacoustic waves,
we also examined how radiative diffusion modifies the magnetorotational
instability (MRI; Balbus \& Hawley 1991).  Our analysis in that paper assumed
pure Thomson scattering opacity, with no heat exchange between the gas and
radiation.  Here we wish to briefly consider the behavior of
the MRI under the more general thermodynamic assumptions used throughout
this paper.

We restrict consideration to axisymmetric perturbations on axisymmetric
equilibria rotating with angular velocity $\Omega(R)$, where $R$ is the
radial distance from the rotation axis.  Apart from the radial dependence
of $\Omega$, which is the source of free energy for the MRI, we completely
neglect all other equilibrium gradients in the flow.  In order to avoid
time-dependent azimuthal fields in the equilibrium that result from the
shear, we assume that the radial component of the magnetic field is zero
in the equilibrium.

Using cylindrical polar coordinates $(R,\phi,z)$, the analysis proceeds
very much as in BS01, with a resulting dispersion relation
\begin{equation}
D_{\rm ms}D_{\rm BH}+k_z^2\vaz^2\vaphi^2\left(k_z^2R
{d\Omega^2\over dR}-k^2\tilde\omega^2\right)-{k_R^2\over k_z^2}
\tilde\omega^2\omega^4=0.
\label{nonstratdisp}
\end{equation}
Here
\begin{equation}
\tilde\omega^2\equiv\omega^2-k_z^2\vaz^2,
\end{equation}
\begin{equation}
D_{\rm ms}\equiv\omega^2-k_z^2\left(C_{\rm s}^2+\vaphi^2\right).
\end{equation}
\begin{equation}
D_{\rm BH}\equiv{k^2\over k_z^2}\tilde\omega^4-\kappa^2\tilde\omega^2-4
\Omega^2k_z^2\vaz^2,
\end{equation}
and $C_{\rm s}^2\equiv{\cal A}\cgas^2+{\cal B}\crad^2+{\cal C}$
in the limit of zero stratification.  From equations (\ref{cala})-(\ref{cald}),
this is
\begin{eqnarray}
C_{\rm s}^2&=&\left\{\left(\omega+{ick^2\over3\kappaf\rho}\right)\left[
\omega+{4i(\gamma-1)\omegaa E\over p}\right]+i\omegaa\omega\right\}^{-1}
\nonumber\\
&\times&\Biggl\{\omega\left(\omega+{ick^2\over3\kappaf\rho}\right)\cgas^2
+\left[\omega^2-9(\gamma-1)\omegaa\left({ck^2\over3\kappaf\rho}\right)
\right]\crad^2\nonumber\\
&+&i\omegaa\omega\left[1+{4(\gamma-1)E\over p}\right]c_{\rm s1}^2
\Biggr\}.
\end{eqnarray}
Here $c_{\rm s1}$ is the total sound speed in the fluid, defined
by
\begin{equation}
c_{\rm s1}^2\equiv{16(\gamma-1)E^2+60(\gamma-1)Ep+9\gamma p^2\over
9[p+4(\gamma-1)E]\rho}={\Gamma_1(p+E/3)\over\rho},
\end{equation}
where $\Gamma_1$ is the first generalized adiabatic exponent commonly used
for matter and radiation (Chandrasekhar 1967).

Equation (\ref{nonstratdisp}) should be compared to equation (39) of BS01,
which it closely resembles.  Not surprisingly, altering
the thermodynamics has produced only one difference: a change in the effective
sound speed to $C_{\rm s}$.  The basic conclusions of BS01 therefore continue
to hold, with only minor quantitative changes:

If the equilibrium azimuthal field is zero ($\vaphi=0$), then the
$k_R=0$ MRI mode (the ``channel solution'') is completely unaltered by
radiation effects.

If radiative diffusion is unimportant on the scale of the MRI critical
wavelength, [$c\Omega/(3\kappaf\rho\vaz^2)\ll1$], then the MRI will be
unaltered provided the azimuthal field energy density is smaller than
the total thermal energy density in the fluid.  To be more precise, the
MRI will be unaltered provided $\vaphi^2\ll\cgas^2+\crad^2$ if the gas
and radiation are not thermally locked together on an orbital time,
$\omegaa[1+4(\gamma-1)E/p]<\Omega$, or $\vaphi^2\ll c_{{\rm s}1}^2$ otherwise.
In either case, the essential
nature of the MRI is unchanged provided the magnetic field is subthermal.

If radiative diffusion is rapid on the scale of an MRI critical wavelength,
then radiation pressure no longer helps enforce incompressibility in
the MRI.  In a radiation pressure dominated medium where the magnetic
field is largely in the azimuthal direction, even subthermal
(with respect to the radiation) fields
can then reduce the growth rate of axisymmetric MRI modes.
Unless $\vaphi^2\ll\cgas^2$
if gas and radiation are not thermally locked,
or $\vaphi^2\ll c_{\rm i}^2$
otherwise, then the growth rate of the MRI will be reduced.

\begin{deluxetable}{cccccc}
\rotate
\tabletypesize{\scriptsize}
\tablecaption{Order of Magnitude Conditions for Radiation
Hydrodynamic Acoustic Wave Instabilities\tablenotemark{a}. \label{tbl-1}}
\tablewidth{0pt}
\tablehead{
\colhead{Thermal} & \colhead{Pressure} &
\colhead{Instability} & \colhead{Asymptotic} & \colhead{Turnover}
& \colhead{Cutoff} \\
\colhead{Regime} & \colhead{Support} & \colhead{Criterion} & 
\colhead{Growth Rate} &
\colhead{Wavenumber\tablenotemark{b}} &
\colhead{Wavenumber}
}
\startdata
$\omega_k>\omegath$
($\delta T_{\rm r}\ne\delta T_{\rm g}$) & $E\gg p$ &
$F\tilde{\Theta}\gta E\times{\rm max}
\left[\cgas,\left({\omegaa\over\kappaf\rho c}\right)\left({c^2\over\cgas}
\right)\right]$ & 
$\tilde{\Theta}\left({g\over\cgas}\right)$ &
$\tilde{\Theta}\left({g\over\cgas^2}\right)$ & $\infty$
 \\
$\omega_k>\omegath$ ($\delta T_{\rm r}\ne\delta T_{\rm g}$) & $E\ll p$ &
$F\tilde{\Theta}\gta E\times{\rm max}
\left[\cgas,\left({\omegaa\over\kappaf\rho c}\right)\left({c^2\over\cgas}
\right)\right]$ &
$\tilde{\Theta}\left({g\over\cgas}\right)\left({E\over p}\right)$ &
$\tilde{\Theta}\left({g\over\cgas^2}\right)\left({E\over p}\right)$ & $\infty$
 \\
$\omega_k<\omegath$
($\delta T_{\rm r}=\delta T_{\rm g}$) & $E\gg p$ &
$F\thetarho\gta Ec_{\rm i}$ &
$\thetarho\left({g\over c_{\rm i}}\right)$ &
$\thetarho\left({g\over c_{\rm i}^2}\right)$ &
$\left({\omegath g\over c_{\rm i}^3}|1+\thetarho|\right)^{1/2}$
 \\
$\omega_k<\omegath$ ($\delta T_{\rm r}=\delta T_{\rm g}$) & $E\ll p$ &
$F\thetarho\gta pc_{\rm i}$ &
$\thetarho\left({g\over c_{\rm i}}\right)$ &
$\thetarho\left({g\over c_{\rm i}^2}\right)$ &
$\left({\omegath g\over c_{\rm i}^3}|1+\thetarho|\right)^{1/2}$
 \enddata

%% Text for table notes should follow after the \enddata but before
%% the \end{deluxetable}. Make sure there is at least one \tablenotemark
%% in the table for each \tablenotetext.

\tablenotetext{a}{Here $\omega_k$ is simply $|k|$ times the phase speed
of the wave ($\cgas$ or $c_{\rm i}$, depending on the thermal regime, for
the hydrodynamic acoustic waves in this table), $\omega_{\rm th}$ is defined
by equation (\ref{omegathermal}), $\omega_{\rm a}$ is defined by equation
(\ref{omegaabs}), $\tilde{\Theta}\equiv|\thetarho+(\gamma-1)
\thetatg|$, and $\thetarho$ and $\thetatg$ are defined in equation
(\ref{thetadefs}).  All the results in this table are valid only in the
regime where the photon diffusion time across a wavelength is faster
than the wave period, but yet the waves are still optically thick so
that photon diffusion still applies.  This requires
$\kappaf\rho(\cgas/c)<k<\kappaf\rho$.}
\tablenotetext{b}{The minimum wavenumber at which the asymptotic growth
rate becomes valid.  (The growth rate typically tends to zero as $k^{1/2}$
as $k\rightarrow0$ below this wavenumber.)  In some cases, the turnover
wavenumber may be so small that
the WKB approximation is violated.  This is an issue primarily for gas
pressure dominated equilibria ($p\gg E$), in which case our analysis is valid
only for $k\gg g/\cgas^2$.  In addition, if $\tilde{\Theta}$ or $|\thetarho|$
are less than unity, as can happen in Thomson scattering dominated media,
the turnover wavenumber is less than the reciprocal of the gas scale height,
and the waves can no longer be purely acoustic in nature at this low
wavenumber.}
%\tablenotetext{b}{Another sample footnote for table~\ref{tbl-1}}

%\tablecomments{Occasionally, authors wish to append a short
%paragraph of explanatory notes that pertain to the entire table, but
%which are different than the caption.  Such notes should be placed in
%a {\tt tablecomments} command like this.}

\end{deluxetable}

\begin{deluxetable}{cccccccc}
\rotate
\tabletypesize{\scriptsize}
\tablecaption{Order of Magnitude Conditions for the Radiation Magnetoacoustic
Wave Instabilities.\tablenotemark{a} \label{tbl-2}}
\tablewidth{0pt}
\tablehead{\colhead{Mode} & \colhead{Thermal}  & \colhead{Magnetic}
&\colhead{Pressure} & \colhead{Instability} & \colhead{Asymptotic}
& \colhead{Turnover} & \colhead{Cutoff}   \\\colhead{}
& \colhead{Regime} & \colhead{Pressure} &\colhead{Support}
& \colhead{Criterion} & \colhead{Growth Rate ($\gamma_d$)}
& \colhead{Wavenumber} & \colhead{Wavenumber}  }
\startdata
$ {\rm SLOW}$ &
$\omega_k>\omega_{\rm th}$ & 
${B^2}/{8\pi}>>p $ &
$E>>p $ &
$F\gta  E\times{\rm max}
\left[\cgas,\left({\omegaa\over\kappaf\rho c}\right)\left({c^2\over\cgas}
\right)\right]   $&
$g/\cgas$&
$g/\cgas^2$&
$\infty$
\\
$ {\rm SLOW}$ &
$\omega_k>\omega_{\rm th}$ & 
$B^2/8\pi<<p $ &
$E>>p $ &
$F\gta \left(\frac{v^2_{\rm A}}{\cgas^2}\right) E\times{\rm max}
\left[v_{\rm A},\left({\omegaa\over\kappaf\rho c}\right)\left({c^2\over
v_{\rm A}}
\right)\right]   $   &
$(g\,v_{\rm A})/\cgas^2$&
$g/\cgas^2$ &
$\infty$
\\ 
$ {\rm SLOW}$ &
$\omega_k>\omega_{\rm th}$ & 
$B^2/8\pi>>p $ &
$E<<p $ &
$ F\gta  E\times{\rm max}
\left[\cgas,\left({\omegaa\over\kappaf\rho c}\right)\left({c^2\over\cgas}
\right)\right]    $&
$\left(\frac{E}{p}\right)g/\cgas$&
$\left(\frac{E}{p}\right)g/\cgas^2$&
$\infty$
\\ 
$ {\rm SLOW}$ &
$\omega_k>\omega_{\rm th}$ & 
$B^2/8\pi<<p $ &
$E<<p $ &
$F\gta \left(\frac{v^2_{\rm A}}{\cgas^2}\right) E\times{\rm max}
\left[v_{\rm A},\left({\omegaa\over\kappaf\rho c}\right)\left({c^2\over
v_{\rm A}}
\right)\right]     $&
$\left(\frac{E}{p}\right)(g\,v_{\rm A})/\cgas^2$&
$ \left(\frac{E}{p}\right) g/\cgas^2$&
$\infty$
\\
$ {\rm FAST}$ &
$\omega_k>\omega_{\rm th}$ & 
${B^2}/{8\pi}>>p $ &
$E>>p $ &
$F\gta  E\times{\rm max}
\left[v_{\rm A},\left({\omegaa\over\kappaf\rho c}\right)\left({c^2\over
v_{\rm A}}
\right)\right]      $&
$g/v_{\rm A}$&
$g/v^2_{\rm A}$&
$\infty$
\\
$ {\rm FAST}$ &
$\omega_k>\omega_{\rm th}$ & 
$B^2/8\pi<<p $ &
$E>>p $ &
$ F \gta\left(\frac{\cgas^2}{v^2_{\rm A}}\right)  E\times{\rm max}
\left[\cgas,\left({\omegaa\over\kappaf\rho c}\right)\left({c^2\over\cgas}
\right)\right]    $&
$(g\,v^2_{\rm A})/\cgas^3$&
$\left(\frac{v^2_{\rm A}}{\cgas^2}\right)g/\cgas^2$&
$\infty$
\\ 
$ {\rm FAST}$ &
$\omega_k>\omega_{\rm th}$ & 
$B^2/8\pi>>p $ &
$E<<p $ &
$ F\gta  E\times{\rm max}
\left[v_{\rm A},\left({\omegaa\over\kappaf\rho c}\right)\left({c^2\over
v_{\rm A}}
\right)\right]     $&
$\left(\frac{E}{p}\right)g/v_{\rm A}$&
$\left(\frac{E}{p}\right)g/v^2_{\rm A}$&
$\infty$
\\ 
$ {\rm FAST}$ &
$\omega_k>\omega_{\rm th}$ & 
$B^2/8\pi<<p $ &
$E<<p $ &
$  F \gta\left(\frac{\cgas^2}{v^2_{\rm A}}\right)  E\times{\rm max}
\left[\cgas,\left({\omegaa\over\kappaf\rho c}\right)\left({c^2\over\cgas}
\right)\right]   $&
$\left(\frac{E}{p}\right) (g\,v^2_{\rm A})/\cgas^3  $&
$\left(\frac{E}{p}\right) \left(\frac{v^2_{\rm A}}{\cgas^2}\right)g/\cgas^2  $&
$\infty$
\\
$ {\rm SLOW}$ &
$\omega_k<\omega_{\rm th}$ & 
${B^2}/{8\pi}>>p $ &
$E>>p $ &
$F\gta E\,c_{\rm i}   $&
$g/c_{\rm i}$&
$g/c^2_{\rm i}$&
$\left(\omega_{\rm th}\gamma_d \right)^{1/2}/c_{\rm i}$
\\
$ {\rm SLOW}$ &
$\omega_k<\omega_{\rm th}$ & 
$B^2/8\pi<<p $ &
$E>>p $ &
$F\gta\left(\frac{v^2_A}{c^2_{\rm i}} \right) E\,v_{\rm A}  $   &
$(g\,v_{\rm A})/c^2_{\rm i}$&
$g/c^2_{\rm i}$&
$c^2_{\rm i}\left(\omega_{\rm th}\gamma_d\right)^{1/2}/v^3_{\rm A} $
\\ 
$ {\rm SLOW}$ &
$\omega_k<\omega_{\rm th}$ & 
$B^2/8\pi>>p $ &
$E<<p $ &
$ F\gta p\,c_{\rm i}  $&
$g/c_{\rm i}$&
$g/c^2_{\rm i}$&
$\left(\omega_{\rm th}\gamma_d \right)^{1/2}/c_{\rm i} $
\\ 
$ {\rm SLOW}$ &
$\omega_k<\omega_{\rm th}$ & 
$B^2/8\pi<<p $ &
$E<<p $ &
$ F\gta\left(\frac{v^2_{\rm A}}{c^2_{\rm i}} \right) p\,v_{\rm A}   $&
$(g\,v_{\rm A})/c^2_{\rm i}  $&
$g/c^2_{\rm i}  $&
$c^2_{\rm i}\left(\omega_{\rm th}\gamma_d\right)^{1/2}/v^3_{\rm A} $
\\
$ {\rm FAST}$ &
$\omega_k<\omega_{\rm th}$ & 
${B^2}/{8\pi}>>p $ &
$E>>p $ &
$F\gta E\,v_{\rm A}       $&
$g/v_{\rm A}$&
$g/v^2_{\rm A}$&
$\left(\omega_{\rm th}\gamma_d\right)^{1/2}/v_{\rm A} $
\\
$ {\rm FAST}$ &
$\omega_k<\omega_{\rm th}$ & 
$B^2/8\pi<<p $ &
$E>>p $ &
$ F \gta \left(\frac{c^2_{\rm i}}{v^2_{\rm A}}\right)E\,c_{\rm i}    $&
$(g\,v^2_{\rm A})/c^3_{\rm i}$&
$\left(\frac{v^2_{\rm A}}{c^2_{\rm i}}\right)g/c^2_{\rm i}$&
$\left(\omega_{\rm th}\gamma_d\right)^{1/2}/c_{\rm i} $
\\ 
$ {\rm FAST}$ &
$\omega_k<\omega_{\rm th}$ & 
$B^2/8\pi>>p $ &
$E<<p $ &
$  F\gta p\,v_{\rm A}     $&
$g/v_{\rm A}$&
$g/v^2_{\rm A}$&
$\left(\omega_{\rm th}\gamma_d\right)^{1/2}/v_{\rm A} $
\\ 
$ {\rm FAST}$ &
$\omega_k<\omega_{\rm th}$ & 
$B^2/8\pi<<p $ &
$E<<p $ &
$  F \gta  \left(\frac{c^2_{\rm i}}{v^2_{\rm A}}\right)p\,c_{\rm i}   $&
$ (g\,v^2_{\rm A})/c^3_{\rm i}  $&
$ \left(\frac{v^2_{\rm A}}{c^2_{\rm i}}\right)g/c^2_{\rm i}  $&
$\left(\omega_{\rm th}\gamma_d\right)^{1/2}/c_{\rm i} $
\enddata

\tablenotetext{a}{Once again, $\omega_k$ is $|k|$ times the phase speed
$v_{\rm ph}$ of the wave.  For simplicity, we have assumed that the flux mean
opacity $\kappaf$ is independent of density and temperature, so that
the hydrodynamic driving terms of the instability vanish.  This could be
the case, for example, if Thomson scattering is the dominant form of
momentum transfer between the gas and radiation.}

\end{deluxetable}

\begin{figure}   
\figurenum{1}
%\epsscale{0.7}
\plotone{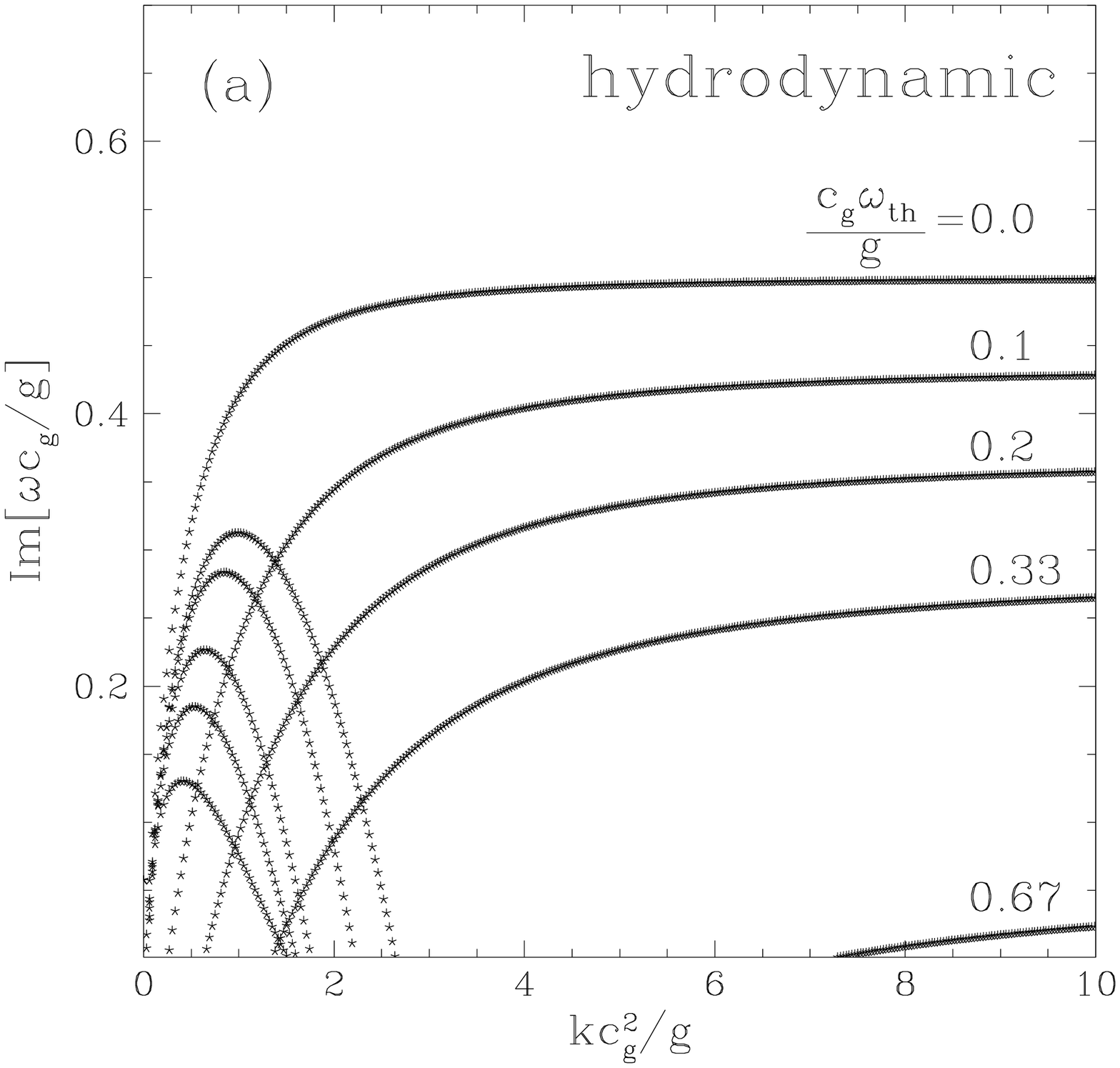}
\vskip0.1truein
\end{figure}

\begin{figure}
\figurenum{1}
\plotone{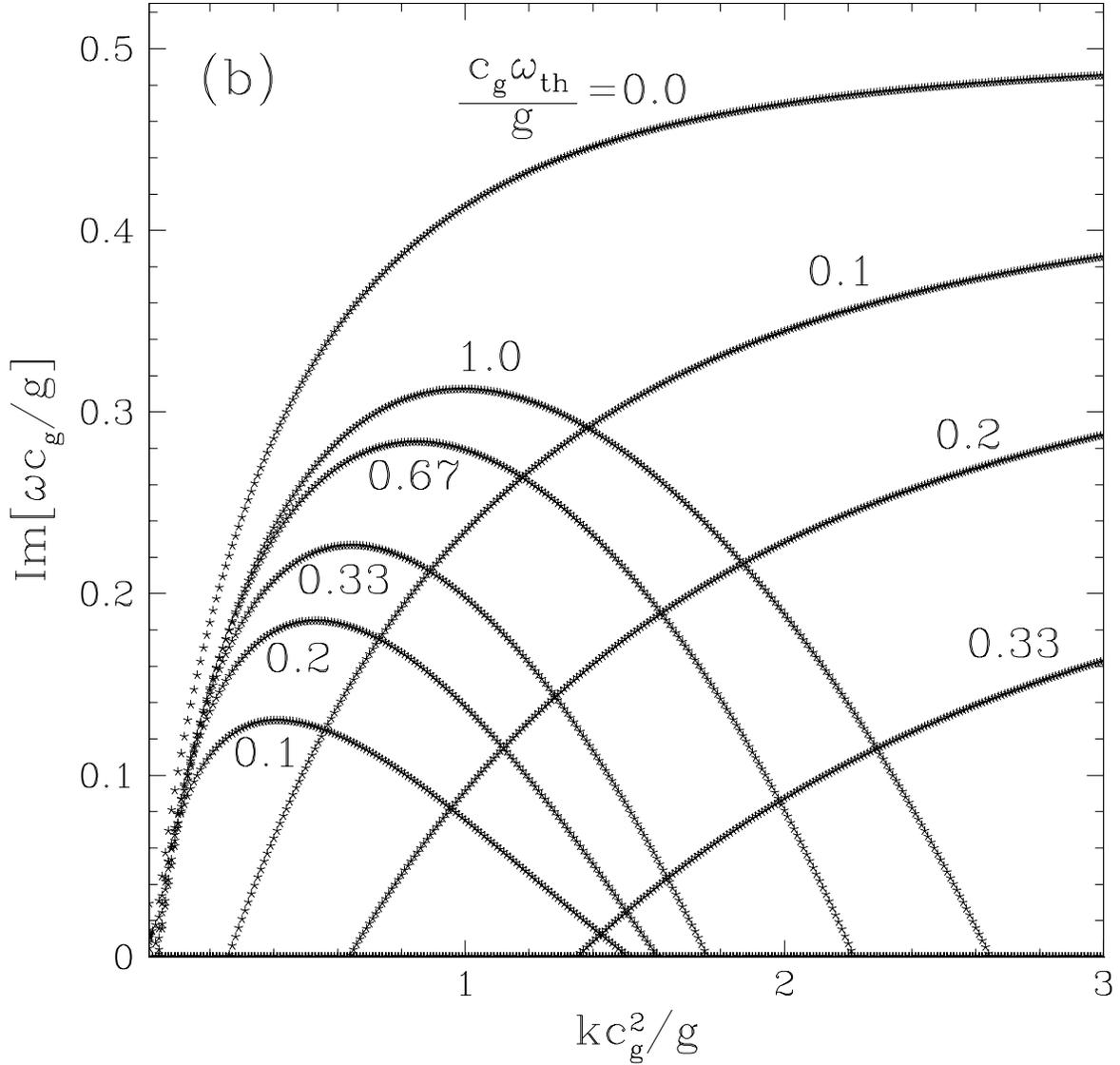}
\vskip0.1truein
\caption{Scaled growth rates of hydrodynamic acoustic wave instabilities as a
function of scaled wavenumber, for different values of the thermal coupling
frequency $\omegath$.
The magnetic field was set to zero, and other equilibrium parameters were
$F=300\rho\cgas^3$ and $\crad/\cgas=10$.
Because of the dominant radiation pressure, we neglected gas pressure gradients
and chose a flux mean opacity that satisfied $\kappaf F/c=g$.
We also chose the flux mean opacity to have derivatives appropriate for
a Kramers type law: $\thetarho=1$
and $\thetatg=-3.5$.  The first plot (a) shows growth rates for the
downward propagating unstable two-temperature waves that exist at high wavenumbers.
In addition, there are
upward propagating one-temperature waves that are also unstable and exist only up to a finite
cutoff wavenumber.  The second plot (b) is a blow up of the left hand region of
plot (a) to show these modes more clearly.  In both plots, the waves are
assumed to be propagating purely vertically, i.e. the wave vector ${\bf k}$
is either parallel or antiparallel to the flux ${\bf F}$.  Such mode directions
have maximal growth rates.}
\end{figure}

\begin{figure}   
\figurenum{2}
%\epsscale{0.7}
\plotone{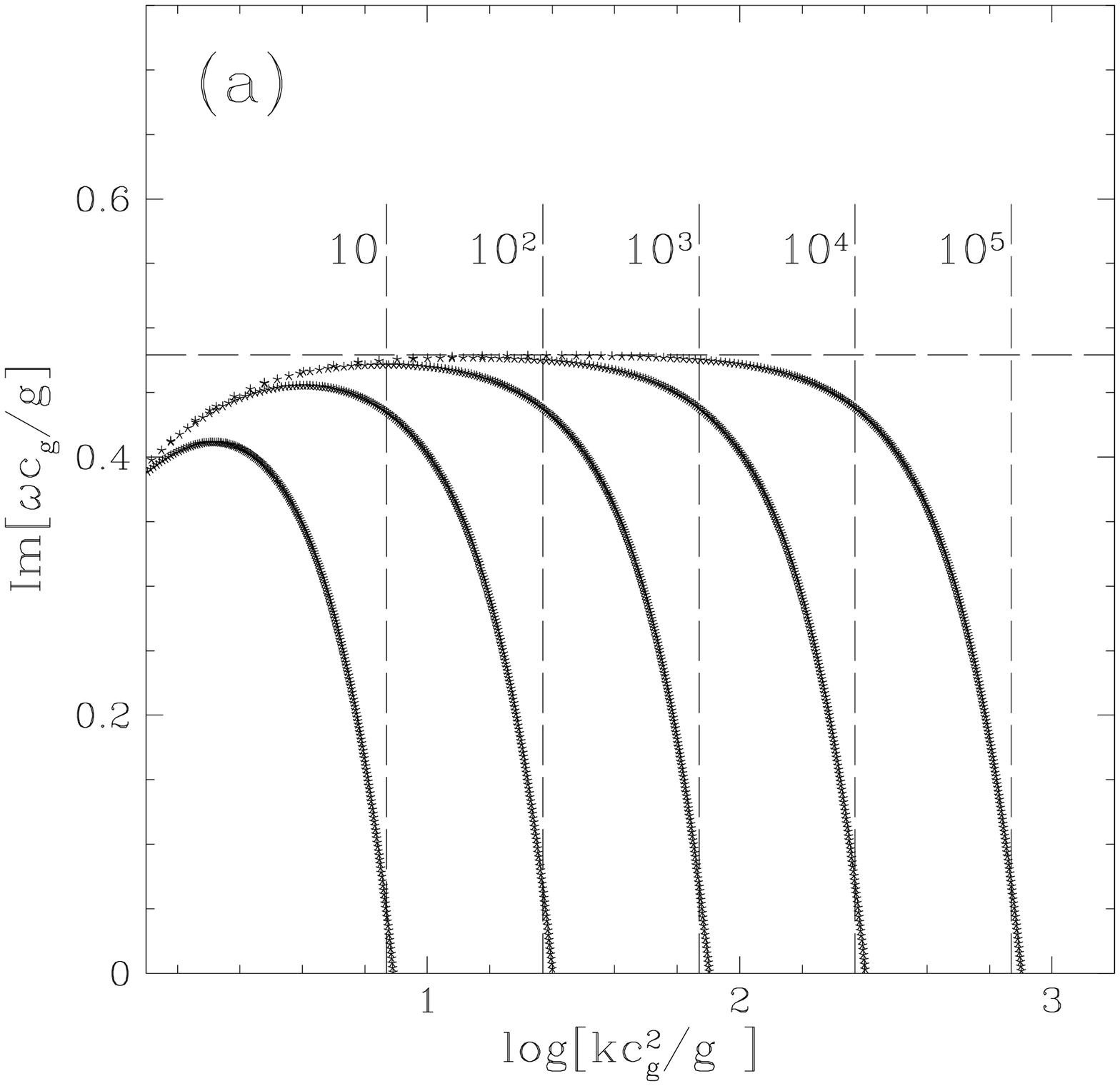}
\vskip0.1truein
\end{figure}

\begin{figure}
\figurenum{2}
\plotone{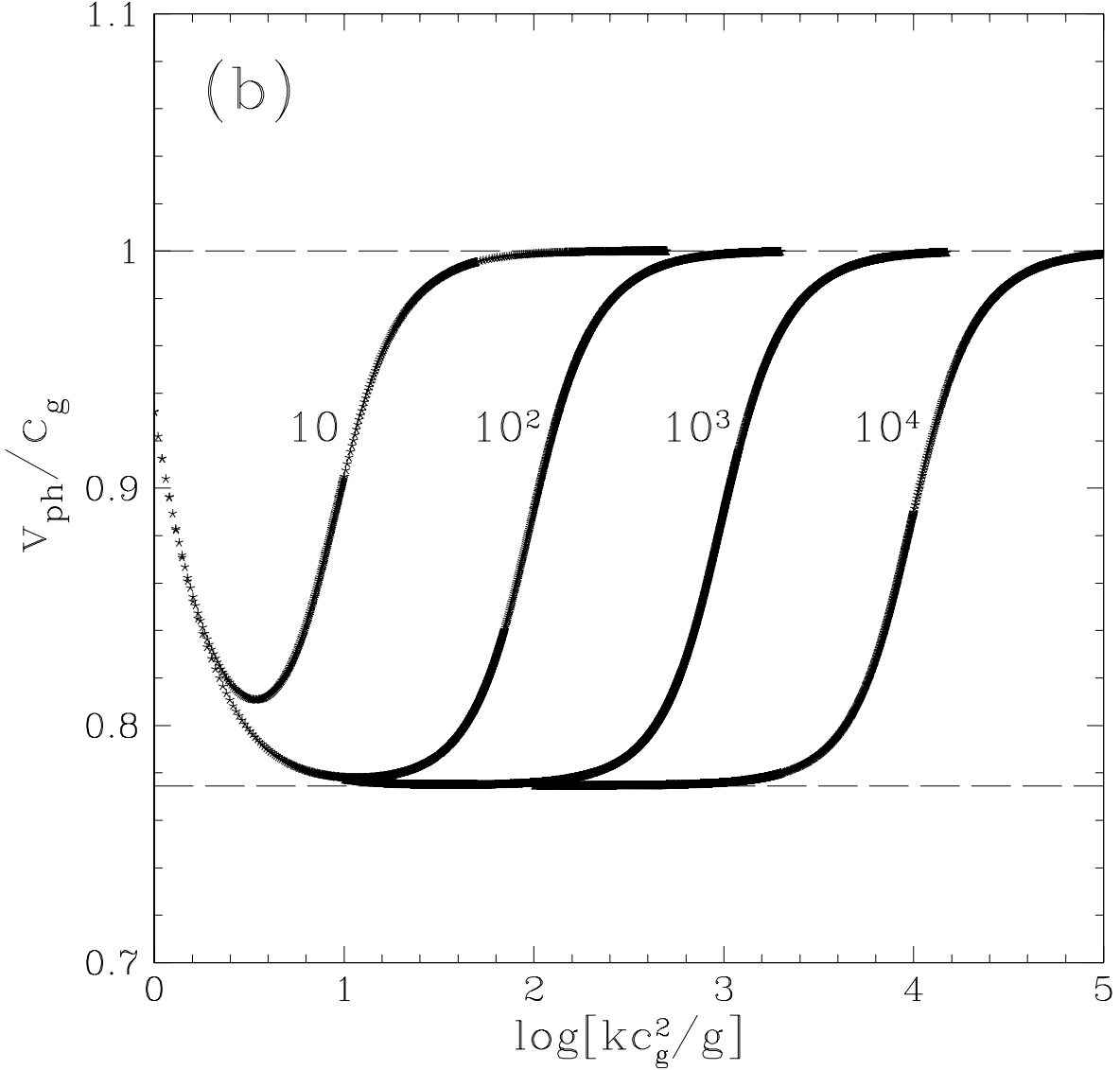}
\vskip0.1truein
\caption{Growth rates (a) and phase speeds (b) for the hydrodynamic
instability at different, high levels of the scaled thermal coupling parameter
$\cgas\omegath/g$, which labels each curve.  Except for $\omegath$, equilibrium parameters are the same as in Figure 1.  The horizontal dashed line in figure (a) indicates
the asymptotic growth rate predicted by equation (\ref{omegalockhydro}), while
the vertical dashed lines indicate the cutoff wavenumber
$(\omegath g|1+\thetarho|/c_{\rm i}^3)^{1/2}$.  The upper and lower
horizontal dashed lines in figure (b) indicate the adiabatic and isothermal
[$c_{\rm i}=(3/5)^{1/2}\cgas\simeq0.77\cgas$] gas sounds speeds,
respectively.}
\end{figure}

\begin{figure}
\figurenum{3}
%\epsscale{0.7}
\plotone{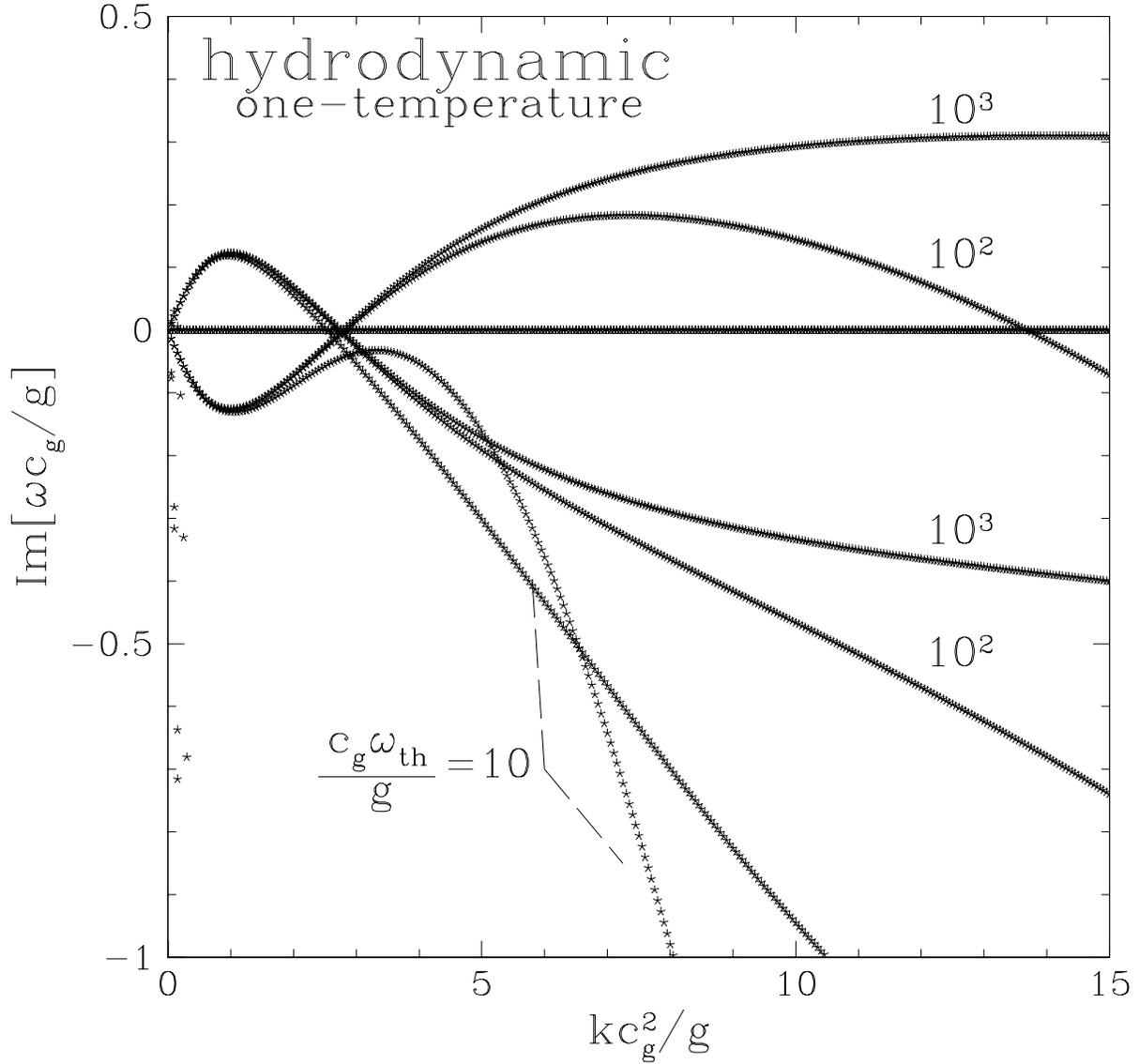}
\vskip0.1truein
\caption{Growth rates for the one-temperature hydrodynamic instability.  
Equilibrium parameters were chosen to be appropriate for an intermediate mass
main sequence stellar envelope.  The equilibrium gas pressure is much larger
than the equilibrium radiation pressure so that $L/L_{\rm Edd}\simeq
10^{-2}$ and $F_z=10\rho c^3_g$.  The flux mean opacity was chosen to be
$\kappa_F=1\,{\rm cm^2\,g^{-1}}$ with $\Theta_{\rho}=1$ and
$\thetatg=-3.5$. The wave vector was chosen to be purely vertical.  
Notice that the growth rates are still dynamical even though the
equilibrium radiation pressure is much smaller than the gas pressure.  
The tight thermal coupling between the gas and radiation elevates the {\it
dynamical} significance of the radiation pressure perturbations, allowing for
relatively large growth rates.  The downward propagating instability at
low $k$ works off of opacity perturbations arising from temperature
fluctuations, while the upward propagating instability at high $k$
is consistent with the one-temperature results of Section 3.2.}
\end{figure}

\begin{figure}   
\figurenum{4}
\epsscale{0.6}
\plotone{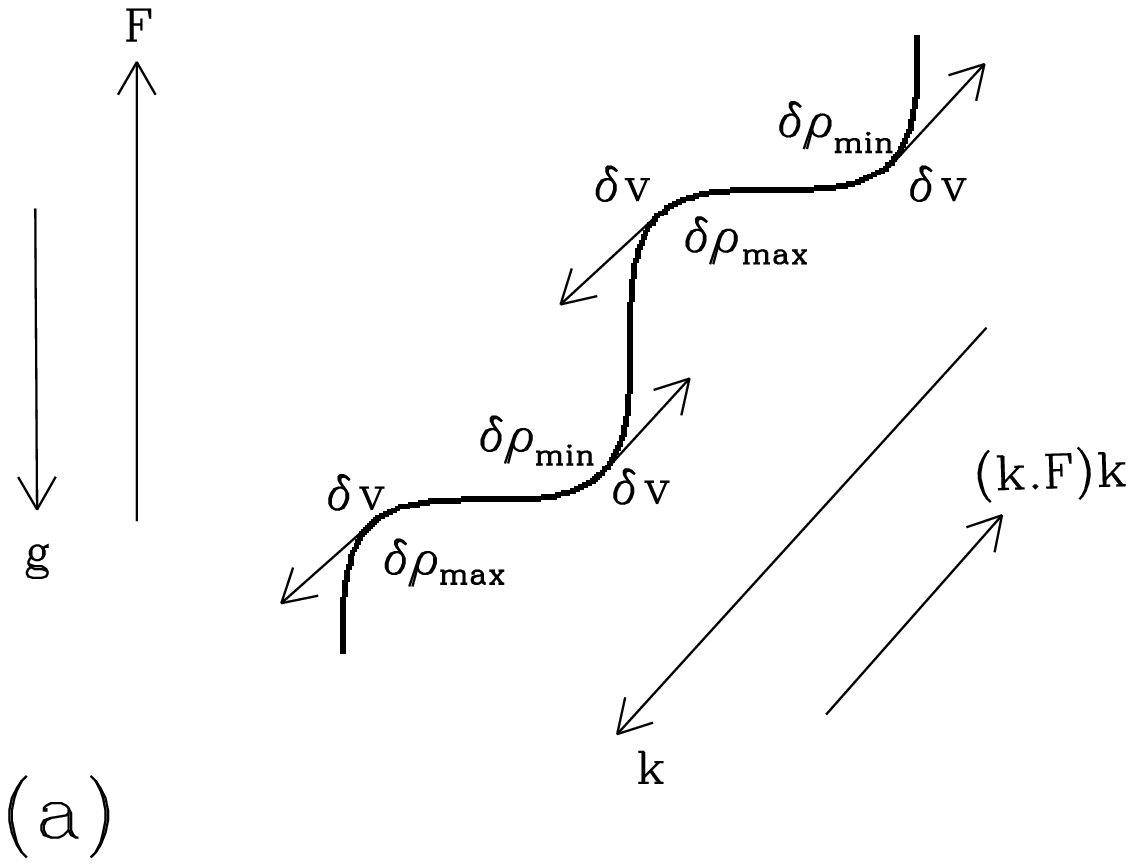}
\end{figure}

\begin{figure}
\figurenum{4}
\epsscale{0.6}
\plotone{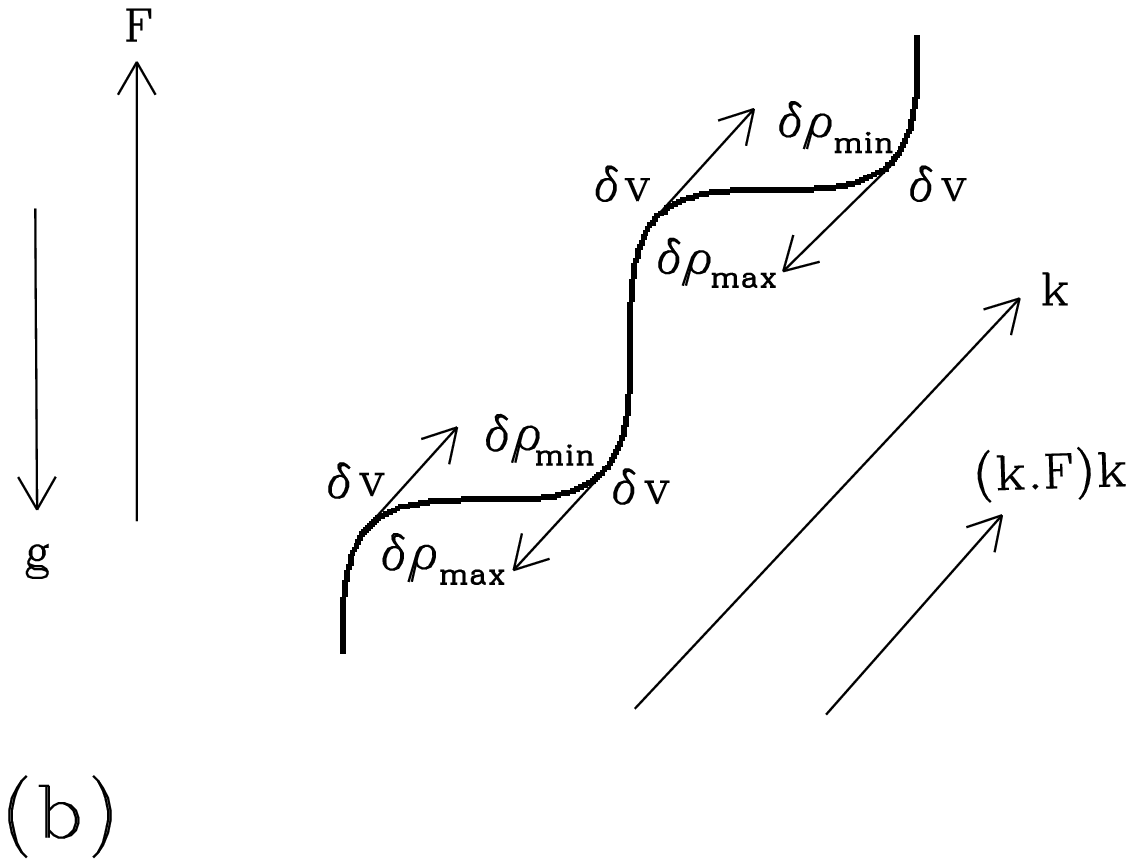}
\vskip0.1truein
\caption{Geometry of radiation pressure force due to density and opacity
fluctuations for (a) downward and (b) upward propagating hydrodynamic
acoustic waves.  Depending
on whether the multiplier of the radiation pressure force is positive or
negative, one direction is damped and the other is unstable, provided the
radiation pressure force is large enough to overcome other forms of damping.}
\end{figure}

\vfill\eject

\begin{figure}   
\figurenum{5}
\epsscale{1.}
\plotone{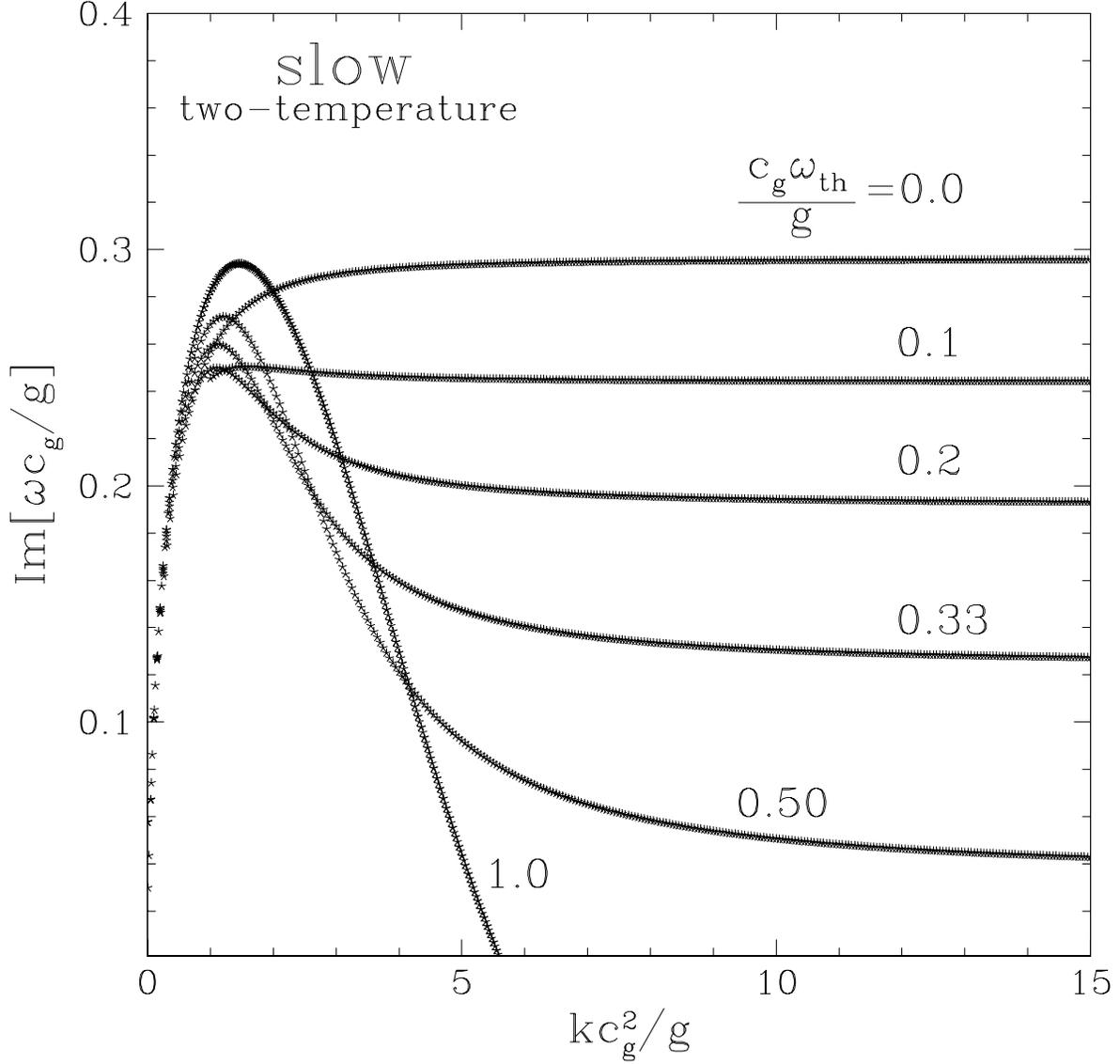}
\vskip0.1truein
\caption{Growth rates for the two-temperature slow MHD instability.
Equilibrium parameters are the same as Figure 1 except for
$\Theta_{\rho}=\thetatg=0$, $k_z/k=0.556$, and a vertical equilibrium
magnetic field with $v_A=5\cgas$.  Again, increasing the 
thermal coupling between the gas and photons leads to damping in this 
two temperature regime.  The fast mode instability does not appear since 
driving from the equilibrium flux does not in this case overcome the
damping produced by radiative diffusion.}
\end{figure}

\vfill\eject

\begin{figure}
\figurenum{6}
\plotone{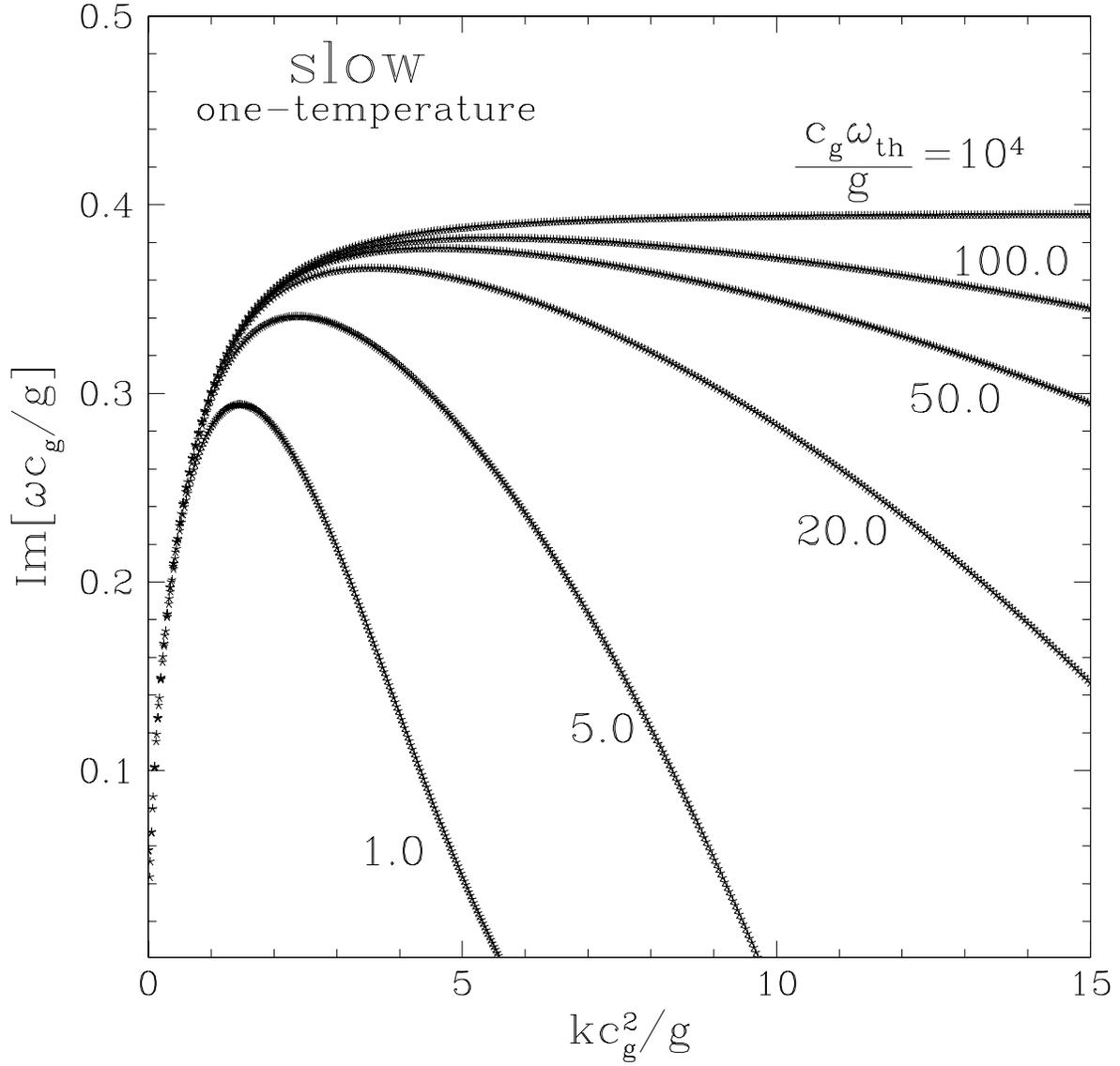}
\vskip0.1truein
\caption{One temperature slow MHD instability for different, high levels of
thermal coupling.  Apart from $\omegath$, parameter values are the same as
Figure 5.}
\end{figure}

\vfill\eject

\begin{figure}
\figurenum{7}
%\epsscale{0.7}
\plotone{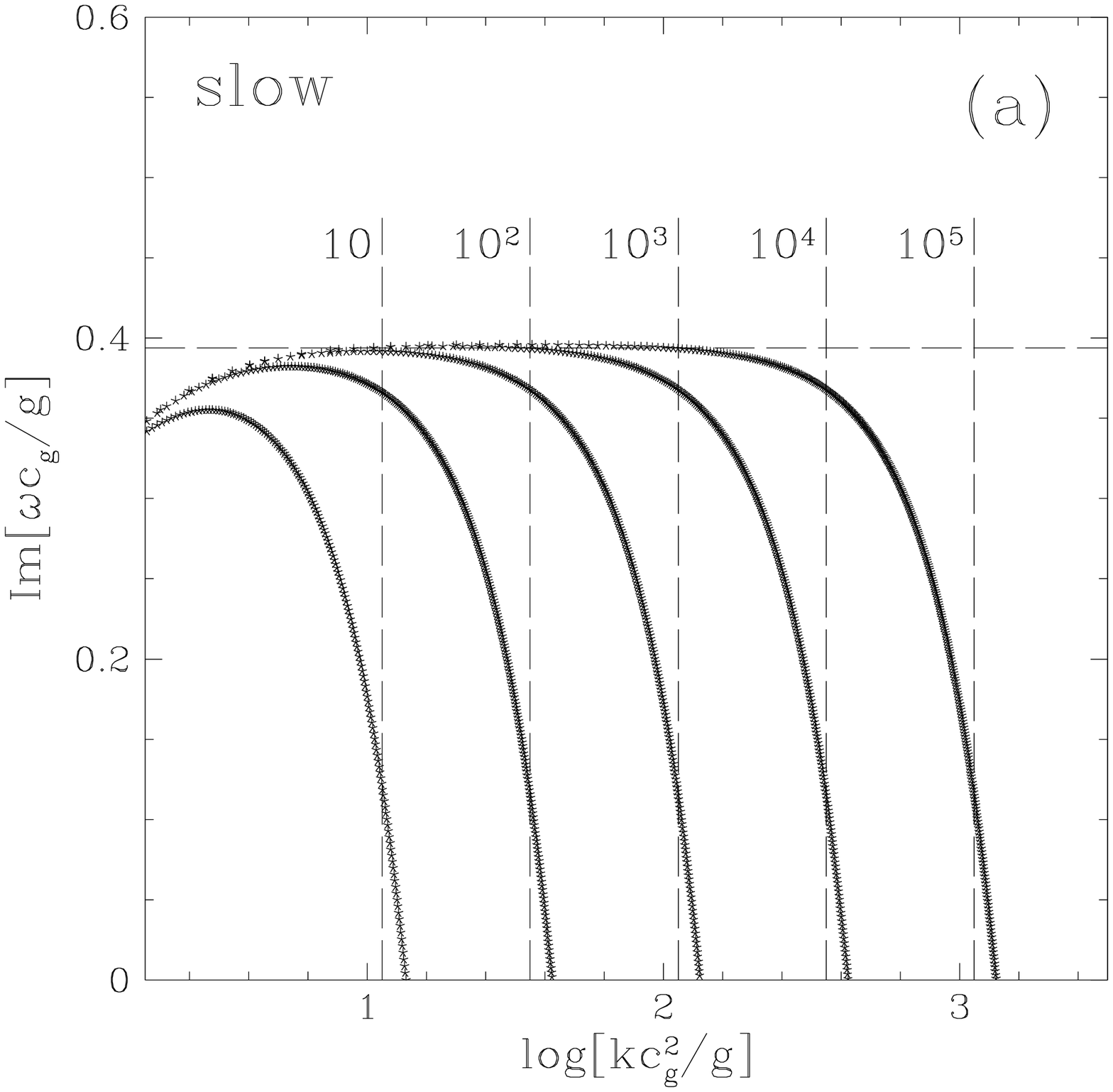}
\end{figure}

\begin{figure}
\figurenum{7}
%\epsscale{0.7}
\plotone{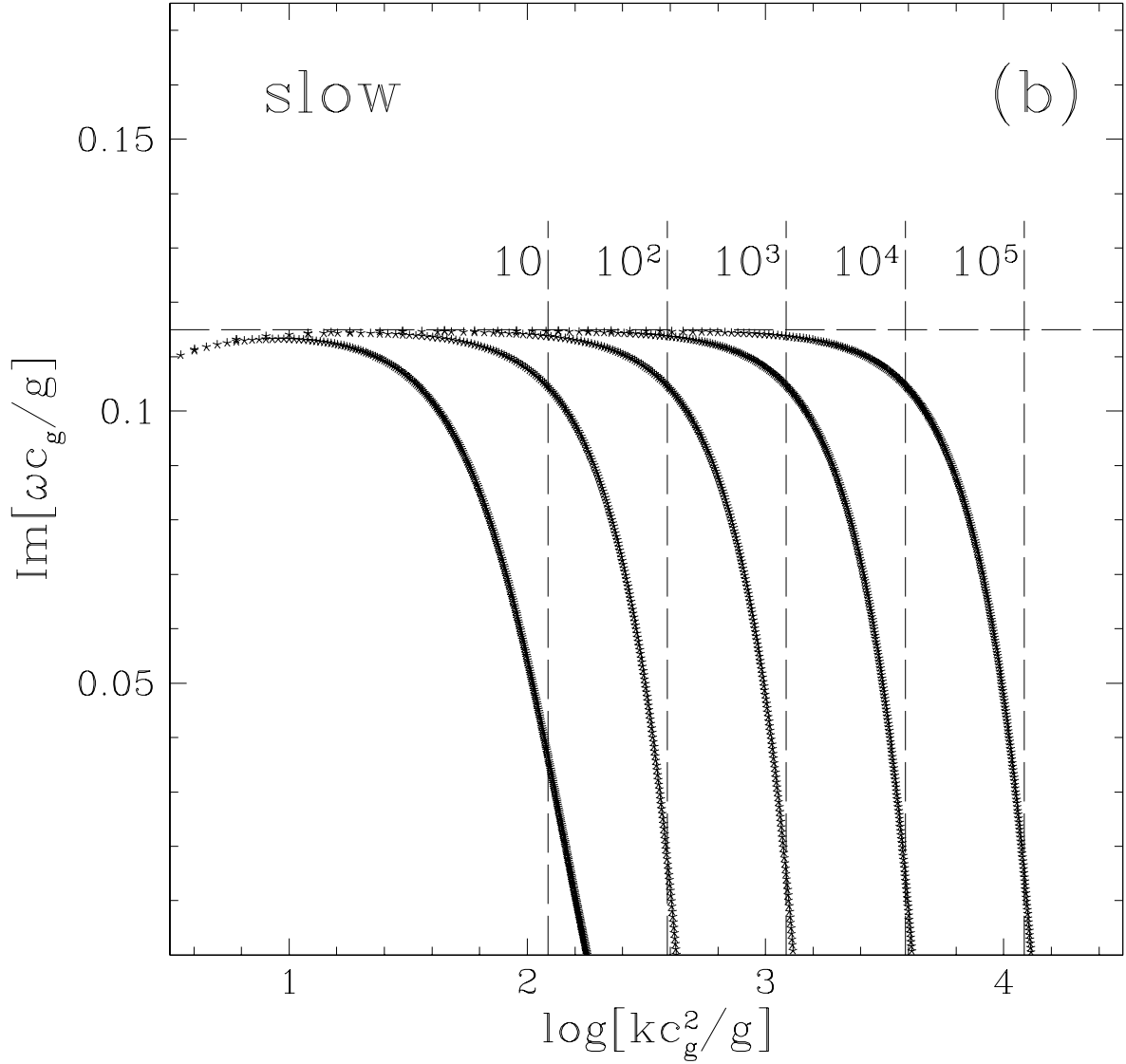}
\caption{One-temperature slow MHD instability with (a) the same parameters
as for Figure 6 except for different values
of $\omegath$, and (b) the same parameters except that a weak equilibrium
magnetic field was chosen with $v_{A}=\cgas/5$.  The horizontal dashed
lines indicate the asymptotic growth rate predicted by equation
(\ref{omegalockmhd}).  The vertical dashed lines indicate the predictions
of our analytic expressions for the cutoff wavenumber in Table 2:  (a)
$(\omegath g/c_{\rm i})^{1/2}/c_{\rm i}$ and (b)
$c_{\rm i}(\omegath g/v_{\rm A})^{1/2}/v_{\rm A}^2$.}
\end{figure}

\vfill\eject

\begin{figure}
\figurenum{8}
\epsscale{0.7}
\plotone{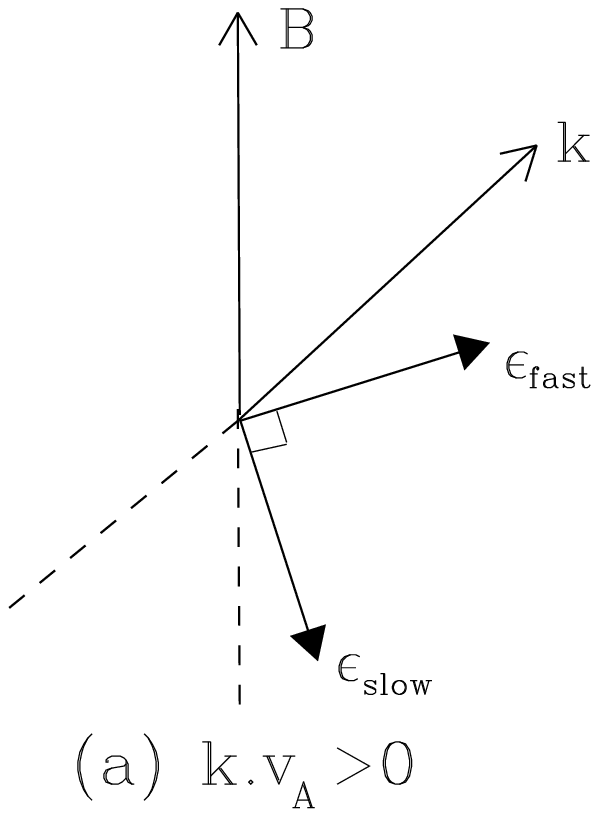}
\end{figure}

\begin{figure}
\figurenum{8}
\plotone{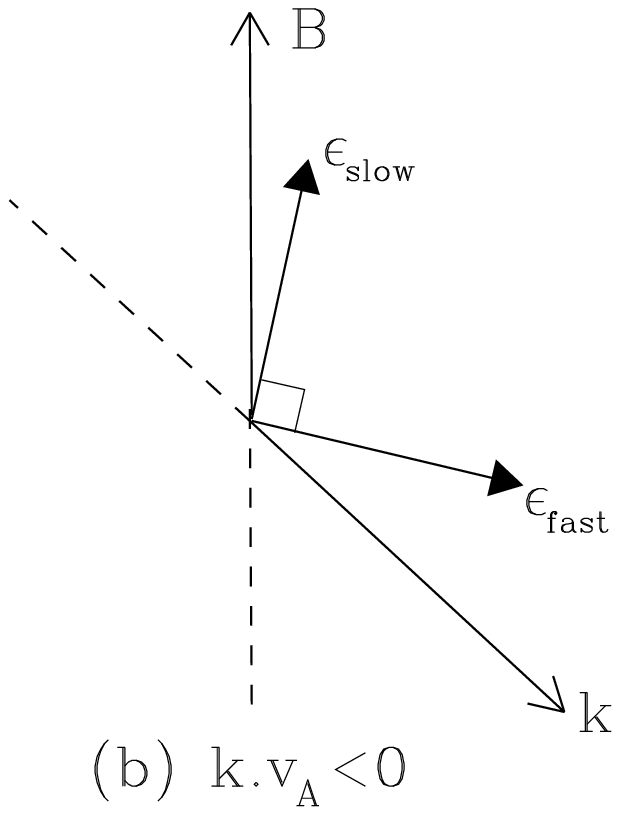}
\vskip0.1truein
\caption{Polarizations of the fast and slow magnetosonic modes.  All vectors
shown lie in the plane of the page.}
\end{figure}

\end{document}